\documentclass{aa}
\usepackage{txfonts}
\usepackage{color, colortbl}
\usepackage[colorlinks, citecolor=blue, linkcolor=blue, urlcolor=blue]{hyperref}

\usepackage{graphicx}
\usepackage{txfonts}
\usepackage{pdflscape}	% Landscape pages
\usepackage{color, colortbl}
\usepackage{dutchcal}
\usepackage{newtxtext,newtxmath}
\usepackage[T1]{fontenc}
\usepackage{ae,aecompl}

\usepackage{pdflscape}	% Landscape pages

\usepackage{listings}
\usepackage{pdflscape}
\usepackage{longtable}
\usepackage{wasysym}
\usepackage{multirow}
\usepackage{threeparttable,booktabs}
\usepackage{graphicx}	% Including figure files
\usepackage{amsmath}	% Advanced maths commands
\usepackage{amssymb}	% Extra maths symbols
\usepackage{tablefootnote}
\usepackage{hyperref}
\usepackage{academicons}
\usepackage{soul}
\usepackage{blindtext}
\usepackage[dvipsnames]{xcolor}

\newcommand{\orcid}[1]{\href{https://orcid.org/#1}{\textcolor[HTML]{A6CE39}{\aiOrcid}}}

\begin{document}

\title{The Galactic Bulge exploration IV.: RR~Lyrae stars as traces of the Galactic bar - 3D \& 5D analysis, extinction variation}

\author{Z.~Prudil\inst{1}, A.~Kunder\inst{2}, L.~{Beraldo~e~Silva}\inst{3}, S.~Gough-Kelly\inst{4}, M.~Rejkuba\inst{1}, S.~R.~Anderson\inst{4}, V.~P.~Debattista\inst{4}, O.~Gerhard\inst{5}, R.~M.~Rich\inst{6}, D.~M.~Nataf\inst{7}, A.~J.~Koch-Hansen\inst{8}, A. Savino\inst{9},  I. D\'ek\'any\inst{8}}

% Victor P. Debattista
% R. M. Rich

\institute{
European Southern Observatory, Karl-Schwarzschild-Strasse 2, 85748 Garching bei M\"{u}nchen, Germany; \email{Zdenek.Prudil@eso.org}
\and Saint Martin's University, 5000 Abbey Way SE, Lacey, WA, 98503 
\and Department of Astronomy \& Steward Observatory, University of Arizona, Tucson, AZ, 85721, USA
\and Jeremiah Horrocks Institute, University of Central Lancashire, Preston PR1 2HE, UK
\and Max-Planck Institut f\"ur extraterrestrische Physik, Giessenbachstra\ss e, 85748 Garching, Germany
\and Department of Physics and Astronomy, UCLA, 430 Portola Plaza, Box 951547, Los Angeles, CA 90095-1547, USA
\and Department of Physics and Astronomy, The Johns Hopkins University, Baltimore, MD 21218
\and Astronomisches Rechen-Institut, Zentrum f{\"u}r Astronomie der Universit{\"a}t Heidelberg, M{\"o}nchhofstr. 12-14, D-69120 Heidelberg, Germany
\and Department of Astronomy, University of California, Berkeley, Berkeley, CA, 94720, USA
}

\date{\today}

\abstract
{RR~Lyrae stars towards the Galactic bulge are used to investigate whether this old stellar population traces the Galactic bar. Although the bar is known to dominate the mass in the inner Galaxy, there is no consensus on whether the RR~Lyrae star population, which constitutes some of the most ancient stars in the bulge and thus traces the earliest epochs of star formation, contributes to the barred bulge. We create new reddening maps and derive new extinction laws from visual to near-infrared passbands using improved RR~Lyrae period-absolute magnitude-metallicity relations, thus enabling distance estimates for individual bulge RR~Lyrae variables. The extinction law is most uniform in $R_{IK_{\rm s}}$ and $R_{JK_{\rm s}}$ and the distances to individual RR~Lyrae based on these colors are determined with an accuracy six and four percent, respectively. Using only the near-infrared passbands for distance estimation we inferred the distance to the Galactic center equal to $d_{JK_{\rm s}}^{\rm cen} = 8217 \pm 1({\rm stat}) \pm 528({\rm sys})$\,pc after geometrical correction. We show that variations in the extinction law toward the Galactic bulge can mimic a barred spatial distribution in the bulge RR~Lyrae star population in visual passbands. This arises from a gradient in extinction differences along Galactic longitudes and latitudes, which can create the perception of the Galactic bar, particularly when using visual passband-based distances. A barred angle in the RR~Lyrae spatial distribution disappears when near-infrared passband-based distances are used, as well as when reddening law variations are incorporated in visual passband-based distances. The prominence of the bar, as traced by RR~Lyrae stars, depends on their metallicity, with metal-poor RR~Lyrae stars ([Fe/H]$< -1.0$~dex) showing little to no tilt with respect to the bar. Metal-rich ([Fe/H]$> -1.0$~dex) RR~Lyrae stars do show a barred/bulge signature in spatial properties derived using near-infrared distances, with an angle $\iota = 18 \pm 5$\,deg, consistent with previous bar measurements from the literature. This also hints at a younger age for this RR~Lyrae subgroup. The 5D kinematic analysis, primarily based on transverse velocities, indicates a rotational lag in RR~Lyrae stars compared to red clump giants. Despite variations in the extinction law, our kinematic conclusions are robust across different distance estimation methods.}

\titlerunning{The bar/bulge traced by RRLs}
\authorrunning{Prudil et al.}
\maketitle

%-------------------------------------------------------------------
\section{Introduction} \label{sec:Intro}

The three-dimensional structure of the Galactic bulge is important for our understanding of the formation and evolution of the Milky Way \citep[e.g.,][]{Blitz1991,Zoccali2016,Barbuy2018}. The structure(s) residing in the inner parts of the Galaxy motivate models to explain these features, following the evolution of the Galaxy over time to reproduce these observed structures \citep[e.g.,][]{Weiland1994,Cao2013,Lim2021,Wylie2022,Khoperskov2023}. By far the most prominent structure in the inner Galaxy is the bar, which is most easily seen and mapped by using red clump (RC) stars as standard candles \citep[e.g.,][]{Stanek1997,Nishiyama2005,Wegg2013,Simion2017,Johnson2022}, but also seen in a number of other studies, such as contour surface brightness maps \citep{Blitz1991,Dwek1995} and from the distribution of the Galactic Center gas \citep[e.g.,][]{Binney1991,Fux1999,Rodriguez2008,Li2022}.

Numerical simulations of disk galaxies have shown that bars and their vertical extensions (bulges) are formed via orbital resonance and instabilities in massive early disks \citep{Combes1981,Combes1990,Raha1991,Merritt1994,Quillen2002,MartinezValpuesta2004,Debattista2004,Debattista2006,Smirnov2019,Sellwood2020}. The vertical metallicity gradient observed in the MW's bar/bulge have also been reproduced in these models \citep{MartinezValpuesta2013,Athanassoula2017,Debattista2017,Fragkoudi2018,Fragkoudi2020,Buck2019}. Galaxies with barred bulges are typically formed secularly out of the disk. 

RC stars map the distribution of the variety of stellar ages of the inner Galaxy stars, and so probe a mix of stellar populations, which may originate from different initial environments \citep[e.g.,][]{Wegg2015,Gonzalez2015,Rojas-Arriagada2020}. Therefore, RR~Lyrae stars are commonly used to focus on the bulge's oldest populations. These are high-amplitude, classical pulsators situated on the horizontal-branch and are considered to belong to exclusively old populations \citep[ages above $10$\,Gyr,][]{Walker1991,Catelan2015,Savino2020}\footnote{Although alternative channels of RR~Lyrae formation has been proposed see \citet{Bobrick2024}.}. Their pulsation characteristics and the shapes of their light curves are used to estimate their luminosities \citep[using period-metallicity-luminosity relations; see][]{Bono2003,Catelan2004,Marconi2015} and photometric metallicities \citep[refer to][]{Jurcsik1996,Sandage2004,Smolec2005}. These features render them crucial for research focusing on the spatial and kinematic properties of the Milky Way \citep[MW, refer to][]{Fiorentino2015,Medina2018,Wegg2019,Prudil2021Orphan,Prudil2022,Ablimit2022} and its neighborhood \citep[see][]{Sarajedini2009,Fiorentino2012M32,MartinezVazquez2015}.

RR~Lyrae stars are classified into subgroups based on their pulsation modes. The predominant group, the RRab subclass, consists of stars pulsating in the fundamental mode. These stars generally exhibit high amplitude variability and asymmetric light curves. The second most populous subclass encompasses the first-overtone pulsators, known as RRc-type stars, is characterized by more symmetric light curves and shorter pulsation periods. The third and rarest subclass includes the double-mode pulsators, RRd, which simultaneously pulsate in both the fundamental and first-overtone modes. In the Milky Way, these three subclasses account for approximately 66, 33, and 1\% of all RR~Lyrae, respectively \citep{Clementini2023}.

Thanks to extensive, long-term photometric surveys targeting the Galactic bulge region, such as the Massive Compact Halo Objects \citep[MACHO,][]{Alcock1998}, the Optical Gravitational Lensing Experiment \citep[OGLE,][]{Udalski2015}, and the Vista Variables in the V\'ia L\'actea survey \citep[VVV,][]{Minniti2010}, a highly comprehensive dataset of RR~Lyrae stars in the direction of the Galactic bulge has been assembled \citep[with completeness above $90\%$,][]{Soszynski2014BulgeRRlyr,Soszynski2023dScutiLMC}. This wealth of data has enabled extensive studies on the spatial distribution and reddening of the Galactic bulge using RR~Lyrae variables \citep[see][]{Kunder2008,Dekany2013,Pietrukowicz2015}.

Surprisingly, it is not clear to what extent the RR Lyrae stars trace the bar, which is in stark contrast to the many other tracers that have confirmed a dominant barred structure. Instead, there is a prevailing discrepancy in the interpretation of distances: distances derived primarily from visual bands suggest that bulge RR~Lyrae stars trace the Milky Way's bar \citep[and therefore are part of the barred bulge morphology,][]{Pietrukowicz2015,Du2020}, while estimates predominantly from infrared passbands imply a unbarred RR~Lyrae population in the Galactic bulge \citep[therefore suggesting a classical bulge morphology,][]{Dekany2013,Prudil2019OOspat,Prudil2019Kin}.

Besides the RR~Lyrae stars, there is also debate on whether the Mira population is also not a clear tracer of the barred bulge. Some studies show Miras trace a bar \citep{Matsunaga2005}, while others see a bar only in young metal-rich (long-period) Miras, but not in the eldest (short-period) Miras, which are comparable in age to the RR~Lyraes \citep{Catchpole2016,Qin2018,Grady2019,Grady2020}.

The relationship between RR~Lyrae variables and the bar has also been extensively investigated through kinematic studies. For instance, radial velocities have been examined in the Bulge Radial Velocity Assay for RR~Lyrae stars \citep[BRAVA-RR,][]{Kunder2016,Kunder2020}, and transverse velocities have been assessed by \citet{Du2020} using proper motions provided by the \textit{Gaia} space mission \citep{Gaia2016,Lindegren2018}. These kinematic approaches have demonstrated that the RR~Lyrae population in the bulge rotates more slowly compared to the majority of bulge giants \citep[refer to][]{Kunder2012,Ness2013IV,Zoccali2017,Sanders2019}. 

One aspect, however, seems clear: a portion of the RR~Lyrae stars in the direction of the Galactic bulge are likely interlopers from other regions of the Milky Way. These stars contribute to, and likely increase, the observed RR~Lyrae velocity dispersion.

There have been several recent improvements in using inner Galaxy RR~Lyrae stars to trace the bulge's structure. First, a common metallicity RR~Lyrae scale defined by \citet{For2011chem}, \citet{Chadid2017}, \citet{Sneden2017}, and \citet{Crestani2021} allows both local and bulge RR~Lyrae stars to be placed on a common $\rm [Fe/H]$ scale \citep{Crestani2021Alpha,Dekany2021}. This is important since [Fe/H] does affect the absolute magnitudes of RR~Lyrae stars, especially in the optical wavelengths. Second, the \textit{Gaia} astrometric mission provides parallaxes of $915$ DR3 RRab stars with parallax uncertainties less than $10$\% \citep{Gaia2016DR1,Clementini2023}. This has led to improved RR~Lyrae absolute magnitude calibrations, which can also be used for bulge RR~Lyrae stars \citep[e.g.,][]{Bhardwaj2023,Prudil2024}. Lastly, the VVV survey has begun releasing photometry that can be used to also study the bulge RR~Lyrae stars \citep[e.g.,][]{Dekany2020,Molnar2022}. The infrared (IR) passbands of VVV are less sensitive to the reddening and extinction in the bulge and, therefore, can be used to compare extinctions based on optical photometry alone to those seen from the longer IR passbands. We capitalize on these improvements to carry out both a 3-dimensional (3D) distance analysis as well as a 5-dimensional (5D) distance and proper motion analysis on the inner Galaxy RR~Lyrae star population.

We present the fourth paper of our series focused on RR~Lyrae stars and the Galactic bulge \citep{Prudil2024,Prudil2024GBEXII,Kunder2024GBX}. Our study is structured as follows. In Section~\ref{sec:data-sets}, we present the compiled photometric and astrometric datasets. Section~\ref{sec:DistancesAll} details our methodology for estimating the reddening, determining the reddening law, and subsequently calculating distances to individual RR~Lyrae stars. The following Section~\ref{sec:RedMapsComp} compares the reddening maps obtained in our study with those existing in the literature. Section~\ref{sec:bulge3D} examines the spatial distribution of single-mode RR~Lyrae stars in the direction of the Galactic bulge and their spatial association with the bar. In Section~\ref{sec:5DBulge}, we analyze the rotation of RR~Lyrae stars in the Galactic bulge using transverse velocities. Section~\ref{sec:Discusion} we discuss our results. Finally, Section~\ref{sec:Summary} summarizes our findings.

% ----------------------- ############# -----------------------
\section{Astro-photometric data set} \label{sec:data-sets}

\begin{figure}
\includegraphics[width=\columnwidth]{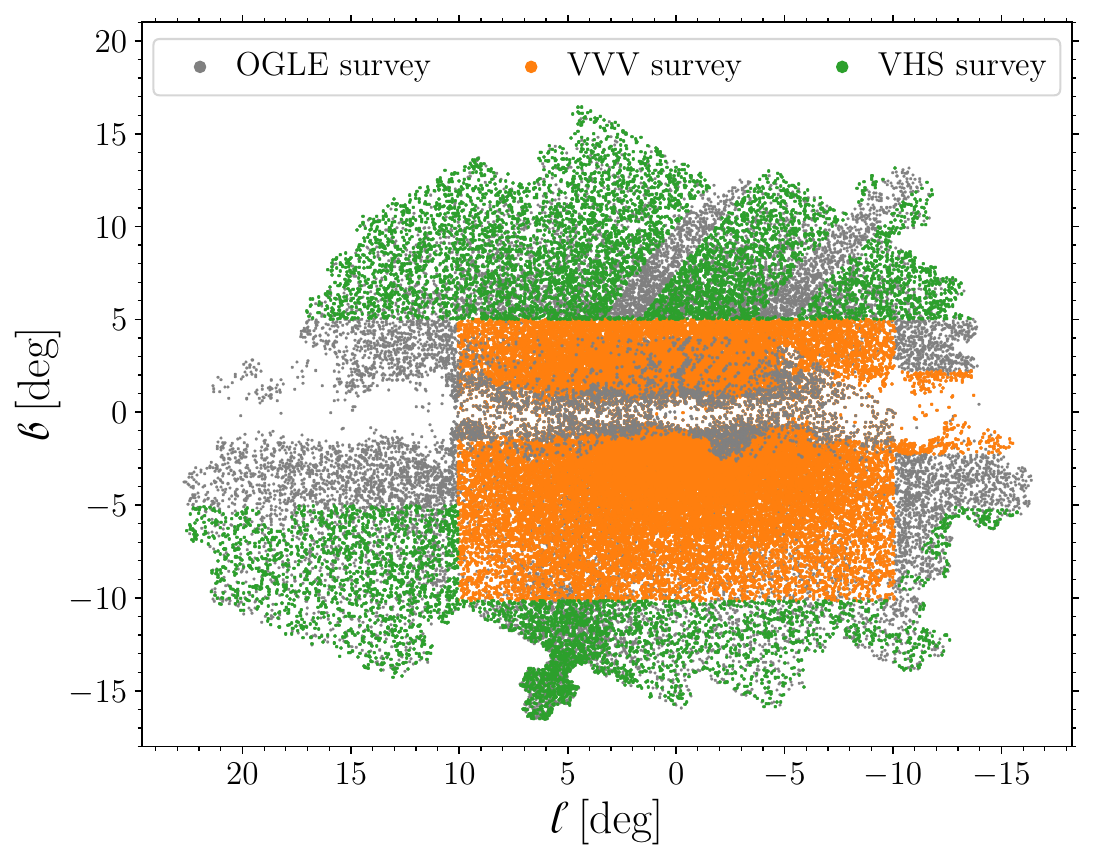}
\caption{The spatial distribution of our entire dataset. The grey points represent RR~Lyrae stars identified by the OGLE survey and in a study by \citet{Dekany2020}. Green points represent OGLE-identified RR~Lyrae stars with VHS photometry. The orange points mark RR~Lyrae stars classified by OGLE with the VVV mean intensity magnitudes from \citet{Molnar2022}.}
\label{fig:MapOfTheSample}
\end{figure}

Our study's data set is derived from four key photometric surveys encompassing the Galactic bulge: Optical Gravitational Lensing Experiment \citep{Udalski2015}, the Vista Variables in the V\'ia L\'actea survey \citep{Minniti2010}, the Vista Hemisphere Survey \citep[VHS,][]{McMahon2013VHS}, and the latest \textit{Gaia} data release \citep[DR3,][]{Gaia2016,GaiaDR32023}. These surveys provide both visual (OGLE and \textit{Gaia}) and near-infrared (VVV and VHS) photometry for a substantial number of RR~Lyrae stars toward the Galactic bulge. We emphasize that each of the above-mentioned surveys contributed mainly by the mean intensity magnitudes. The classification of RR~Lyrae variables was primarily based on the OGLE survey. The other RR~Lyrae catalogs were added as an intersection (X-match) with OGLE data. The wide range in wavelength of the selected surveys and improved procedures to estimate absolute magnitudes of individual RR~Lyrae stars open the possibility of countering the severe reddening toward the Galactic bulge and accurately estimating distances of individual variables in our data set. The summary of used data products from individual surveys is listed in Table~\ref{tab:SummaryUsedData}, our entire data set is displayed in Figure~\ref{fig:MapOfTheSample}, and in total consists of $72165$ single mode (RRab and RRc) RR~Lyrae variables.

\subsection{OGLE-IV photometry} \label{subsec:OGLEphot}

The fourth data release of OGLE \citep{Soszynski2014BulgeRRlyr,Soszynski2019Disk} presents abundant photometry (with some stars having more than $15000$ observations) in $V$ and $I$-passbands for more than $60000$ reliably classified RR~Lyrae stars ($68127$, in total used in this study). They provide the basic information on the variability, including mean apparent magnitudes in the $I$-band, ephemerides (pulsation period $P$, time of maximum brightness $M_{\rm 0}$), the amplitude of light changes ($\text{Amp}_{I}$) and some of the Fourier coefficients derived from the photometric light curves ($R_{21}$, $R_{31}$, $\varphi_{21}$, $\varphi_{31}$)\footnote{The aforementioned coefficients can be described by the following equations $R_{i1} = A_{i} / A_{1}$ and $\varphi_{i1} = \varphi_{i} - i\varphi_{1}$, where the $R_{i1}$ represent the amplitude ratio, and the $\varphi_{i1}$ represent differences in phase.}. The Fourier coefficients and pulsation periods can be used to estimate the photometric metallicities of individual RR~Lyrae stars \citep[e.g.,][]{Jurcsik1996,Smolec2005,Dekany2021,Mullen2022} and consequently absolute magnitudes. 

While the OGLE survey provides extensive data, it does not include uncertainties for these values, except for pulsation periods. Consequently, to accurately estimate photometric metallicities \citep[using relations from][]{Dekany2021} and absolute magnitudes, we recalculated the Fourier coefficients to obtain amplitudes $A_{1}$ and $A_{2}$ and phase differences $\varphi_{31}$ and their uncertainties. 

We decided to derive the mean intensity $I$-band magnitudes and Fourier coefficients based on an approach described in \citet{Petersen1986}. We optimized the following Fourier light curve decomposition:
\begin{equation} \label{eq:FourSer}
m\left ( t \right ) = m_{I} + \sum_{k=1}^{n} A_{k} \cdot \cos \left (2\pi k \vartheta + \varphi_{k} \right ) \\.
\end{equation}
In equation~\ref{eq:FourSer}, $m_{I}$ represents the mean intensity magnitudes, $A_{k}$ and $\varphi_{k}$ stand for amplitudes and phases. The $n$ denotes the degree of the fit which we adapted for each light curve. We use the same approach as in \citet{Prudil2019OOspat}, with the minor modification that we enforced the fourth degree ($n=4$) as a minimum (to obtain Fourier coefficients and their covariances for $\varphi_{31}$). Due to the sparsity of the data, this condition was not applied to the $V$-band. The $\vartheta$ represents the phase function defined as:
\begin{equation} \label{eq:phasing}
\vartheta = \left(\text{HJD}-M_{0}\right)/P \\,
\end{equation}
where HJD is a Heliocentric Julian Date of the observation and $M_{0}$ stands for the time of brightness maximum. The calculated Fourier coefficients and their uncertainties subsequently were used in the estimate of the photometric metallicities ([Fe/H]$_{\rm phot}$) using procedures described in \citet{Dekany2021}\footnote{The photometric metallicities are aligned with the metallicity scale introduced by \citet{Crestani2021}. This scale is based on the studies used in its development, including \citet{For2011chem}, \citet{Chadid2017}, \citet{Sneden2017}, and \citet{Crestani2021}.}. Based on the Fourier decomposition, we also obtained uncertainties on the mean intensity magnitudes, $\sigma_{m_{I}}$, that we subsequently used in distance determination. 

Furthermore, we also performed Fourier light curve decomposition on the less abundant data for the $V$-band and obtained mean intensity magnitudes, $m_{V}$ with their associated uncertainties. The newly derived $m_{V}$ and $m_{I}$ and their associated uncertainties do not significantly differ from those provided by OGLE. Mean intensity magnitudes from multiple passbands for RR~Lyrae stars can serve as a reddening indicator and thus improve distance estimations \citep[e.g.,][]{Kunder2010,Haschke2011}. Finally, we emphasize that OGLE-IV served as the main source of RR~Lyrae identification in this work.

\subsection{Gaia photometry} \label{subsec:Gaia}

To obtain additional colors for distance estimation, we cross-matched our OGLE-identified single mode RR~Lyrae stars (RRab and RRc) with the RR~Lyrae variables identified in the third data release of the \textit{Gaia} catalog \citep[\textit{Gaia} DR3,][]{Clementini2023}. Thus, we obtained \texttt{source\_id}'s together with photometric (mostly between $25$ to $45$ observations in $G$-band) and astrometric data for most of our RR~Lyrae data set (approximately $80$ percent). This catalog also contains information on pulsation properties of identified RR~Lyrae pulsators, but only the peak-to-peak magnitude in the $G$-band is used in our analysis (see Table~1). We decided to keep OGLE pulsation properties as they are derived from a larger number of observations compared to \textit{Gaia}.

For the remaining RR~Lyrae variables identified by OGLE without a \textit{Gaia} RR~Lyrae counterpart ($\approx 20$ percent), we \text{cross-matched} the \texttt{gaia\_source} data set \citep{GaiaDR32023} to obtain their associated mean flux magnitudes ($m_{G_{\rm BP}}$), proper motions ($\mu_{\alpha}^{\ast}$ and $\mu_{\delta}$), and their associated uncertainties and correlations (in case of proper motions). We note that the \texttt{gaia\_source} catalog does not provide an error on the mean magnitudes due to asymmetric error distribution in magnitude space, and only uncertainties on flux are provided. We converted their fluxes to magnitude space since we do not use the unidentified RR~Lyrae stars in the \textit{Gaia} catalog in the reddening estimation. We assigned them an uncertainty based on an error on the flux and variation in the zero-point\footnote{See Table~5.4 in \url{https://gea.esac.esa.int/archive/documentation/GDR3/Data_processing/chap_cu5pho/cu5pho_sec_photProc/cu5pho_ssec_photCal.html}.}. In addition to photometric and astrometric properties, we also obtained some of the \textit{Gaia} flags on the quality of photometry and astrometry, namely re-normalized unit weight error (RUWE\footnote{The RUWE parameter estimates the quality of the $Gaia$ astrometric solution.}) and \texttt{ipd\_frac\_multi\_peak} that refers to the detection of a double peak in image processing of a given object, possibly identifying sources of binarity or blending \citep{Gaia2016,GaiaDR32023}.

\subsection{VVV photometry} \label{subsec:VVVphot}

We also collected infrared photometry for bulge RR~Lyrae stars from the VVV survey. VVV provides observations in the near-infrared passbands ($Z$, $Y$, $J$, $H$, and $K_{\rm s}$), where the most numerous are observations in the $K_{\rm s}$-band. VVV provides aperture photometry with five different apertures for each object (with, on average, $172$ observations in $K_{\rm s}$-band). Therefore, it is up to the user to select an appropriate aperture. We collected individual observations provided in the photometric catalogs by the Cambridge Astronomy Survey Unit (CASU). As shown by \citet{Hajdu2020}, the VVV survey exhibits some issues with photometric zero-point calibration. Thus, before we proceeded with our analysis, we recalibrated the obtained $K_{\rm s}$ and $J$-band photometry using a procedure described in \citet{Hajdu2020} and implemented in the \texttt{correct\_vvv\_zp} code\footnote{The module can be found here: \url{https://github.com/idekany/correct_vvv_zp}.}.

\begin{figure}
\includegraphics[width=\columnwidth]{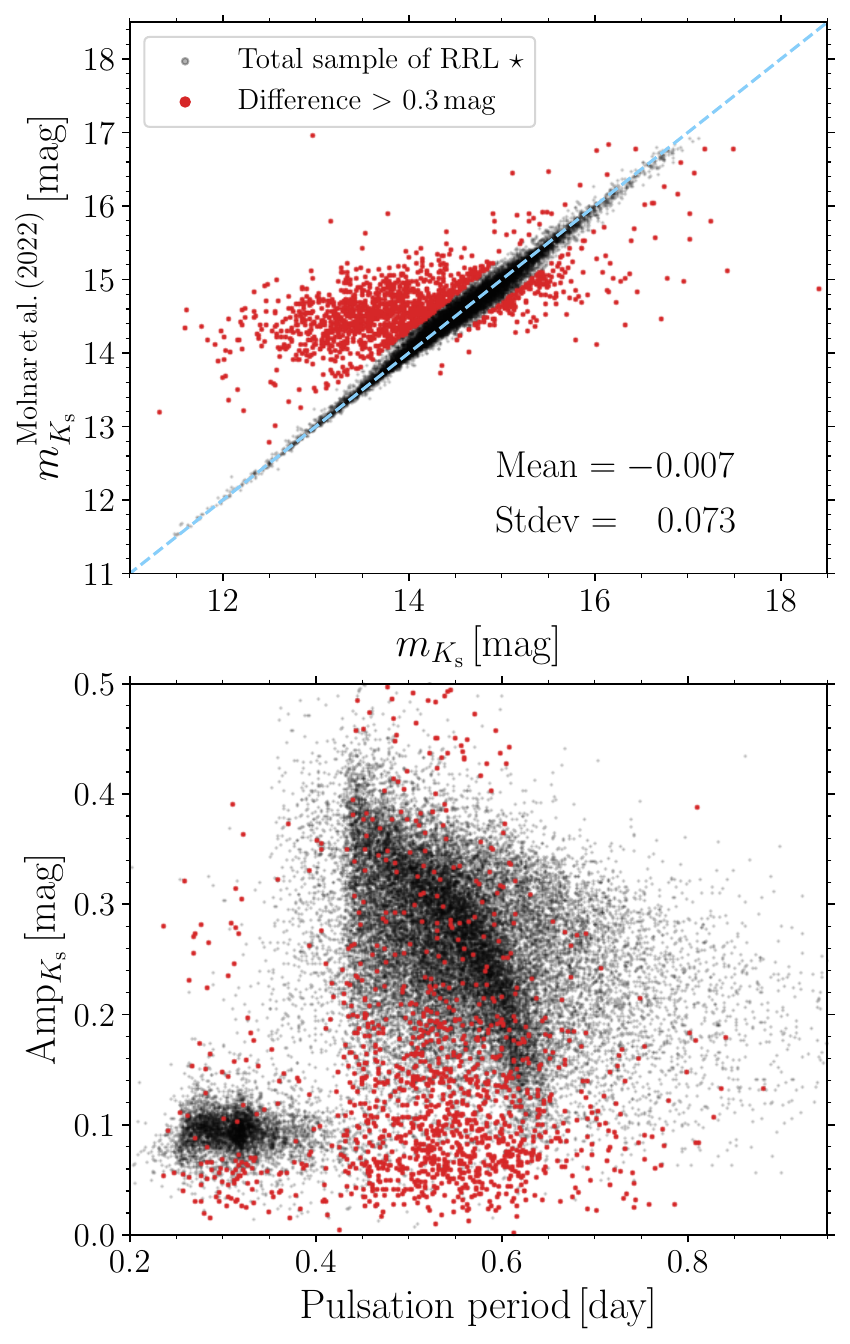}
\caption{Comparison of matched RR~Lyrae stars in mean intensity magnitude space (top panel) and period-amplitude plane (bottom panel). The grey dots in both panels represent the total cross-matched sample \citep[between our aperture VVV data and sample provided in][]{Molnar2022}, and red points stand for stars where the absolute difference between both mean intensity magnitudes exceeded $0.3$\,mag. To assess both photometric sources, we listed the mean and standard deviation between both samples (with discrepant RR~Lyrae stars removed). The blue line in the top panel represents the identity line.}
\label{fig:ComparisonVVV}
\end{figure}

To select the appropriate aperture, we proceeded as follows: we performed Fourier decomposition of the $K_{\rm s}$-band observations using ephemerides and variability identification from the OGLE survey and measured the signal-to-noise ratio, SNR, as a cost function and estimated mean intensity magnitudes for individual apertures \citep[see][for details]{Dekany2021}. We defined the SNR using the following equation:
\begin{equation}
\text{SNR} = \frac{\text{Amp}_{K_{\rm s}} \cdot \sqrt{\text{N\_data}}}{\text{RMSE}} \\,
\end{equation}
where $\text{Amp}_{K_{\rm s}}$ represents total amplitude (peak-to-peak, between $0.1$ to $0.5$\,mag) of the light curve, $\text{N\_data}$ is the number of observations, and \text{RMSE} represents the root-mean-square error. The appropriate aperture and subsequent mean intensity magnitudes were selected based on the maximum value of SNR for a given RR~Lyrae light curve. Using this approach, we obtained mean intensity magnitudes and their uncertainties ($m_{K_{\rm s}}$, $\sigma_{m_{K_{\rm s}}}$) for individual RR~Lyrae variables observed by the VVV survey.

In addition to VVV $K_{\rm s}$, we also obtained VVV mean intensity magnitudes for the $J$-band. For individual RR~Lyrae pulsators, we used the same apertures as for the $K_{\rm s}$-passband. We utilized the \texttt{pyfiner}\footnote{\url{https://github.com/idekany/pyfiner}.}
\citep{Hajdu2018} routine to estimate the corrected weighted mean $J$-band magnitudes.

To increase the number of RR~Lyrae stars toward the Galactic bulge and to cover lower Galactic latitudes, we included in our catalog RR~Lyrae stars identified by \citet{Dekany2020}, in addition to the OGLE RR~Lyrae dataset. The \citet{Dekany2020} data set provides photometry in $J$, $H$, and $K_{\rm s}$-passbands, and to estimate intensity mean magnitudes, we proceeded in the same way as in the case of the VVV photometry. Therefore, we obtained an additional $\sim 4000$ fundamental-mode RR~Lyrae stars with mean intensity magnitudes in $J$, $H$, and $K_{\rm s}$-passbands, and pulsation properties (pulsation period and amplitude in $K_{\rm s}$ band). The newly acquired data set was cross-matched with the \textit{Gaia} catalog in the same way as described in Section~\ref{subsec:Gaia}. Lastly, using the \texttt{rrl\_feh\_nn}\footnote{\url{https://github.com/idekany/rrl_feh_nn}.}
\citep{Dekany2022} module, we obtained photometric metallicities that were subsequently used for estimating the absolute magnitudes of individual RR~Lyrae stars. Here we note that \texttt{rrl\_feh\_nn} is calibrated using essentially the same data set as for relations used for $I$-band data from OGLE \citep{Dekany2021}.

%\subsubsection{Blends in VVV photometry} \label{subsubsec:Blends}

Despite our best efforts to minimize possible blends by selecting a suitable aperture for a given RR~Lyrae star in the VVV footprint, there was still a possibility of contamination by a nearby star. We compared our mean intensity magnitudes with a recent study by \citet{Molnar2022}, which utilized VVV point spread function (PSF) photometry to search for variable stars toward the Galactic bulge. PSF photometry reaches a deeper limiting magnitude as well as improves the number of sources detected, which is helpful in crowded regions like the bulge \citep{Surot2019}. The bulge RR~Lyrae stars have $K \sim15$~mag at the highly reddened low-latitude regions, which is still $\sim3$ magnitudes brighter compared to where PSF photometry has the most effect and where differences between PSF and aperture photometry are seen \citep[e.g.,][]{Zhang2019}. We matched our VVV data set together with \citet[][comparison of $\sim 32000$ RR~Lyrae single-mode pulsators]{Molnar2022} using \textit{Gaia} \texttt{source\_id}, and looked for differences between our mean intensity magnitudes and those determined in \citet{Molnar2022}. 

The comparison is depicted in Fig.~\ref{fig:ComparisonVVV} where approximately four percent of matched pulsators exhibit systematically brighter apparent mean intensity magnitudes in our analysis. We separated this discrepant sub-population by using the absolute difference between both mean intensity magnitudes, where stars with values above $0.3$ were marked with red points in Fig.~\ref{fig:ComparisonVVV}. Selected variables based on the absolute difference also exhibit very low amplitudes, hinting toward blending with a nearby star that contributes its light to the selected aperture. We tried to mitigate the discrepancy by using a smaller aperture, but that only solved the problem in approximately one-third of the discrepant cases. 

We noticed that the number of discrepant cases could be reduced by more than half if we remove stars with RUWE~$>1.4$ or with \texttt{ipd\_frac\_multi\_peak}~$>4$. Therefore, we implemented both aforementioned criteria (see Eq.~\ref{eq:Condition1}) to clean our data set of these problematic cases. In addition, the remaining stars that pass the two conditions but still deviate from \citet[][$41310$ and $39517$ stars with $K_{\rm s}$ and $J$ band data, respectively]{Molnar2022} photometry have, on average, uncertainties on mean intensity magnitudes six times higher than are average uncertainties for a sample of RR~Lyrae variable with difference below $0.3$\,mag. Despite that, we decided to use mean intensity magnitudes for $J$, $K_{\rm s}$ from \citet{Molnar2022} for stars matched with our data set. We used only stars where pulsation periods matched those in our sample ($\Delta P < 0.0001$\,day). We used the VVV aperture photometry described in the previous subsection \citep[data from][]{Dekany2020} for the remaining variables. Lastly, for stars from the \citet{Molnar2022} catalog, we used estimated photometric [Fe/H]$_{\rm phot}$ from OGLE and VVV photometry described in Sections~\ref{subsec:OGLEphot} and \ref{subsec:VVVphot}.

Considering the significance and impact of blending in the direction of the Galactic bulge, we chose to apply the following criteria to our dataset for all subsequent analyses in this paper (unless otherwise specified):
\begin{equation} \label{eq:Condition1}
\text{RUWE} < 1.4 \hspace{0.5cm} \text{and} \hspace{0.5cm} \texttt{ipd\_frac\_multi\_peak} < 5 \hspace{0.35cm}.
\end{equation}
RR~Lyrae variables that did not meet these conditions or lacked the specified \textit{Gaia} flags were excluded from any analysis presented henceforth.

\subsection{The VISTA Hemisphere Survey}

Lastly, to utilize the available data completely in order to counter the severe reddening, we explored matches (based on equatorial coordinates) between our RR~Lyrae data set and the VHS survey. The aim of the VHS is to map out the whole southern celestial hemisphere (excluding areas covered by other VISTA surveys) using the near-infrared passbands (approximately 20000 deg$^2$). Contrary to the VVV survey, the VHS survey contributes single-exposure photometry in $J$, $H$, and $K_{\rm s}$, complete with the corresponding modified Julian dates (subsequently converted to the Heliocentric Julian date). Due to the longer exposure in $K_{\rm s}$-band, the VHS survey is approximately two magnitudes deeper ($5\sigma$ at $\approx 20$\,mag) than the VVV survey ($5\sigma$ at $\approx 18$\,mag). The difference in coverage between VVV and VHS surveys is displayed in Figure~\ref{fig:MapOfTheSample}.

We utilized publicly available data from the fifth data release of the VHS survey\footnote{Available at the VISTA Science Archive \url{http://horus.roe.ac.uk/vsa/}.}. The DR5 contains only aperture photometry, which, as shown in the previous Subsection~\ref{subsec:VVVphot}, possesses a potential for blending of sources toward the Galactic bulge. To minimize this problem, we vetted our data in the following way. We required that none of our matched sources had a companion within $2$\,arcsec (due to the selected aperture) and that they fulfilled the criteria in Eq.~\ref{eq:Condition1}.

These conditions reduced our matched catalog (over $25000$ objects) to approximately $11000$ stars with single epoch $J$, $H$, and $K_{\rm s}$ photometry. Due to the pulsation nature of our objects, to determine precise distances, we needed to estimate mean intensity magnitudes for our VHS data set which sample our targets at random phase with a single exposure. We used the procedure described in \citet{Braga2018Ome,Braga2019} that utilizes photometric templates and amplitude scaling relations to estimate mean intensity magnitudes for matched stars. The ephemerides necessary for this calculation were taken from data provided by the OGLE survey. Since only single exposures were provided for matched RR~Lyrae stars, the uncertainties on mean intensity magnitudes were approximately a factor of ten larger in comparison with VVV mean intensity magnitudes. This translated into an increased average error in the distance by approximately $200$\,pc than for stars with VVV data. 

\setlength{\tabcolsep}{4.pt}
\begin{table}
\footnotesize
\caption{An overview of the surveys used here. The photometric surveys referenced in this study and their key parameters are listed. We note that \textit{Gaia} $m_{G_{\rm BP}}$ comes from RR~Lyrae and \textit{Gaia} astrometric catalog $\otimes$ with OGLE-IV.}
\label{tab:SummaryUsedData}
\begin{tabular}{l|cccc}
\hline \hline
Surveys & mean mag. & pulsation prop. & Additional info. & Num. stars\\
\hline 
OGLE & $m_{I}$ & $P$, $M_{0}$, Amp$_{I}$ & [Fe/H]$_{\rm phot}$ & $68127$\\ \hline 
OGLE & $m_{V}$ & Amp$_{V}$ & [Fe/H]$_{\rm phot}$ & $28665$\\ \hline 
\textit{Gaia} & $m_{G_{\rm BP}}$ & Amp$_{G}$ & $\mu_{\alpha}$, $\mu_{\delta}^{\ast}$, flags$^{\tnote{1}}$ & $52432$ \\ \hline 
VVV & $m_{J}$, $m_{K_{\rm s}}$ & Amp$_{K_{\rm s}}$ & [Fe/H]$_{\rm phot}^{\tnote{2}}$ & $43284$ \\ \hline 
VHS & $m_{J}$, $m_{K_{\rm s}}$ & -- & -- & $10153$\\ \hline 
\end{tabular}
\begin{tablenotes}
\item[1] $^{\tnote{1}}$ RUWE and \texttt{ipd\_frac\_multi\_peak}.
\item[2] $^{\tnote{2}}$ For stars without [Fe/H]$_{\rm phot}$ from OGLE-IV data we used module developed in \citet{Dekany2022} to estimate [Fe/H]$_{\rm phot}$ from $K_{\rm s}$ light curves.
\end{tablenotes}
\end{table}

\section{Distances to individual RR~Lyrae stars} \label{sec:DistancesAll}

In the following section, we estimate distances to individual RR~Lyrae pulsators toward the Galactic bulge in our data set. The severe reddening hampers the effort to calculate distances at the RR~Lyrae population in the Galactic bulge. Fortunately, the near-infrared photometry of our data set and intrinsic properties of RR~Lyrae variables (such as the period-absolute magnitude-metallicity relations, PMZ) can help counter the extinction. RR~Lyrae stars are efficient for tracing reddening, enabling the creation of detailed extinction maps, provided there is adequate quantity and spatial distribution of these stars \citep[see, e.g.,][]{Kunder2010,Haschke2011,Prudil2019OOspat}.

Previous studies of the RR~Lyrae population toward the Galactic bulge used mainly PMZ relations from $V$ and $I$-passbands \citep{Pietrukowicz2015,Kunder2016,Kunder2020,Du2020} combined with reddening laws and maps based on VVV photometry \citep{Gonzalez2012,Nataf2013} to estimate distances. In this study, we follow an approach used in one of our previous works where we used RR~Lyrae stars themselves to trace extinction \citep{Prudil2019OOspat}, particularly the reddening vector. The baseline for estimating reddening toward the Galactic bulge using RR~Lyrae stars are reliable PMZ relations for available photometry ($G_{\rm BP}$, $V$, $I$, $J$, and $K_{\rm s}$-passbands in our case). Therefore, we utilized re-calibrated PMZ relations for our passbands from our previous work \citep{Prudil2024} and estimated absolute magnitudes and their uncertainties for all RR~Lyrae stars in our sample.

To estimate the reddening vector, we used all single-mode RR~Lyrae stars that have $G_{\rm BP}$, $V$, $I$, $J$, and $K_{\rm s}$ mean intensity magnitudes (at least two out of five) together with determined [Fe/H]$_{\rm phot}$ based on their $I$ and $K_{\rm s}$-band photometric properties. We note that stars with the VHS $J$, and $K_{\rm s}$-magnitudes did not enter in the reddening law estimation for $J - K_{\rm s}$ and $I - K_{\rm s}$ colors. This was done due to the possibility of blending, which would severely affect available aperture photometry. It is important to emphasize that values for [Fe/H] used in the calibration of PMZ relations \citep{Prudil2024} were on the same metallicity scale as [Fe/H]$_{\rm phot}$ derived for our bulge sample. Therefore, no further conversion is necessary, and we could use metallicities directly estimated from photometry. 

Unfortunately, not all RR~Lyrae stars in our data set have $K_{\rm s}$ mean intensity magnitudes. Thus, we also created reddening maps using $G_{\rm BP}$, $V$, and $I$ passbands to fully utilize the available data, and provide up to four estimates of the color excess $E\left ( J - K_{\rm s} \right )$, $E\left ( I - K_{\rm s} \right )$, $E\left ( V - I \right )$, and $E\left ( G_{\rm BP} - I \right )$. Further, we note that for estimating the reddening vector for the color excess $E\left ( G_{\rm BP} - I \right )$, we used only RR~Lyrae stars identified as variables by \textit{Gaia} \citep{Clementini2023}.

\begin{figure*}
\includegraphics[width=2\columnwidth]{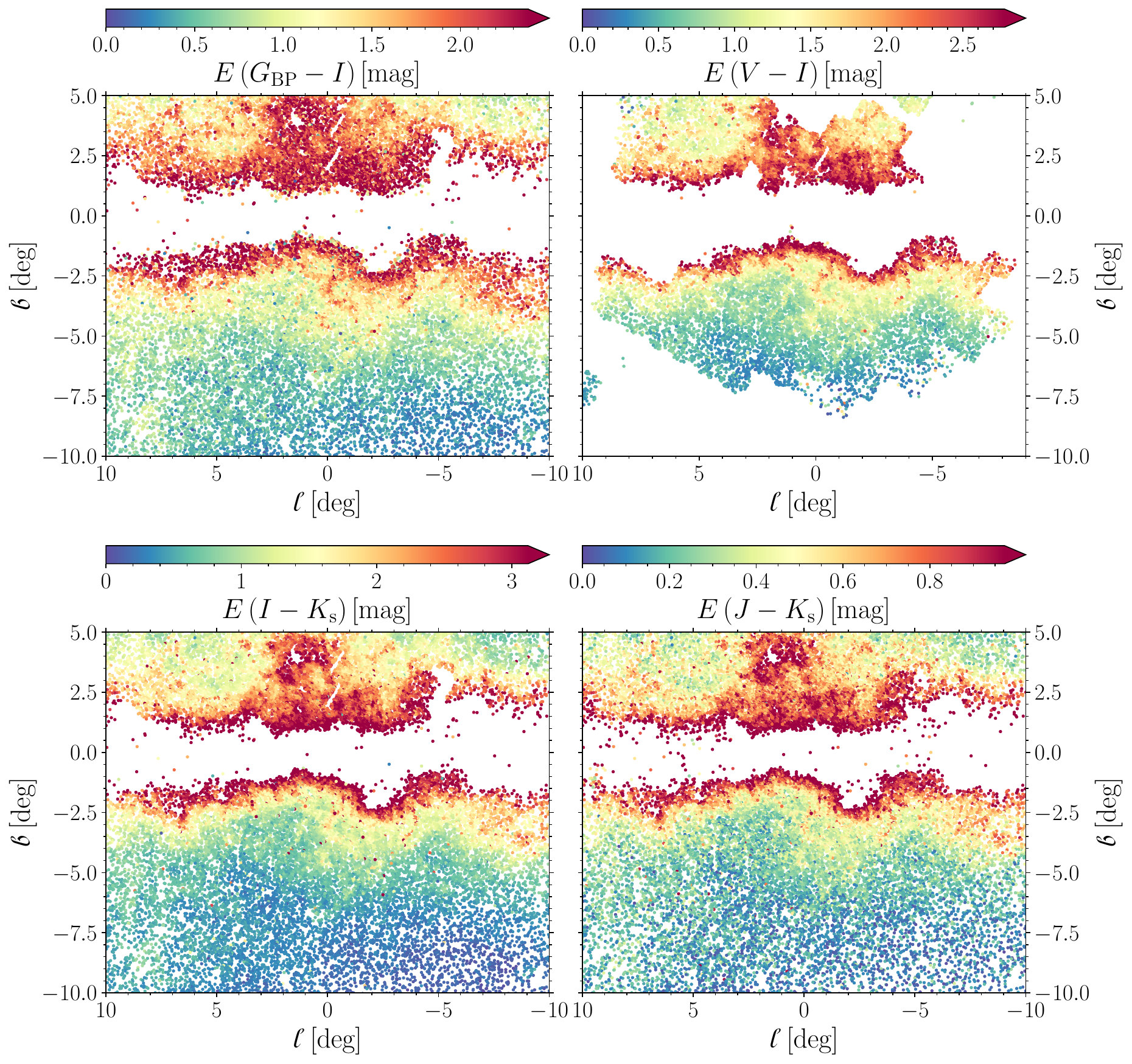}
\caption{The spatial distribution of color excesses estimated in this study, $E\left ( G_{\rm BP} - I \right )$, $E\left ( V - I \right )$, $E\left ( I - K_{\rm s} \right )$, and $E\left ( J - K_{\rm s} \right )$. Each point represents a given single-mode RR~Lyrae star color-coded based on its associated reddening. Note the different color scales in each panel.}
\label{fig:ReddedningMaps}
\end{figure*}

Pulsation periods, together with [Fe/H]$_{\rm phot}$, allowed us to estimate absolute magnitudes $M_{G_{\rm BP}}$, $M_{V}$, $M_{I}$, $M_{J}$ and $M_{K_{\rm s}}$ \citep[using relations from][see their Eq.~16,~18,~19,~22, and~23]{Prudil2024}, which in turn we used to estimate the color excesses $E\left ( J - K_{\rm s} \right )$, $E\left ( I - K_{\rm s} \right )$, $E\left ( V - I \right )$, and $E\left ( G_{\rm BP} - I \right )$ described by the following equations:
\begin{gather} \label{eq:ColorExcess}
E\left ( J - K_{\rm s} \right ) = \left ( m_{J} - m_{K_{\rm s}} \right ) - \left ( M_{J} - M_{K_{\rm s}} \right ) , \\
E\left ( I - K_{\rm s} \right ) = \left ( m_{I} - m_{K_{\rm s}} \right ) - \left ( M_{I} - M_{K_{\rm s}} \right ) , \\
E\left ( V - I \right ) = \left ( m_{V} - m_{I} \right ) - \left ( M_{V} - M_{I} \right ) , \\ 
E\left ( G_{\rm BP} - I \right ) = \left ( m_{G_{\rm BP}} - m_{I} \right ) - \left ( M_{G_{\rm BP}} - M_{I} \right ) .
\end{gather}
For determining distances to individual RR~Lyrae stars, we assume that reddening is uncorrelated with distance modulus, e.g., $\left< \left ( m_{K_{\rm s,0}} - M_{K_{\rm s,0}} \right ) \right> = \left< \left ( m_{K_{\rm s}} - M_{K_{\rm s}} \right ) \right> + R_{JK_{\rm s}} \cdot E\left ( J - K_{\rm s} \right )$ where index $0$ denotes dereddened quantities and $R_{JK_{\rm s}}$ is the slopes of the reddening law. In Figure~\ref{fig:ReddedningMaps}, we present the spatial distribution of four derived color excess maps. These maps focus solely on the central parts of our dataset, closely aligning with the VVV footprint. This approach is specifically chosen to emphasize the central regions, where we have estimated the reddening laws and conducted the majority of our spatial and kinematic analyses.

The color excess and distance modulus (e.g., $m_{K_{\rm s}} -  M_{K_{\rm s}}$ and $m_{I} - M_{I}$) then follow nearly linear dependence, where the $R_{JK_{\rm s}}$, $R_{IK_{\rm s}}$, $R_{VI}$, and $R_{G_{\rm BP}I}$ are the slopes of the reddening laws:
\begin{gather}
A_{K_{\rm s}} = R_{JK_{\rm s}} \cdot E\left ( J - K_{\rm s} \right ) , \\
A_{K_{\rm s}} = R_{IK_{\rm s}} \cdot E\left ( I - K_{\rm s} \right ) , \\
A_{I} = R_{VI} \cdot E\left ( V - I \right ) , \\
A_{I} = R_{G_{\rm BP}I} \cdot E\left ( G_{\rm BP} - I \right ) ,
\end{gather}
where $A_{K_{\rm s}}$ and $A_{I}$ are extinction values in respective bands toward a given RR~Lyrae star. This assumption is discussed further below and consistently used in Figures~\ref{fig:ReddLawJK},~\ref{fig:ReddLawIK},~\ref{fig:ReddLawVI},~\ref{fig:ReddLawGbpI} in the Appendix. The $R_{JK_{\rm s}}$, $R_{IK_{\rm s}}$, $R_{VI}$, and $R_{G_{\rm BP}I}$ were then estimated through a linear fit between color excess and distance modulus. In the first step, we needed to define bulge RR~Lyrae stars because our RR~Lyrae data set covers the Sagittarius dwarf galaxy \citep{Ibata1994}, and the Sagittarius stream behind the Galactic bulge \citep[see, e.g.,][]{Hamanowicz2016}. RR~Lyrae variables in the Sagittarius dwarf galaxy are further away and will be more reddened than stars in the bulge \citep[see, e.g.,][]{Kunder2009Sag}. We used two linear relations to separate the Galactic bulge RR~Lyrae stars from the foreground and background RR~Lyrae population \citep[similar to the selection in][]{Pietrukowicz2015}. The linear relations (in mean intensity magnitude and color space, see Table~\ref{tab:CutsOnMagAndColor}) had the following form:
\begin{gather}
m_{\rm low} = a_{\rm low} \cdot \text{color} + b_{\rm low} , \\
m_{\rm upp} = a_{\rm upp} \cdot \text{color} + b_{\rm upp} .
\end{gather}
We also included a cap for maximum mean intensity magnitude for individual reddening vector estimation to ensure the high completeness of our analyzed sample (marked with the red horizontal line in Figures~\ref{fig:ReddLawJK},~\ref{fig:ReddLawIK},~\ref{fig:ReddLawVI},~\ref{fig:ReddLawGbpI}). In addition, to emphasize the point on blending (see Subsection~\ref{subsec:VVVphot}), we again used criteria in Eq.~\ref{eq:Condition1}. The values for $a_{\rm low}$, $a_{\rm upp}$, $b_{\rm low}$, and $b_{\rm upp} $ are listed for individual color-magnitude combinations in Table~\ref{tab:CutsOnMagAndColor}. In the second step, we binned the color excesses over which we wanted to estimate reddening laws. We used a moving weighted average with a window size equal to $750$ and a step size equal to $500$ (stars) for all four passband combinations. We estimated the weighted average and the error on the mean of the respective distance modulus for each binned color excess region.

\setlength{\tabcolsep}{3.pt}
\setlength\extrarowheight{2pt}
\begin{table}
\caption{List of linear relations and conditions used to remove foreground and background RR~Lyrae stars from our data set for reddening vector estimation. The first column lists passbands for the mean intensity magnitudes and colors used. The second and third columns contain coefficients for linear relations used as boundary conditions. The last column represents maximum mean intensity magnitudes to preserve the high completeness of our sample.}
\label{tab:CutsOnMagAndColor}
\begin{tabular}{l|cc|l}
\hline \hline 
Passbands & $a$ & $b$ & Magnitude cap. \\
\hline 
\multirow{2}{*}{$m_{K_{\rm s}}$, $m_{J}$-$m_{K_{\rm s}}$}     & $a_{\rm low} = 1.4$  & $b_{\rm low} = 13.25$  & $m_{K_{\rm s}} < 18$  \\
                            & $a_{\rm upp} = 1.4$ & $b_{\rm upp} = 15.00$ &  \& $m_{J} < 16.5$                        \\ \hline 
\multirow{2}{*}{$m_{I}$, $m_{I}$-$m_{K_{\rm s}}$}     & $a_{\rm low} = 1.2$  & $b_{\rm low} = 13.25$  & $m_{K_{\rm s}} < 18$  \\
                            & $a_{\rm upp} = 1.2$ & $b_{\rm upp} = 15.00$ &  \&  $m_{I} < 18$                             \\ \hline 
\multirow{2}{*}{$m_{I}$, $m_{V}$-$m_{I}$}     & $a_{\rm low} = 1.2$  & $b_{\rm low} = 13.50$  & $m_{I} < 18$     \\
                            & $a_{\rm upp} = 1.2$ & $b_{\rm upp} = 15.25$ &  \&  $m_{V} < 22$                        \\ \hline 
\multirow{2}{*}{$m_{I}$, $m_{G_{\rm BP}}$ - $m_{I}$} & $a_{\rm low} = 1.2$  & $b_{\rm low} = 13.00$  & $m_{I} < 18$ \\
                            & $a_{\rm upp} = 1.2$ & $b_{\rm upp} = 15.50$ & \& $m_{G_{\rm BP}} < 22$  \\ \hline                       
\end{tabular}
\end{table}
%%%%%%%%%%%%%%%

The results of linear fits, including their uncertainties and correlations between slope and intercept, and selection criteria are depicted in the Appendix~\ref{sec:AppFigures} (see Figures~\ref{fig:ReddLawJK},~\ref{fig:ReddLawIK},~\ref{fig:ReddLawVI}, and~\ref{fig:ReddLawGbpI}). From the intercepts of our linear fits, we can approximately estimate the distance to the Galactic center. We note that in our method, we assume no physical correlation between distance and reddening \citep[similarly to the approach by][using the red clumps]{Alonso-Garcia2017}. This condition is not strictly met here, but since the majority of our targets are located away from the Galactic plane, where most of the dust resides, we consider a typical Bulge RR~Lyrae to be behind the vast majority of the dust.

\begin{figure}
\includegraphics[width=\columnwidth]{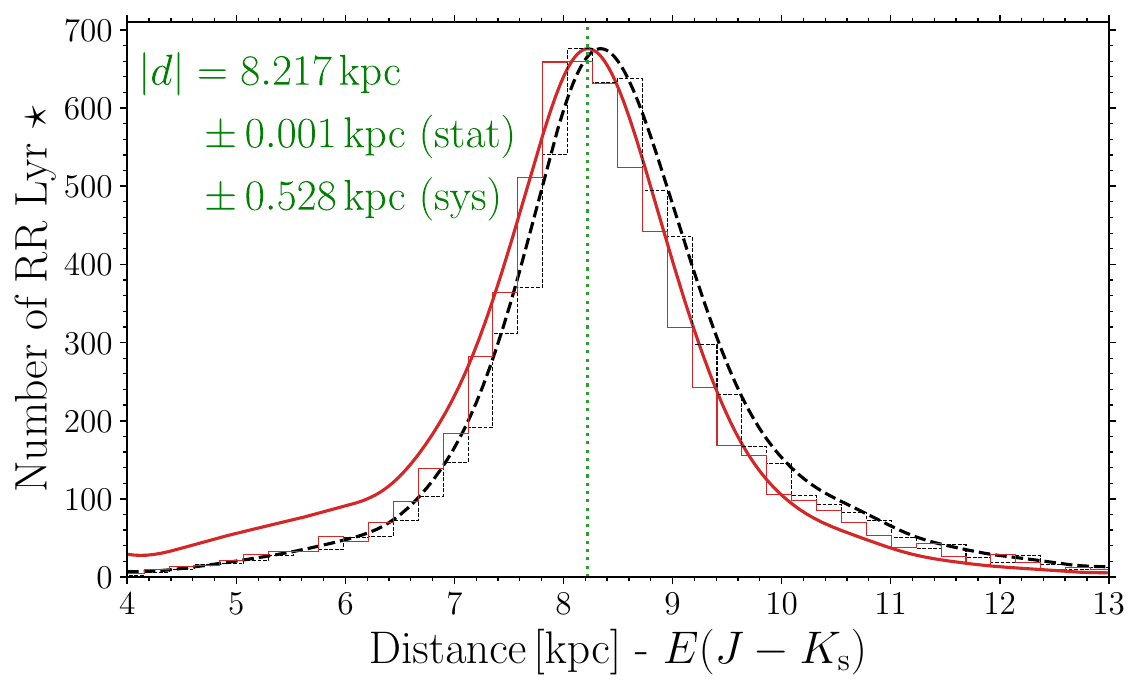}
\caption{The distance distribution for $d_{JK_{\mathrm{s}}}$ (black histogram), its kernel density estimate (KDE, black dashed line), and KDE for the distance distribution corrected for the cone effect (red line). The corrected distance to the Galactic center is marked with a green dotted line.}
\label{fig:DistanceToCenter}
\end{figure}

To estimate the distance to the Galactic center using our derived distances we need to apply two geometrical corrections: (i) such as projecting onto the Galactic plane via $\cos\mathcal{b}$ and (ii) correcting for the cone effect by scaling the distance distribution by a $d^{-2}$ factor. Once we apply these geometric corrections we focus on the region where our dataset has the highest completeness (within $\mathcal{b} = (-2, -6)$\,deg and $\left| \mathcal{l} \right| < 5$\,deg). We then estimate the distance to the Galactic center by fitting the cone-effect corrected kernel density estimate with a Gaussian function: $d_{JK_{\rm s}}^{\rm cen} = 8217 \pm 1({\rm stat}) \pm 528({\rm sys})$\,pc, $d_{IK_{\rm s}}^{\rm cen} = 8230 \pm 1({\rm stat}) \pm 379({\rm sys})$\,pc, $d_{VI}^{\rm cen} = 8058 \pm 1({\rm stat}) \pm 974({\rm sys})$\,pc, $d_{G_{\mathrm{BP}I}}^{\rm cen} = 7759 \pm 4({\rm stat}) \pm 861({\rm sys})$\,pc. The statistical and systematic uncertainties of individual distances to the Galactic center were estimated based on the uncertainties calculated for each distance in this section. The overall statistical and systematic uncertainties are the average values of these individual uncertainties. The statistical uncertainty is additionally scaled by the square root of the number of RR~Lyrae stars used in the distance estimation. The estimated values are well within the accepted distance measurements for the Galactic center \citep{GRAVITY2021,Leung2023}. The cone-effect correction is depicted in Figure~\ref{fig:DistanceToCenter} for $d_{JK_{\mathrm{s}}}$.

\begin{table}
\caption{The list of estimated reddening laws in this study and the number of RR~Lyrae stars ($N^{\star}$) used for each reddening law.}
\begin{tabular}{cccc}
\hline \hline
$R_{JK_{\rm s}}$ & $R_{IK_{\rm s}}$ & $R_{VI}$ & $R_{G_{\rm BP}I}$ \\
$0.487 \pm 0.014$ & $0.153 \pm 0.005$ & $1.205 \pm 0.006$ &  $1.108 \pm 0.01$ \\ 
$N^{\star} = 27448$ & $N^{\star} = 22754$ & $N^{\star} = 15956$ & $N^{\star} = 32538$ \\ \hline
\end{tabular}
\label{tab:ReddedningLawTable}
\end{table}

Derived mean reddening laws based on single-mode RR~Lyrae stars toward the Galactic bulge are listed in Table~\ref{tab:ReddedningLawTable}. The reddening law derived for $R_{JK_{\rm s}}$ is a bit higher but still consistent within $1.5\sigma$ to those estimated in previous studies, e.g., $R_{JK_{\rm s}} = 0.428 \pm 0.04$ and $R_{JK_{\rm s}} = 0.443 \pm 0.036$ for \citet{Alonso-Garcia2017}, and \citet{Wang2019Ext}, respectively. On the other hand $R_{JK_{\rm s}}$ agrees well with the reddening law estimated by \citet{Nishiyama2006} equal to $R_{JK_{\rm s}} = 0.494 \pm 0.006$. Our estimate for $R_{IK_{\rm s}}$ is slightly smaller than estimates of \citet[][$R_{IK_{\rm s}} = 0.164$]{Dekany2013} and \citet[][$R_{IK_{\rm s}} = 0.160$]{Nataf2013} and to a degree higher than the estimate from \citet[][$R_{IK_{\rm s}} = 0.140$]{Prudil2019OOspat}. In addition, our $R_{VI}$ value matches with the estimate by \citet[][$R_{VI}=1.215$]{Nataf2013}. Newly defined reddening laws subsequently served in estimating distances toward bulge RR~Lyrae stars through the distance modulus:
\begin{gather}
d_{JK_{\rm s}} = 10^{1 + 0.2 \cdot \left ( m_{K_{\rm s}} - M_{K_{\rm s}} - R_{JK_{\rm s}} \cdot E\left ( J - K_{\rm s} \right ) \right )} ,\\
d_{IK_{\rm s}} = 10^{1 + 0.2 \cdot \left ( m_{K_{\rm s}} - M_{K_{\rm s}} - R_{IK_{\rm s}} \cdot E\left ( I - K_{\rm s} \right ) \right )} ,\\
d_{VI} = 10^{1 + 0.2 \cdot \left ( m_{I} - M_{I} - R_{VI} \cdot E\left ( V - I \right ) \right )} ,\\
d_{G_{\rm BP}I} = 10^{1 + 0.2 \cdot \left ( m_{I} - M_{I} - R_{G_{\rm BP}I} \cdot E\left ( G_{\rm BP} - I \right ) \right )} . 
\end{gather}
The uncertainties on individual distances were calculated by including all sources of errors (from the mean intensity and absolute magnitudes, reddening law, and color excess). We also divided our sources of uncertainty into statistical\footnote{Individual uncertainties in the mean intensity magnitudes, pulsation periods, and photometric metallicities.} and systematical categories\footnote{Uncertanities in the period-absolute magnitude-metallicity relation, relation for the photometric metallicities.} where systematic uncertainties dominate the error budget ($\approx 95$ of the total uncertainty on distance).

In our approach, uncertainties on distances purely from VVV photometry uncertainties constitute approximately six percent of the total distance. When we combine VVV $K_{\rm s}$-band and OGLE $I$-band, we reach down to four percent uncertainty on distance. The increase in precision is due to removing the need for using the $J$-band, which has a larger scatter in the PMZ relation, and due to fewer photometric epochs in VVV. Distance uncertainties using purely visual passbands contribute to an uncertainty of around $10-11$ percent. The derived uncertainties on distances are slightly higher when compared with previous studies \citep[particularly with the precision of ten and three percent for optical and infrared data, respectively,][]{Neeley2017}. The full table with derived photometric metallicities, absolute magnitudes, color excesses, and distances for all sample RR~Lyrae stars is included in Appendix~\ref{tab:CompleteList} and as a supplementary material for this paper. The verification of our distances is included in the Appendix~\ref{sec:VerfDistGC}.

Lastly, we emphasize that using a single reddening law for the entire Galactic bulge region is a sub-optimal solution. Variations from the commonly assumed \citet{Cardelli1989} extinction law have been reported at low Galactic latitudes \citep{Nishiyama2005,Nishiyama2006,Nishiyama2009}. The decision to use a single reddening law was imposed by the low stellar density of RR~Lyrae variables in this region, which prohibits the estimation of the reddening law in binned coordinate space.

\section{Comparison of reddening maps} \label{sec:RedMapsComp}

\subsection{Comparison of derived reddening maps in this study} \label{subsec:ReddeningMapThiswork}

In the following analysis, we compare the estimated extinction toward RR~Lyrae stars derived from our reddening maps. Our focus is on comparing the predicted $A_{I}$ values, which are based on different color excesses derived using both visual and infrared passbands. This approach is motivated by the studies of \citet{Dekany2013} and \citet{Pietrukowicz2015}, which employed different passbands for distance estimation and reported somewhat contradictory results.

To achieve this, we need to transform reddening laws for $E\left( J - K_{\rm s} \right)$ and $E\left( I - K_{\rm s} \right)$ to obtain $A_{I}$. For these transformations, we utilize the results for $R_{IK_{\rm s}}$ and $R_{JK_{\rm s}}$ from Section~\ref{sec:DistancesAll}. For $E\left( I - K_{\rm s} \right)$, we calculate the extinction in the $I$-band as follows:
\begin{gather}
E\left ( I - K_{\rm s} \right ) = A_{I} - A_{K_{\rm s}} ,\\
E\left ( I - K_{\rm s} \right ) = A_{I} - R_{IK_{\rm s}} \cdot E\left ( I-K_{\rm s} \right ) ,\\
E\left ( I - K_{\rm s} \right ) + R_{IK_{\rm s}} \cdot E\left ( I-K_{\rm s} \right ) = R_{VI} \cdot E\left ( V - I \right ) ,\\
E\left ( I - K_{\rm s} \right ) \cdot \left ( 1 + R_{IK_{\rm s}} \right ) = A_{I} ,\\
A_{I} = 1.153 \cdot E\left ( I - K_{\rm s} \right ) . \label{eq:AiUsingIK}
\end{gather}
For $E\left ( J-K_{\rm s} \right )$, we use the following approach:
\begin{gather}
E\left ( J - K_{\rm s} \right ) = A_{J} - A_{K_{\rm s}} ,\\
E\left ( J - K_{\rm s} \right ) = 1.487 \cdot E\left ( J - K_{\rm s} \right ) - R_{IK_{\rm s}} \cdot E\left ( I - K_{\rm s} \right ) ,\\
E\left ( J - K_{\rm s} \right ) = 1.487 \cdot E\left ( J - K_{\rm s} \right ) - R_{IK_{\rm s}} \cdot A_{I}/1.153 ,\\
A_{I} = 3.67 \cdot E\left ( J - K_{\rm s} \right ) . \label{eq:AiUsingJK}
\end{gather}
Using the derived Equations~\ref{eq:AiUsingIK} and \ref{eq:AiUsingJK}, we calculated the extinction through $E\left( I - K_{\rm s} \right)$ and $E\left( J - K_{\rm s} \right)$ for the $I$-band. 

\begin{figure*}
\includegraphics[width=2\columnwidth]{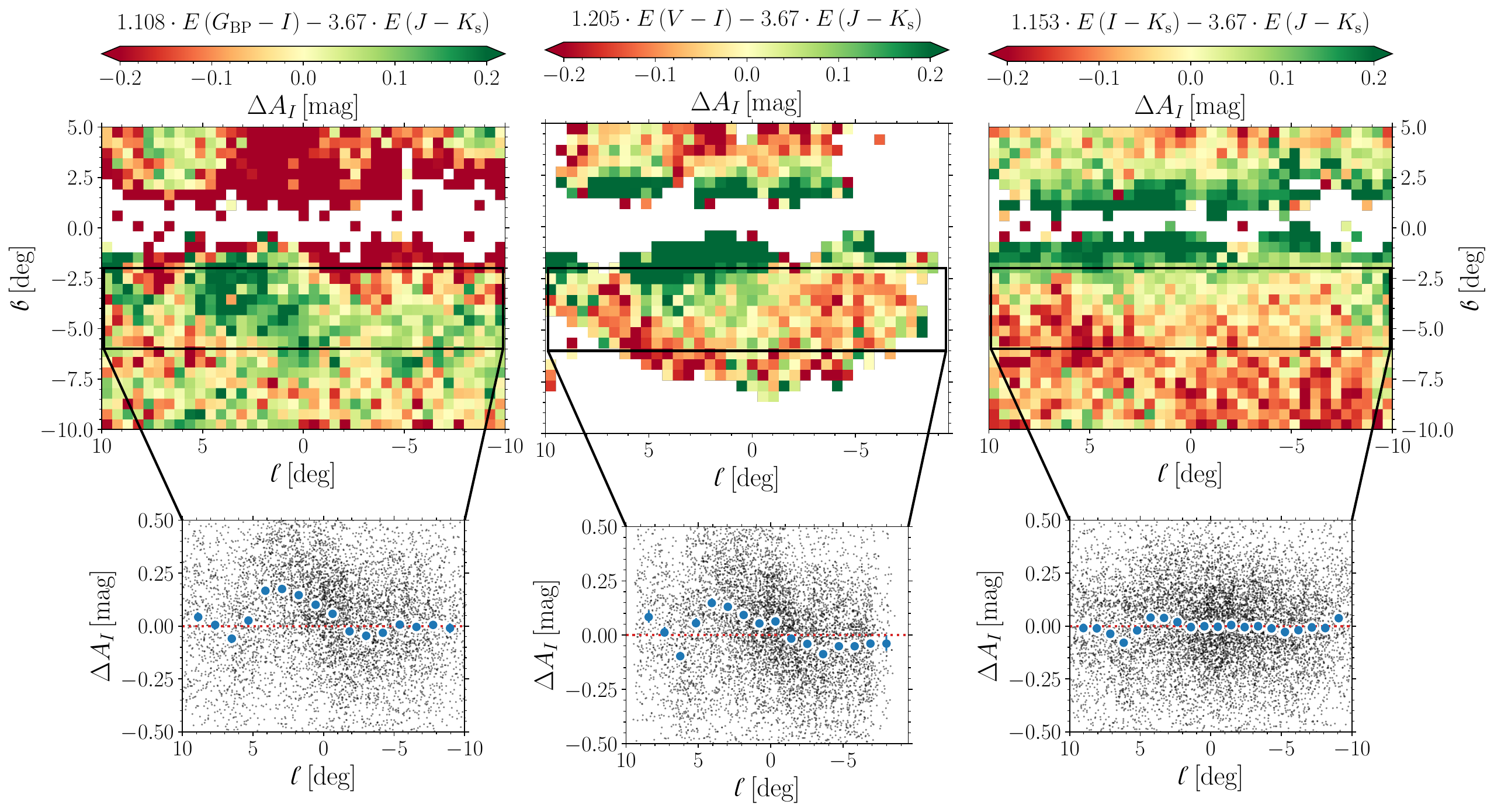}
\caption{The binned spatial distribution (step equal to $0.5$\,deg) of differences in estimated extinction $\Delta A_{I}$. Each panel depicts the differences between the extinction ($A_{I}$) estimated from purely optical or near IR-optical color excesses ($E\left ( G_{\rm BP} - I \right )$, $E\left ( V - I \right )$ and $E\left ( I - K_{\rm s} \right )$), and the extinction estimated from the $E\left ( J - K_{\rm s} \right )$ color excess. The insets focus on an area below the Galactic plane where we binned (blue circles, seventeen steps between $-9.5$ to $9.5$\,deg) the $\Delta A_{I}$ as a function of $\mathcal{l}$ with grey dots representing individual RR~Lyrae stars.}
\label{fig:AICompThisStudy}
\end{figure*}

In Figure~\ref{fig:AICompThisStudy}, we present the binned spatial maps (with a bin step equal to $0.5$\,deg in both directions) showing the difference in $A_{I}$ (denoted as $\Delta A_{I}$) and its variation with respect to Galactic longitude. Significant variation in $A_{I}$ as a function of both Galactic longitude and latitude is evident. Focusing initially on the Galactic latitude variation, the middle and right-hand panels indicate that closer to the Galactic plane, $E\left( J - K_{\rm s} \right)$ predicts smaller $A_{I}$ values than $E\left( V - I \right)$ and $E\left( I - K_{\rm s} \right)$. In contrast, for $A_{I}$ based on $E\left( G_{\rm BP} - I \right)$, the trend reverses near the Galactic plane, with $E\left( J - K_{\rm s} \right)$ predicting higher $A_{I}$ values. These fluctuations underscore the challenges in estimating reddening near the Galactic plane, especially with broad visual passbands.

To investigate the Galactic longitude variation, we examine the region $\mathcal{b} = (-6, -2)$ degrees, previously studied for indications of the bar traced by RR~Lyrae stars \citep{Pietrukowicz2015}. We observe a notable divergence in $A_I$ in Galactic longitude, especially evident in the insets of Figure~\ref{fig:AICompThisStudy}. A distinct gradient is observed, particularly in the visual passbands (left-hand and middle panels). Notably, at positive $\mathcal{l}$, there is a bump in $\Delta A_{I}$ with an amplitude of approximately $0.25$ mag for both visual reddenings. For $E\left( V - I \right)$, a decrease in $\Delta A_{I}$ at negative $\mathcal{l}$ is also noticeable. Assuming the $A_{I}$ based on $E\left( J - K_{\rm s} \right)$ is accurate, these discrepancies could lead to distance underestimation at positive $\mathcal{l}$ and overestimation at negative $\mathcal{l}$. Such $A_{I}$ variations could result in a distance difference of about $1$ kpc for Galactic bulge stars. Referring to Figure 3 in \citet{Gonzalez2011}, the distance difference between $\mathcal{l} = (-5, 5)$ degrees is around $1$ kpc. Therefore, it is crucial to acknowledge that a shift of several hundred parsecs, as observed in visual passbands, could mimic the appearance of the bar. In the right-hand panel, the bump is less pronounced, with an amplitude less than half that for visual passbands. The $\Delta A_{I}$ variation is not as significant as in the visual passbands, highlighting the importance of using infrared passbands for estimating extinction and distances towards the Galactic bulge.

\subsection{Comparison of derived reddening maps with literature} \label{subsec:ReddeningMapLit}

Following up on the previous subsection, we replaced the $E\left( J - K_{\rm s} \right)$ values derived in this study with data from the literature, specifically the reddening map from \citet{Surot2020}\footnote{Available at \url{http://basti-iac.oa-teramo.inaf.it/vvvexmap/}.} and the \citet[][using the Bulge Extinction and Metallicity Calculator, BEAM2\footnote{\url{https://www.oagonzalez.net/beam-calculator}.}]{Gonzalez2012} studies. We matched our dataset with the reddening maps using equatorial and Galactic coordinates. Both cited studies utilized red clump stars identified in the VVV survey to assess reddening towards the Galactic bulge. We performed the same comparison as in Fig.~\ref{fig:AICompThisStudy}, employing Equation~\ref{eq:AiUsingJK}. We selected these two studies specifically because they provide direct $E\left( J - K_{\rm s} \right)$ measurements. Both studies are based on VVV data, the main difference is in the use of aperture photometry \citep{Gonzalez2012} vs PSF photometry for the more recent study by \citep{Surot2020} and the bin sizes of provided maps.

The comparison based on the \citet{Surot2020} reddening map is depicted in Figure~\ref{fig:AICompSurot}, while the comparison based on the \citet{Gonzalez2012} reddening map is shown in Figure~\ref{fig:AICompGonza} and included in the Appendix. Both reddening maps present a picture similar to the analysis in Figure~\ref{fig:AICompThisStudy}. For both visual-based color excesses, we observe mostly larger $\Delta A_{I}$ at positive $\mathcal{l}$ and lower $\Delta A_{I}$ at negative $\mathcal{l}$. The $A_{I}$ gradient in these reddening maps is shallower than ours but still shows a total amplitude of about $0.15$ mag, equivalent to approximately $600$ pc at the distance to the Galactic bulge. Assuming the extinction in the $I$-band estimated using $E\left( J - K_{\rm s} \right)$ is correct, this could result in a bar-like tilt \citep[of a smaller angle than found using red clump stars by][]{Wegg2013} in the visual distances for RR~Lyrae stars. 

In the right-hand panel of Figures~\ref{fig:AICompSurot} and~\ref{fig:AICompGonza}, we note that the extinction $A_{I}$ estimated using $E\left( I - K_{\rm s} \right)$ is underestimated by approximately $0.05$ mag and $0.08$ mag, based on the reddening maps from \citet{Surot2020} and \citet{Gonzalez2012}, respectively. This results in a shift in distances of about $200$ pc and $300$ pc at the distance of the Galactic bulge, respectively, and remains more or less consistent across the majority of Galactic longitudes. The reason for this shift for both references is probably the absolute magnitude calibration of the photometric VVV system. This calibration is based on the catalog that is a result of cross-correlation between the \textit{Two-Micron Sky Survey} \citep[2MASS,][]{Cutri2003,Skrutskie2006} and VVV survey performed by the Cambridge AstronomySurvey Unit (CASU\footnote{\url{http://casu.ast.cam.ac.uk/}}).

\subsection{Using reddening $E\left ( V - I \right )$ from literature} \label{subsec:Nataf2013OGLE}

The previous two subsections explored the variation between reddening maps derived in this study and those available in the literature for the Galactic bulge. We used $E\left( J - K_{\rm s} \right)$ as a reference for our comparisons. In the following analysis, we will use the $E\left( V - I \right)$ value from \citet{Nataf2013}\footnote{The data are available at \url{https://ogle.astrouw.edu.pl/cgi-ogle/getext.py}.} to estimate $A_{I}$ and compare the obtained values with those calculated from our $E\left( J - K_{\rm s} \right)$, as well as $E\left( J - K_{\rm s} \right)$ values from \citet{Gonzalez2012} and \citet{Surot2020}. We again utilize Eq.~\ref{eq:AiUsingJK} to derive $A_{I}$.

Figure~\ref{fig:AICompNataf} presents the comparison. Observing the $A_{I}$ variation with respect to Galactic longitude, we note a consistent gradient along $\mathcal{l}$, as seen in the previous two comparisons. There is an overestimation of $A_{I}$ from visual color excess at positive Galactic longitudes and a decrease in $A_{I}$ towards negative Galactic longitudes. For literature-based reddening maps, we obtained amplitude variations around $0.13$\,mag ($0.5$\,kpc at the distance to the Galactic bulge), while for $A_{I}$ estimates based on our $E\left( J - K_{\rm s} \right)$, we found an amplitude of $0.28$\,mag (approximately $1.1$\,kpc difference in distance).

The consistently observed gradient for $5 > \mathcal{l} > -5$ further supports the conclusions from this section. Assuming the obtained color excess $E\left( J - K_{\rm s} \right)$ in this study and in studies focused on the Galactic bulge is correct, we conclude that the usage of visual passbands for RR~Lyrae stars toward the Galactic bulge, as a means to obtain reddening, leads to potential spurious distance gradients along Galactic longitude. These gradients, despite uncertain total amplitude, result in a shift in distance ranging from $0.5$ to $1.0$\,kpc between positive and negative longitudes. Depending on the amplitude, such a shift in distance can mimic the tilt of the bar as observed in red clump stars \citep[assuming a $1$\,kpc difference between the red clumps at $\mathcal{l} = -5$ and $\mathcal{l} = 5$\,deg,][]{Gonzalez2012,Wegg2013}. We observe a shift between extinction estimated in this study compared to literature values; this shift appears systematic and does not exceed $300$\,pc (see the previous Subsection~\ref{subsec:ReddeningMapLit} on the probable cause of this shift). Therefore, it is fully covered by the error budget, contributing only to the average distance shift to the Galactic bulge.

\section{Galactic bulge in 3D} \label{sec:bulge3D}

The following section is inspired by and provides a comparison with earlier studies that focused on the spatial characteristics of the bulge RR~Lyrae population, primarily those by \citet{Dekany2013} and \citet{Pietrukowicz2015}. In what follows, we examined the spatial distribution of RR~Lyrae variables towards the Galactic bulge, represented in Heliocentric Cartesian coordinates (X, Y, and Z). To assess the spatial properties of the Galactic bulge, we aimed to test the presence and orientation of the bar. Our methodology for estimating the angle of the bar was inspired by the approach outlined in \citet{Pietrukowicz2015}. In our approach the distance distribution is described by a Gaussian mixture model \citep[GMM, from \texttt{scikit-learn} Python library][]{Pedregosa2012}. Based on testing the GMM using the Bayesian Information Criterion (BIC) and the Akaike Information Criterion (AIC), we utilized three GMM components in all cases. The AIC and BIC did not show significant changes with an increased number of GMM components. Using the GMM we estimate density levels, specifically using the half-maximum of the distribution as a reference. Similar to \citet{Pietrukowicz2015}, we projected our distances onto the Galactic plane using $\cos\mathcal{b}$. We divided our data into eleven bins based on Galactic longitude, ranging from $-5$ to $5$ degrees. For each bin, comprising measured distances and their uncertainties, we estimated the half maxima of the distribution using the GMM. To account for uncertainties in distance, we employed a Monte Carlo simulation (with $1000$ iterations) to vary distances within their error margins (assuming a normal distribution). The individual half-maxima bins and their uncertainties were determined using the median and absolute median deviation.

The obtained half-maxima distances (and their uncertainties), along with median Galactic longitudes and latitudes, were used to estimate X and Y coordinates. Errors on X and Y were derived from the uncertainties in the distances. For distances obtained using methods and reddening maps from the literature, we used a single value for distance uncertainty, $\sigma_{d}$, set at $600$\,pc. To determine the inclination angle of the bar, $\iota$, we fitted the binned Cartesian coordinates to an ellipse \citep[as per the approach used by][]{Pietrukowicz2015}.

\begin{figure}
\includegraphics[width=\columnwidth]{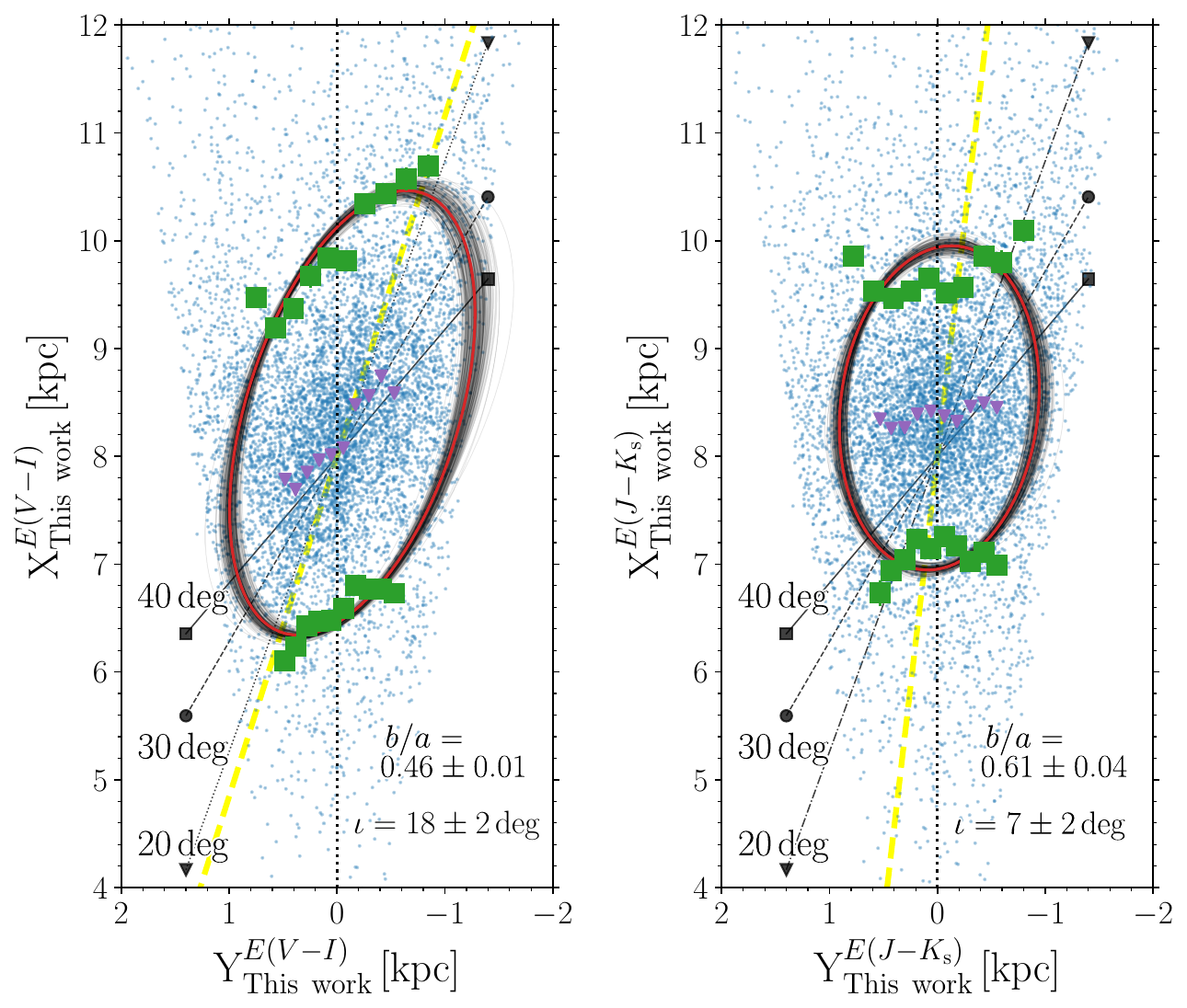}
\caption{Example of ellipse fitting (red line) to the spatial distribution for RR~Lyrae stars (blue points). We used distances from this study estimated based on $E\left ( V - I \right )$ (left-hand panel) and $E\left ( J-K_{\rm s} \right )$ (right-hand panel) color excesses. The green squares (used to derive ellipse parameters) and purple triangles represent the half-maximum and maximum of the GMM in a given bin (eleven bins between $\mathcal{l} \geq -5$ and $\mathcal{l} \leq 5$\,deg). The different bar angles are outlined with solid black dashed and dotted lines. The yellow dashed and black dotted lines mark the measured and zero angles, respectively. The fit quality expressed using $\chi^{2}$ are $0.36$, and $0.40$ for left-hand and right-hand panels, respectively.}
\label{fig:EllipseAnalysis}
\end{figure}

An ellipse is a particular form of a conic section, that can be defined by a quadratic polynomial equation:
\begin{equation} \label{eq:Ellipse}
\textrm{Ellipse}(x,y)=a \cdot x^{2} + b \cdot x \cdot y + c \cdot y^{2} + d \cdot x + e \cdot y + f = 0 ,
\end{equation}
where $a, b, c, d, e, f$ are the Cartesian conic coefficients of the fit. This definition has a specific constraint on parameters $a, b, c$; where $b^{2} - 4ac < 0$. Our method is somewhat similar to the fitting method described in \citet{Halir1998}, which involves fitting an ellipse to given data points by minimizing the algebraic distance using a least squares approach (for Equation~\ref{eq:Ellipse}). It forms a model matrix from the data points, solves a generalized eigenvalue problem to enforce the ellipse-specific constraint, and extracts the ellipse parameters from the eigenvector corresponding to the smallest positive eigenvalue\footnote{For interested readers, we provide our fitting code \url{https://github.com/ZdenekPrudil/Ellipse-fitting}}. To incorporate uncertainties in the individual X bins, we conducted a Monte Carlo error analysis, varying X values within their uncertainties and recalculated ellipse parameters for each iteration. The inclination angle of the bar was then obtained using the following relation:
\begin{equation} \label{eq:LikeHood_angle}
\iota = \textrm{atan2}\left ( b, a - c \right ) / 2.
\end{equation}
Besides the $\iota$ between the semi-major axis from the x-axis we also obtained values for semi-major ($a$) and semi-minor ($b$) axes\footnote{Using the formulas from \url{https://mathworld.wolfram.com/Ellipse.html}.} which we used to estimate the $b/a$ ratio. We report both the angle and axis ratio as the average and standard deviation of the resulting parameter distribution. Figure~\ref{fig:EllipseAnalysis} illustrates an example of our analysis, showcasing the measurement of the bar inclination angle (for conditions on distance, please refer to Subsection~\ref{subsec:VIJKangles} and conditions in Eq.~\ref{eq:condiForGMM},~\ref{eq:condiForGMM1}, and~\ref{eq:condiForGMM2}). In the following sections, we present the optimal values of $\iota$ and $\sigma_{\iota}$ as measurements of the bar angle relative to the line-of-sight for our dataset. Lastly, we report the quality of the ellipse fit for each angle measurement. Although the ellipse may not always be the optimal model for the underlying distance distribution, we use it here to facilitate comparison with the work of \citet{Pietrukowicz2015} and to obtain an approximate estimate of the bar parameters. Each following spatial plot where we measure the bar angle is also accompanied by contours (marked with green lines) based on the Kernel Density estimate (KDE). The contour levels were selected to avoid our spatial distribution's center and edges to evade distortion.

\subsection{3D Spatial Distribution} \label{subsec:3Dbulge}

To analyze the spatial distribution, we use different approaches in estimating distance (combination of visual and near-infrared passbands), particularly the distances estimated in the previous Section~\ref{sec:DistancesAll}. Moreover, we look at the distance distributions derived in previous studies, particularly the seminal work of \citet{Pietrukowicz2015} and \citet{Molnar2022}. The former work utilizes PMZ relations from \citet{Catelan2004} together with reddening maps by \citet{Gonzalez2012} and reddening law from \citet{Nataf2013}. The latter study used infrared passbands in combination with Wesenheit magnitudes \citep{Madore1982}, period-metallicity-Wesenheit magnitude relations from \citet{Cusano2021}, and $R_{JK_{\rm s}}$ from \citet{Alonso-Garcia2017}. We also applied the same criteria as in \citet{Pietrukowicz2015} to create a data set with identical stars that differ in distance estimation method. Lastly, we calculated Cartesian coordinates for each set of distance estimates (X, Y, and Z centered on the Sun). Therefore, we compare the same stars with different distance estimates (in total, more than $16000$ RR~Lyrae variables). To compare their distribution with the position of the bar, we used data provided by \citet{Gonzalez2011}. We note that we shifted the bar distance to match our \citet{Pietrukowicz2015} and \citet{Molnar2022} distances. Lastly, in this case, we did not use \textit{Gaia} flags to remove possible blended objects (conditions in Eq.~\ref{eq:Condition1}) since the aforementioned studies did not use these criteria either. On the other hand, when estimating the tilt of the bar, we focused on the area below the Galactic plane ($-6 \leq \mathcal{b} \leq -2$), similar to the study by \citet{Pietrukowicz2015}.

\begin{figure}
\centering
\includegraphics[width=1.\columnwidth]{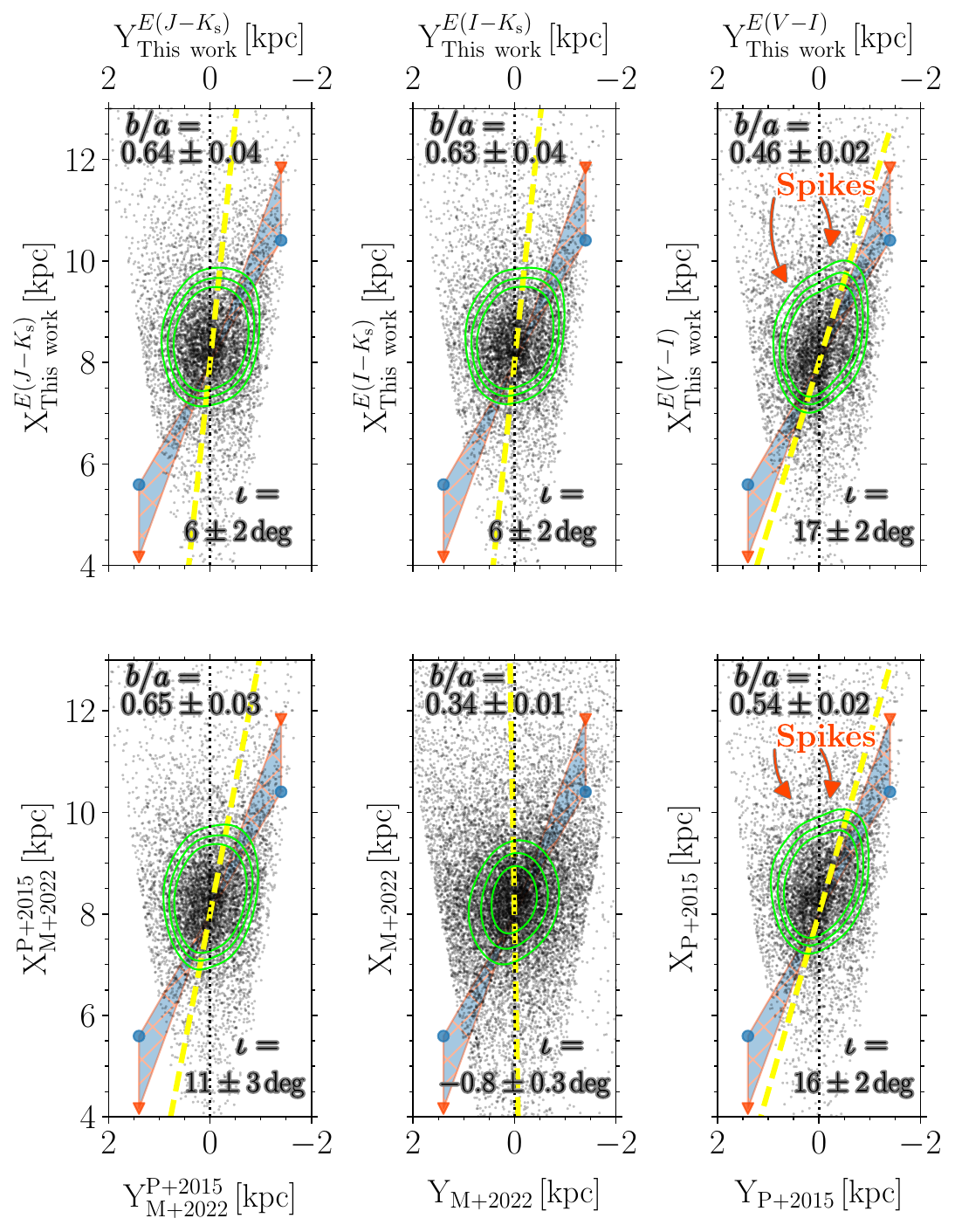}
\caption{The spatial distribution of RR~Lyrae pulsators used for comparison in the Cartesian space. Individual subplots display different distance estimates based on near-infrared (left column), a combination of near-infrared and visual, and only visual reddening. The top panels show Cartesian coordinates derived based on distances estimated in this study. The bottom panels display the distances derived based on procedures from \citet{Pietrukowicz2015} (P+2015, bottom right-hand panel) and \citet{Molnar2022} (M+2022, middle and left-hand bottom panel). We also show two tentative bar angles with red triangles ($20$\,degrees) and blue circles ($30$\,degrees) and a shaded region marks the angles in between. Note that the Sun is at X~$ = 0$ and Y~$ = 0$\,kpc. As depicted in Fig.~\ref{fig:EllipseAnalysis}, the measured and zero angles are marked with yellow dashed and black dotted lines. Lastly, the $\chi^2$ of each fit from top left to bottom right are: $0.56$, $0.57$, $0.35$, $0.51$, $15$, and $0.39$.}
\label{fig:SpatialComparisonCNG}
\end{figure}

In Figure~\ref{fig:SpatialComparisonCNG}, we present our spatial comparison. In the top three panels, we show the distance distribution in Cartesian coordinates derived in this study. For the distances derived based on $E\left ( J-K_{\rm s} \right )$ and $E\left ( I - K_{\rm s} \right )$, we see much smaller values for the bar angle. There is no significant tilt in the direction of the bar as observed by \citet{Gonzalez2011}. Using the method described at the beginning of this section, we found the bar angle for distances estimated using $E\left ( J-K_{\rm s} \right )$ and $E\left ( I-K_{\rm s} \right )$; $\iota = 6 \pm 2$\,deg and $\iota = 6 \pm 2$\,deg, respectively. For the distances estimated purely from visual passbands ($E\left ( V - I \right )$, top right-hand plot), we see a tilt with the bar ($\iota = 17 \pm 2$\,deg), and additional substructure, \textit{spikes}, in the bulge spatial distribution. A similar structure is also visible in the bottom right-hand panel, where we show the Cartesian spatial distribution based on distances derived in \citet{Pietrukowicz2015}. This panel shows the tilt ($\iota = 16 \pm 2$\,deg) that suggests the bar's position. We also see a similar substructure variation as for the distances derived using $E\left ( V - I \right )$. The aforementioned spikes are also visible in the top panel of Figure~1 in \citet[][see their Y vs. X plane]{Du2020}. The origin of these spikes does not appear to be physical but perhaps associated with the reddening and reddening law itself. These spikes are most prominent in regions above the Galactic plane ($\mathcal{b} > 0$\,deg where the reddening is higher than below the plane). 

The middle and left-hand bottom panels of Fig.~\ref{fig:SpatialComparisonCNG} show Cartesian coordinates for RR~Lyrae variables with distances derived through the procedure described in \citet{Molnar2022}. The middle panel shows the entire \citeauthor{Molnar2022} data set, and the left-hand plot shows only stars in common between \citet{Molnar2022} and \citet{Pietrukowicz2015}. We do not find a strongly tilted bar in either of the two panels (middle and left-hand bottom). The angles found are $\iota = -0.9 \pm 0.3$\,deg and $\iota = 10 \pm 3$\,deg, for the middle and left-hand panels, respectively. We see a rather smooth distribution, similar to our distances estimated based on $E\left ( J-K_{\rm s} \right )$ and $E\left ( I - K_{\rm s} \right )$, without any substructure seen in the distances derived through visual passbands. 

To summarize, we spatially recover the bar-like feature and its angle from the use of distances based on visual passbands. The measured angle agrees reasonably well with previous bar angle measurements using red clump giants, $\iota = 20 - 30$\,degrees \citep[see, e.g.,][]{Wegg2013,Simion2017,Leung2023,Vislosky2024}. When using distances obtained from near-infrared data, we detect only a negligible tilt for a bar-like structure. The appearance of the bar in visual passbands is likely connected to the gradient in $\Delta A_{I}$ observed for visual passbands in Section~\ref{sec:RedMapsComp}. In previous studies using optical photometry only, this gradient in $A_{I}$ is not correctly accounted for in distance determinations to RR~Lyrae stars.

\subsection{Visual vs. Infrared investigation of the discrepancy} \label{subsec:VIJKangles}

In the previous subsection, we conducted a comparative analysis between the distance estimates for RR~Lyrae stars towards the Galactic bulge found in the existing literature and our own estimates. We observed a notable discrepancy: distances derived primarily from infrared passbands present a different perspective on the substructure of the bulge compared to those based on visual passbands \citep[mirroring the divergence found in][]{Dekany2013,Pietrukowicz2015}. Our distance estimates using the $E\left ( V - I \right )$ color-excess reproduce very well the substructure traced by \citet{Pietrukowicz2015} even though \citet{Pietrukowicz2015} used different period-luminosity relations and their approach in the treatment of extinction also differed from ours. In this subsection, we attempt to rectify this discrepancy by using alternative sources of reddening for visual passbands and examining the resulting spatial distribution.

\begin{figure}
\centering
\includegraphics[width=1.\columnwidth]{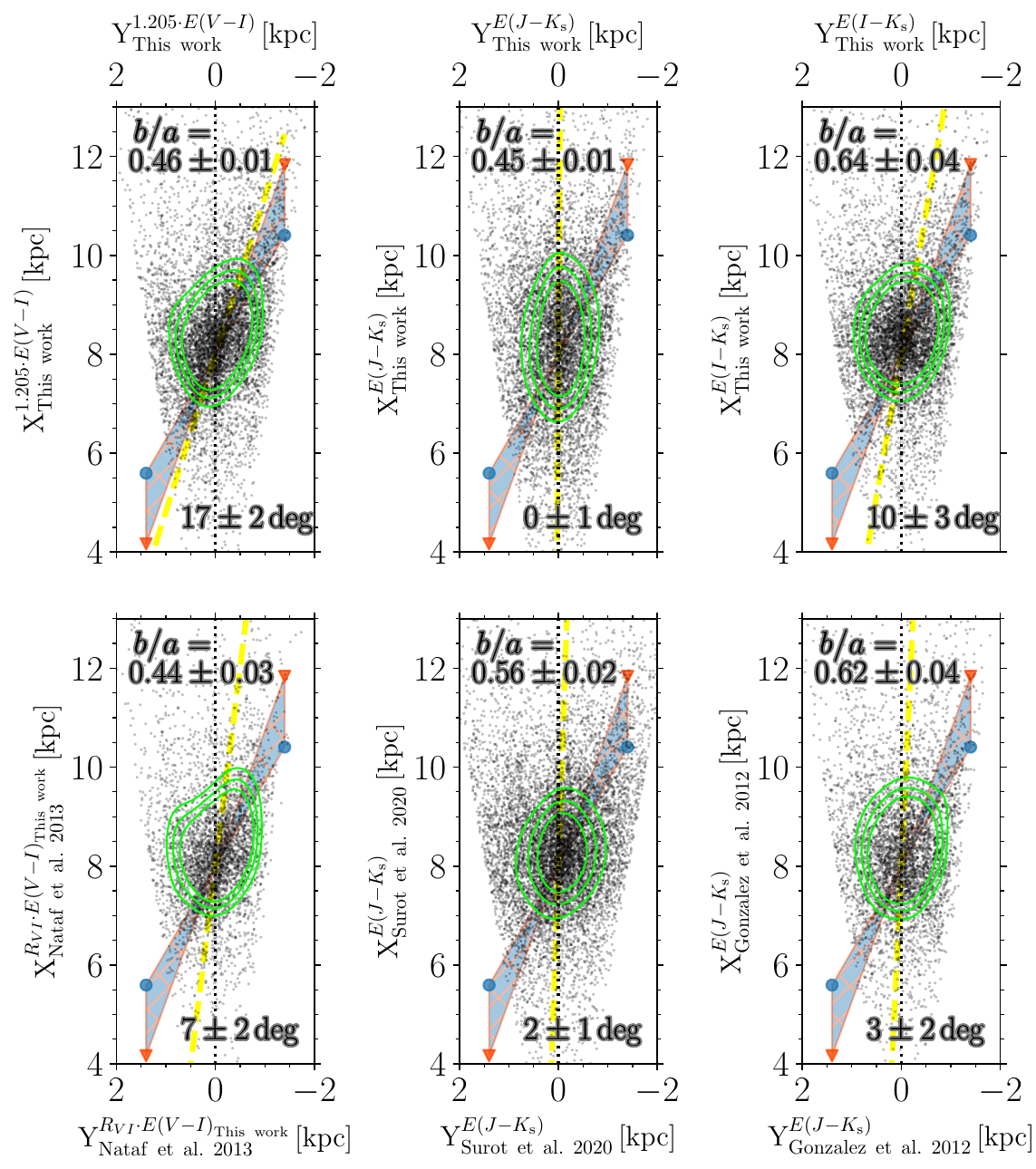}
\caption{Similar to Figure~\ref{fig:SpatialComparisonCNG} we depict spatial distribution bulge RR~Lyrae variables using absolute and mean intensity magnitudes (in $I$-band) together with different methods to account for extinction toward the Galactic bulge. The top three panels show spatial distributions derived using color excesses calculated in this work. The bottom two panels show the same spatial distributions but for distances estimated using literature reddening maps toward the Galactic bulge \citep[from left to right,][]{Nataf2013,Surot2020,Gonzalez2012}. To measure individual bar angles we binned spatial dataset into eleven bins between $\mathcal{l} \geq -5$ and $\mathcal{l} \leq 5$\,deg). As in Fig.~\ref{fig:EllipseAnalysis} we depict measured and zero angles with yellow dashed and black dotted lines, respectively. The $\chi^2$ for each fit from top left to bottom right are $0.39$, $3.4$, $0.37$, $1.6$, $1.4$, and $0.43$.}
\label{fig:SpatialComparisonWithDiffRedd}
\end{figure}

We first use $E\left ( J-K_{\rm s} \right )$ and $E\left ( I - K_{\rm s} \right )$ to estimate the extinction in $I$-band. Using the derived equations~\ref{eq:AiUsingIK} and \ref{eq:AiUsingJK}, we can calculate the extinction through $E\left ( I - K_{\rm s} \right )$ and $E\left ( J-K_{\rm s} \right )$ for $I$-band. In this comparison, we also use bulge extinction maps from \citet{Gonzalez2012} and \citet{Surot2020}, which provide $E\left ( J-K_{\rm s} \right )$ independent of our measurements, and through Eq.~\ref{eq:AiUsingJK}, we can transform them into $A_{I}$. We also utilize $R_{VI}$ from a study by \citet{Nataf2013}, which is based on OGLE passbands and red clump stars. In the following comparison, we use Eq.~\ref{eq:Condition1}, and we also imposed the following conditions on the data set to focus on the Galactic center region with the highest completeness and purity of the OGLE RR~Lyrae sample:
\begin{gather} \label{eq:condiForGMM}
-10 < \mathcal{l}\,[\text{deg}] < 10 , \\ \label{eq:condiForGMM1}
\hspace{0.3cm} \text{and} \hspace{0.3cm} -6.0 < \mathcal{b} < -2.0 , \\ \label{eq:condiForGMM2}
\hspace{0.3cm} \text{and} \hspace{0.3cm} 1.0 < d\,\text{[kpc]} < 20.0 \hspace{0.5cm}.
\end{gather}
The newly calculated spatial distributions using various sources for reddening for estimation $A_{I}$ are depicted in Figure~\ref{fig:SpatialComparisonWithDiffRedd}. In this figure, we see a clear distinction between the spatial distribution estimated using our $E\left ( V - I \right )$ color-excess, and those with reddening estimated through near-infrared passbands ($E\left ( I-K_{\rm s} \right )$ and $E\left ( J-K_{\rm s} \right )$). The disparity is also quantified in the measured bar angle, where our distances based on visual passbands yield an apparent tilt angle of the bar around $18 \pm 2$\,deg, while all other near-infrared reddening estimates provide shallower and/or insignificant angles. In addition, using the varying reddening law from \citet[][$R_{VI}$]{Nataf2013}, we also observe a shallower angle for the bar. Based on this comparison, it appears that the reason behind the discrepancy in studies like \citet{Dekany2013} and \citet{Pietrukowicz2015} is the use of $E\left ( V - I \right )$ reddening and a single reddening law. Therefore, the observed gradient along the Galactic longitude in extinction difference $\Delta A_{I}$ found in Section~\ref{sec:RedMapsComp} for visual passbands appears to mimic the tilt of the bar but is an artifact of the dust properties.

\begin{figure}
\includegraphics[width=\columnwidth]{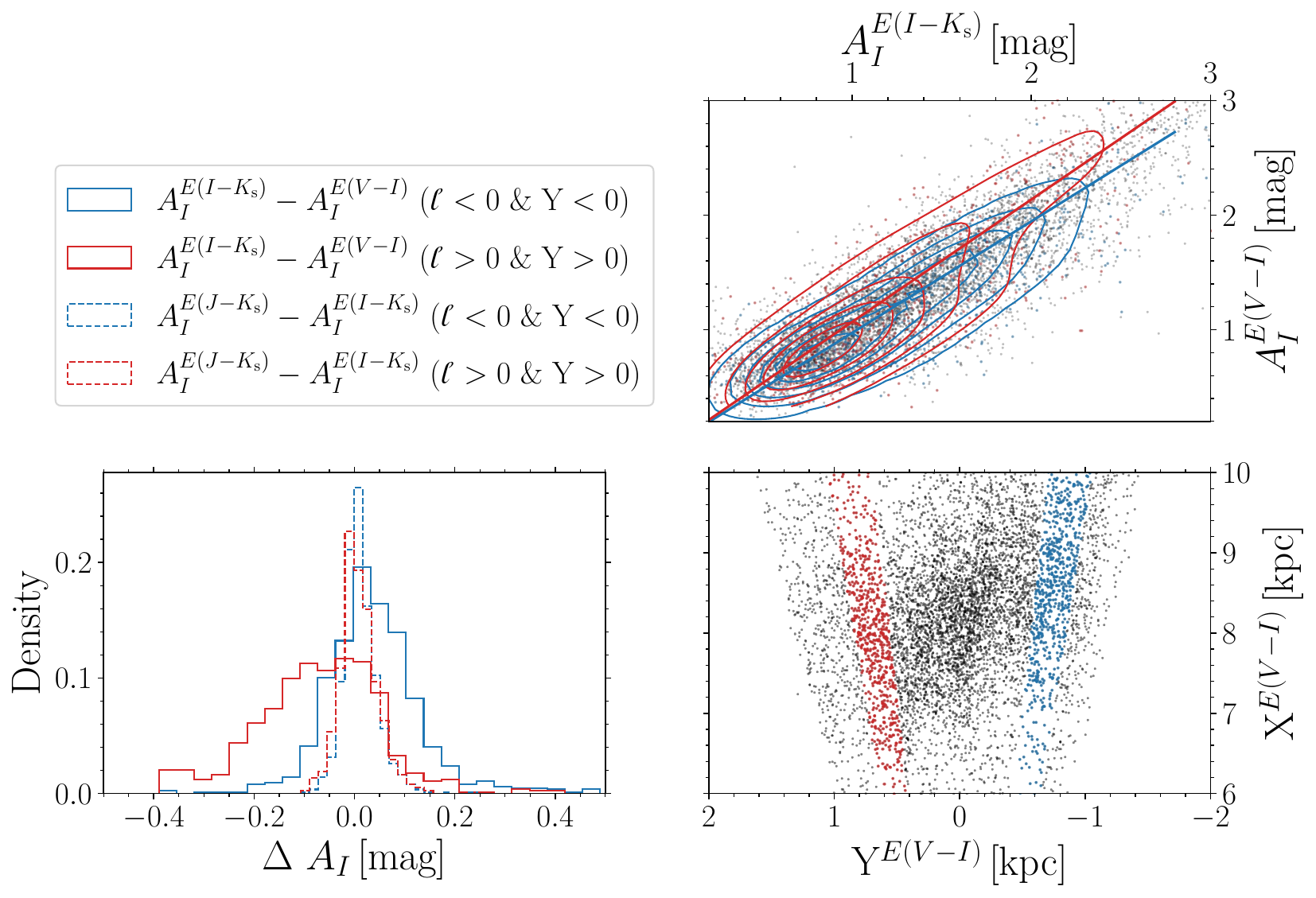}
\caption{Comparison of $A_{I}$ extinction estimates in two Galactic longitude bins. The red lines and points represent the bin at positive Galactic longitudes, while the blue points and lines represent stars at negative Galactic longitudes. The dashed lines in the histogram represent the difference in $A_{I}$ for near-infrared-based color-excesses, while solid lines show the difference for mainly visual passband-based color-excesses.}
\label{fig:ComparisonAi}
\end{figure}

To further explore the discrepancy between visual and near-infrared sources of extinction, we compared two Galactic longitude bins where the difference between the bar inclination is the most significant among visual and near-infrared-based color-excesses. In this comparison (depicted in Figure~\ref{fig:ComparisonAi}), we did not use any reddening source from the literature, but, we used solely color-excess and reddening laws estimated in this work. We focused on the following two bins:
\begin{gather}
\text{Bin-1} = -6 < \mathcal{l} < -4 \\
\text{Bin-2} = 4 < \mathcal{l} < 6 \hspace{0.5cm}.
\end{gather}
We compared the estimated $A_{I}$ based on $E\left ( V - I \right )$ and $E\left ( I - K_{\rm s} \right )$ (solid blue and red lines in the bottom left histogram). We also included a comparison for $A_{I}$ estimated using $E\left ( I - K_{\rm s} \right )$ and $E\left ( J - K_{\rm s} \right )$ (dashed blue and red lines in the bottom left histogram) which served as a baseline for the comparison. 

The assessment showed that the average difference between $A_{I}$ calculated using $E\left ( I - K_{\rm s} \right )$ and $E\left ( J - K_{\rm s} \right )$ in both bins is negligible (accounts to $0.02$\,mag which translates to approximately $80$\,pc at a distance to the Galactic bulge). On the other hand the average difference in $A_{I}$ estimated from $E\left ( V - I \right )$ and $E\left ( I - K_{\rm s} \right )$ is almost a factor of six larger ($0.12$\,mag in $\Delta A_{I}$, in distance approximately $470$\,pc). This disparity only grows larger for more reddened stars and reaches above $1$\,kpc. This analysis is in agreement with the results presented in Subsection~\ref{subsec:ReddeningMapThiswork}.

\subsection{Spikes} \label{subsec:Spikes}

To investigate the nature of the spikes and substructure variation, we explored the possibility of variation in the reddening law by selecting three regions in the Galactic bulge based on the Galactic longitude. There, we independently determined reddening laws for $E\left ( V - I \right )$ and for comparison also for $E\left ( J-K_{\rm s} \right )$ in the same way as in Section~\ref{sec:DistancesAll} but with smaller bins ($50$) and step size ($20$). We note, that we did not restrict our data set in the Galactic latitude. The results of this analysis are listed in Table~\ref{tab:ModifyRedLaw} and displayed in Figure~\ref{fig:ReddLawModified}. We found a significant variation in $R_{VI}$ based on Galactic longitude and also small changes in $R_{JK_{\rm s}}$. When we apply the modified $R_{VI}$ based on the star's Galactic longitude, it leads to a small decrease in the substructure visible in Fig.~\ref{fig:SpatialComparisonCNG} (see panel with modified distances in Appendix~\ref{fig:ReddLawModified}). The spikes do not fully disappear since the variation in the reddening law is most likely on sub-degree levels, and we did not consider changes in reddening law in the Galactic latitude direction \citep{Nataf2013,Schlafly2016}. We also report a change in reddening law for $R_{JK_{\rm s}}$ albeit smaller than for $R_{VI}$. Quantitatively speaking, the variation in $R_{VI}$ results in changes in distances on average of $0.6$\,kpc while for $R_{JK_{\rm s}}$ the changes are on average below $0.2$\,kpc.

\begin{table}
\caption{Table listing the variation in the reddening law. The first column lists the analyzed regions in the Galactic bulge. The second and third columns list the measured variation in the reddening law.}
\label{tab:ModifyRedLaw}
\begin{tabular}{lcc}
\hline \hline
       & $R_{VI}$ & $R_{JK_{\rm s}}$ \\ \hline
Spike-1 $\mathcal{l}=(-2.0; -1.0)$ & $1.24 \pm 0.02$ & $0.45 \pm 0.04$ \\
Center $\mathcal{l}=(0.0; 1.0)$ & $1.18 \pm 0.02$ & $0.42 \pm 0.03$ \\
Spike-2 $\mathcal{l}=(2.0; 3.0)$ & $1.28 \pm 0.02$ & $0.52 \pm 0.04$ \\ 
Global & $1.205 \pm 0.006$ & $0.487 \pm 0.014$ \\ \hline
\end{tabular}
\end{table}

To further explore the origin of these spikes, we compared for this data set our reddening values for $E(J-K_{\rm s})$ and $E(V-I)$ with values estimated based on work by \citet{Pietrukowicz2015}. In Figure~\ref{fig:ComparisonRedd}, we displayed our comparison and noticed a clear linear trend with a minimum offset and scatter lower than our average uncertainty on $E(J-K_{\rm s})$. We also displayed stars approximately associated with the two spikes shown in Figure~\ref{fig:SpatialComparisonCNG} (selected based on their X and Y coordinates, particularly in regions with X~$>8.5$\,kpc). Variables in these spikes do not deviate from the overall trend in $E(J-K_{\rm s})$ and $E(V-I)$ reddening comparisons. Their reddening does not appear to be over/under-estimated. We emphasize that we are comparing $E(J-K_{\rm s})$ derived here with $E(J-K_{\rm s})$ from \citet{Gonzalez2012}, and $E(V-I)$ estimated in this work with $E(V-I)$ calculated using the method outlined in \citet{Pietrukowicz2015}. We do not compare the extinction derived from $E(J-K_{\rm s})$ from $E(V-I)$ as we did in Section~\ref{sec:RedMapsComp}.

Thus, the probable reason behind the spikes lies more on the visual side of the reddening and reddening law itself. This is supported by the spikes being more prominent in the spatial map derived using our estimated $E(V-I)$ but not the spatial map derived by $E(J-K_{\rm s})$. The visual passbands are more affected by extinction, and varying the reddening law in the Galactic latitude and longitude \citep[as seen, e.g., in][]{Nishiyama2006,Nataf2013,Schlafly2016} could result in a suitable mix for artificial structures to appear in the spatial distribution. For example, in our work, we use a single universal reddening law for all variables in our data set. This might work well for the near-infrared data (smooth distribution in spatial properties, see top panels of Fig.~\ref{fig:SpatialComparisonCNG}), where the reddening variations are subtler, thus leading to a much lower impact on the results. On the other hand, such a universal \textit{single reddening law} approach probably fails in the visual, where even if the reddening values agree with those determined by different methods (see Figure~\ref{fig:ComparisonRedd}), the likely variation in the reddening law in the Galactic longitude and latitude in a reddening sensitive part of the spectrum creates artificial substructures. Lastly, the severity of the difference between reddening treatment for distances from purely $J$ and $K_{\rm s}$, and those derived only from $V$ and $I$ is that reddening is, on average, three times higher for the latter. This emphasizes the importance of proper extinction treatment toward the Galactic bulge, especially when using visual passbands where the variation in reddening law should be included. Lastly, the known issue of the VVV photometric calibration \citep[see][]{Hajdu2020} is most likely not responsible for these spikes since we also see them in our distance calibration for $E\left ( V - I \right )$ where we did not use VVV photometry.

\section{Metallicity and spatial distribution} \label{subsection:metal3d}

We explore the differences in spatial properties of our RR~Lyrae dataset by dividing it into two groups based on estimated photometric metallicity. We categorized the dataset into metal-poor ([Fe/H]$_{\rm phot} < -1.0$\,dex) and metal-rich ([Fe/H]$_{\rm phot} > -1.0$\,dex) RR~Lyrae stars with distances estimated using near-infrared ($E(J-K_{\rm s})$) passbands. The boundary between the metallicity bins was selected based on Figure~\ref{fig:BarAngleMET} and by the work of \citet{Crestani2021Alpha} and \citet{Prudil2020Disk}, where we see that metal-rich RR~Lyrae stars with spectroscopic [Fe/H]~$>-1.0$\,dex have, in general, very lower [$\alpha$/Fe] abundance, and thus the halo contamination in the metal-rich bin should be minimal.

\subsection{Different approach in estimating bar angle} \label{subsec:TestingTensor}

In this Section, we implemented a different approach for estimating the bar angle for the metal-rich RR~Lyrae population due to their scarcity, as well as for tracing the bar angle across different metallicity bins. Specifically, we use the inertia tensor (assuming all masses are equal to one) and its eigenvalues to estimate the bar angle and the axis $b/a$ ratio. In this method, we calculate the cylindrical radius ($R_{\rm cyl}$) using the Cartesian coordinates X and Y, and use only stars within a given limit, $R_{\rm cyl}^{\rm lim}$.

To test this method, we first conducted a simple simulation in which we generated two sets of spatial (X, Y, and Z) distributions. The first set consisted of a uniform (unbarred) distribution across all three Cartesian coordinates:
\begin{gather}
p({\rm X, Y, Z}) = \mathcal{U}(4.0 < \textrm{X} < 12.0)\,\textrm{kpc}\\
\hspace{1.9cm} \mathcal{U}(-4.0 < \textrm{Y} < 4.0)\,\textrm{kpc}\\
\hspace{1.9cm} \mathcal{U}(-4.0 < \textrm{Z} < 4.0)\,\textrm{kpc} \hspace{1cm}.
\end{gather}
The second set followed an ellipsoidal distribution with the spatial properties of the Galactic bar \citep[$\iota = 27$\,deg, $b/a = 1.0/2.1$, and $c/a = 0.82/2.1$, based on][]{Wegg2013}. Both distributions were convolved with our estimated distance uncertainties (six percent for $E(J-K_{\rm s})$ distances) and restricted based on our \textit{cone-view} of the Galactic bulge (Fig.~\ref{fig:EllipseAnalysis}) to better simulate our observational dataset. The two distributions were then combined, keeping the size of the ellipsoidal population fixed while increasing the size of the axisymmetric population. We also varied the limiting $R_{\rm cyl}^{\rm lim}$ and examined how these changes affected the recovered bar angle and $b/a$ axis ratio. For comparison, in addition to the cone-view, we included the full simulated dataset, unconstrained by the cone.

\begin{figure}
\includegraphics[width=\columnwidth]{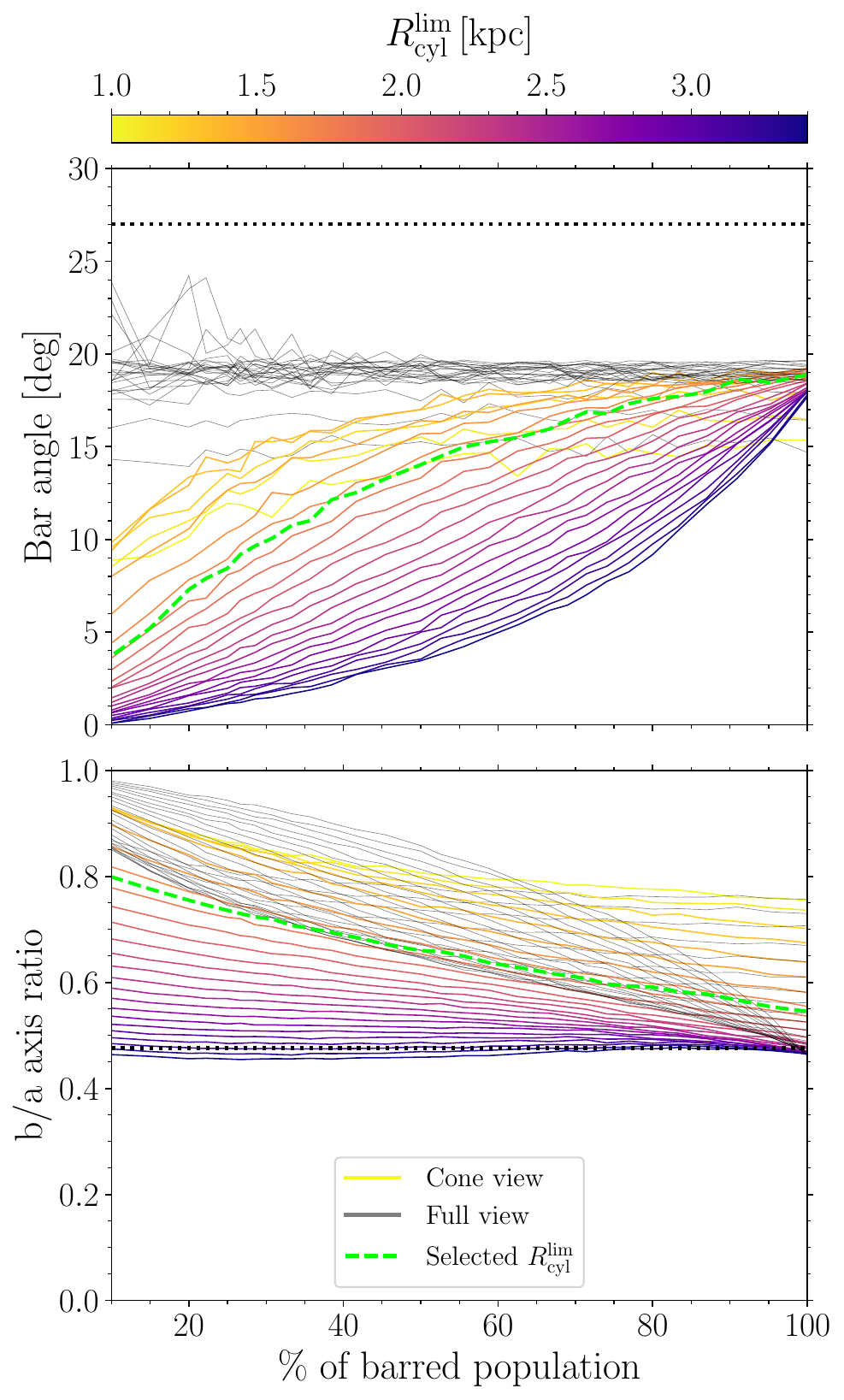}
\caption{The test of the inertia tensor method for estimating the bar angle (top panel) and the $b/a$ axis ratio (bottom panel) for our simulated dataset. The black solid lines represent the simulated, unrestricted view of the Galactic bar for different values of $R_{\rm cyl}^{\rm lim}$. The color-coded solid lines show measurements for different $R_{\rm cyl}^{\rm lim}$ in the cone-view of the Galactic bar. The dashed green line indicates the selected value of $R_{\rm cyl}^{\rm lim}=1.75$\,kpc used in our observational dataset. The dotted black line in both panels denotes the true value of the bar angle and axis ratio set for the ellipsoidal distribution.}
\label{fig:TestTensor}
\end{figure}

In Figure~\ref{fig:TestTensor}, we present our test results. From the unconstrained view, we observe that the true bar angle is never accurately estimated. Despite variations in the limiting $R_{\rm cyl}^{\rm lim}$ and the size of the axisymmetric population, the bar angle consistently fluctuates around $19$\,deg in nearly all cases. This effect is due to the convolution of the simulated distributions with our distance uncertainties, which cause the bar angle to appear smaller \citep[][]{Hey2023,Vislosky2024,Zoccali2024}. The situation is similar for the cone view; the true bar angle is never fully recovered. Only for the smallest values of $R_{\rm cyl}^{\rm lim}$, and with minimal contamination from the axisymmetric population, does the measured bar angle reach $18$-$19$\,deg, constrained by the distance uncertainties. On the other hand, for the axis ratio, higher values of $R_{\rm cyl}^{\rm lim}$ better capture the overall axis ratio despite the presence of distance errors.

In the end, we decided to use $R_{\rm cyl}^{\rm lim} = 1.75$\,kpc for our observational dataset. This decision was driven by the test above and by the narrowness of our dataset in Y coordinate (Fig.~\ref{fig:EllipseAnalysis}). By including stars from larger cylindrical radii we would preferentially select variables in the foreground and background of the Galactic bulge while not increasing the amount of stars in the bar region.

\subsection{Bar angle as a function of metallicity} \label{subsec:MetalBaryAng}

We again applied conditions in Eq.~\ref{eq:Condition1} and selected stars within $\mathcal{b} = (-2, -6)$ and $\left | \mathcal{l} \right | < 10 $\,degrees.

\begin{figure}
\includegraphics[width=\columnwidth]{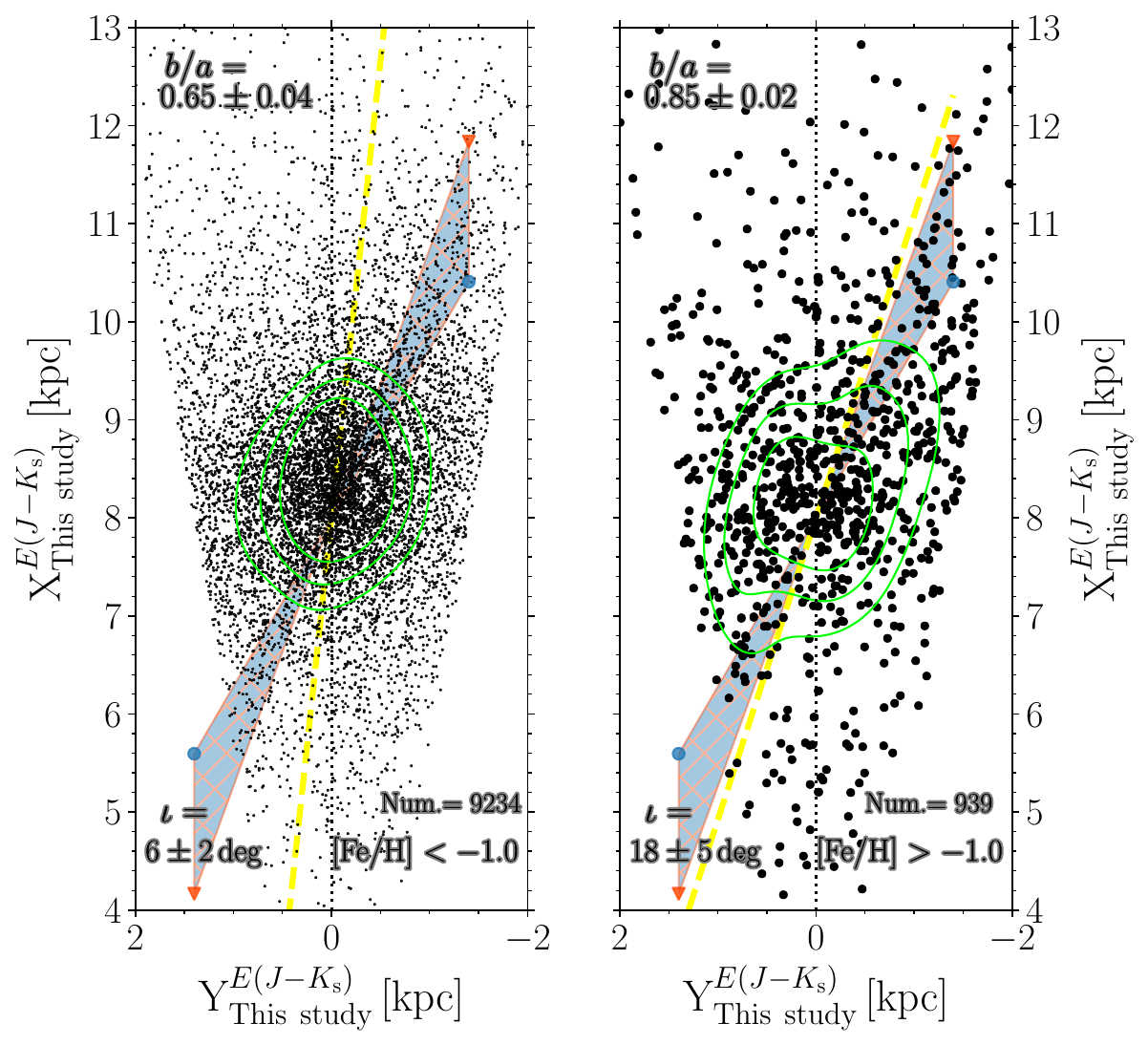}
\caption{The spatial distribution analysis for near-infrared distances derived using $E(J-K_{\rm s})$. The right-hand and left-hand panel show metal-poor ([Fe/H]$_{\rm phot} < -1.0$\,dex) and metal-rich ([Fe/H]$_{\rm phot} > -1.0$\,dex) RR~Lyrae variables in our data set (black dots). Two approximate bar angles ($20$ and $30$\,degrees) are shown with red triangles and blue circles, respectively with shading marking the angles in between. Similarly to Fig.~\ref{fig:EllipseAnalysis} we show measured and zero angles with yellow dashed and black dotted lines. Lastly, the quality of the ellipse fit for the left-hand panel is $\chi^{2} = 0.45$}
\label{fig:SpatialFEHcomparison}
\end{figure}

In Figure~\ref{fig:SpatialFEHcomparison}, we display comparisons for both metal-rich and metal-poor samples. We see that the prominence of the bar angle changes differs for both metallicity groups. Specifically, the metal-poor population shows little to no indication of the bar, as evidenced by the low bar angle ($\iota = 6 \pm 2$\,deg). We note that the value of the bar angle in this case was derived using the ellipse fitting method (described in Section~\ref{sec:bulge3D}). We also used the inertia tensor approach to verify this value, obtaining $\iota = 9 \pm 1$\,deg and $b/a = 0.82 \pm 0.01$. Furthermore, we implemented a modified 2D version of the inertia tensor method as described in \citet[][see Section~4 in later]{Zemp2011,Pulsoni2020}. This method iteratively determines the bar angle and axis ratio. The initial selection of stars is adjusted iteratively to better match the ellipsoidal shape of the bar. After each iteration, the axis ratios and bar angle are updated based on the newly selected set of stars, according to the current estimate of the bar's orientation and ellipticity. This process is repeated until a specified convergence criterion is met (in our case, when the change in the axis ratio between iterations is smaller than one percent). Using this approach, we obtained $\iota = 5 \pm 1$\,deg and $b/a = 0.58 \pm 0.01$, once again confirming the low bar angle for the metal-poor RR~Lyrae population. 

Conversely, the bar signature is clear in the metal-rich population, with a tilt of $\iota = 18 \pm 5$\,deg\footnote{The uncertainty on the angle was derived by variation of X and Y coordinate within their errors.}. We note that the derived bar angle is lower but within the errors in agreement with what is typically assumed for the Milky Way bar based on red clump stars. This discrepancy may be attributed to the low number of RR~Lyrae stars in our metal-rich bin (suggesting that the Galactic bulge does not contain many metal-rich RR~Lyrae pulsators) or to kinematic fractionation. According to \citet{Debattista2017}, kinematic fractionation occurs when initially co-spatial stellar populations with different in-plane random motions (radial velocity dispersions) separate during bar formation and evolution. Metal-poor populations, which are generally older and have higher radial velocity dispersions, tend to form a weaker bar and become vertically thicker, leading to a boxy bulge structure. On the other hand, metal-rich populations, with lower initial radial velocity dispersions, form a stronger bar and a more pronounced peanut-shaped bulge. This difference in bar strength and shape could potentially lead to a tilt between the metal-poor and metal-rich populations. However, as indicated by the simulations, this effect may not be universally applicable and could depend on specific initial conditions of the bulge and the evolutionary history of the bar \citep{Debattista2017,Fragkoudi2017}. Lastly, we note that the difference between the bar angles derived from metal-poor and metal-rich is on the borderline for being statistically significant ($p_{\rm value} = 0.03$).

\begin{figure}
\includegraphics[width=\columnwidth]{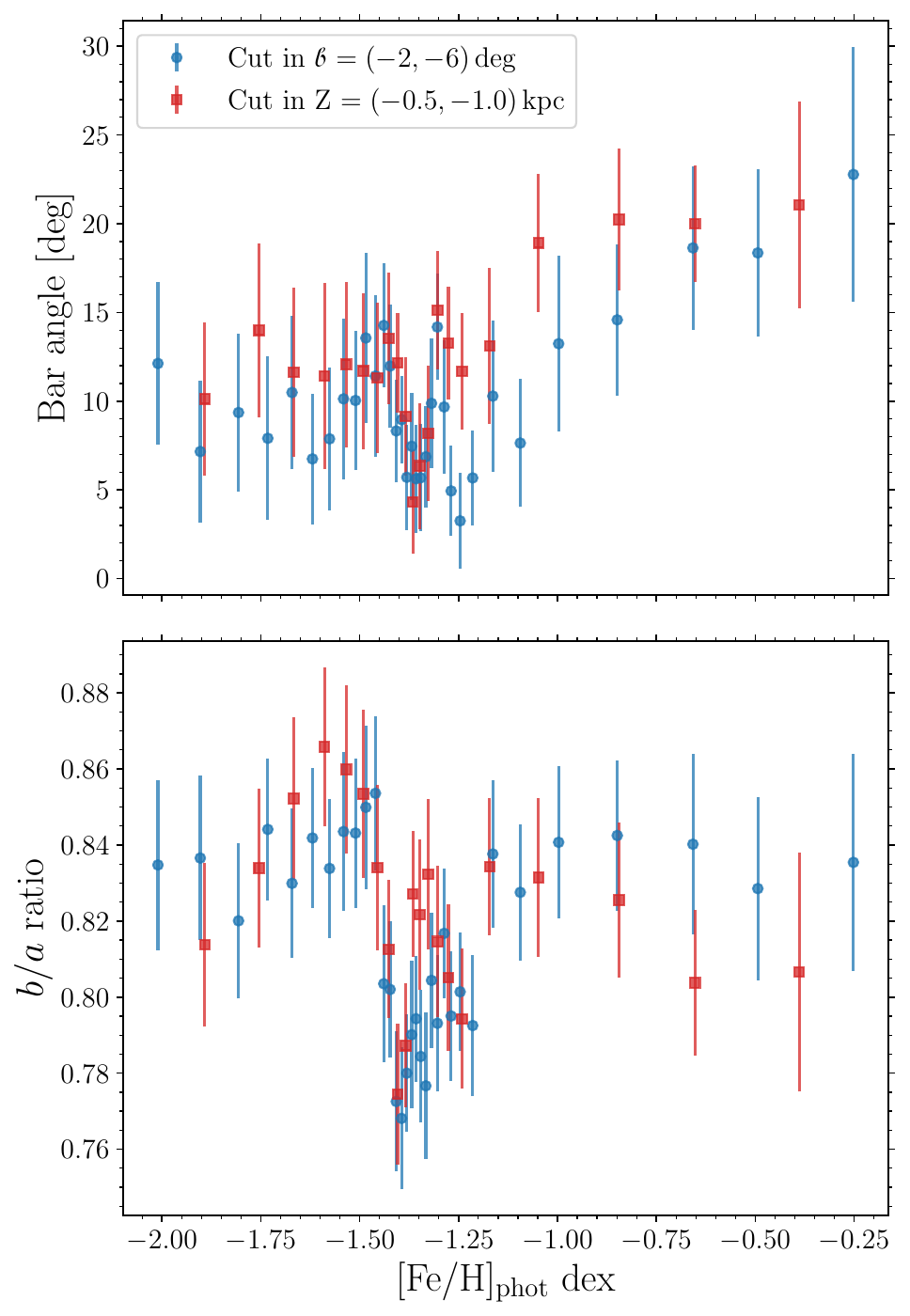}
\caption{The dependence of the estimated bar angle (top panel) and axis ratio $b/a$ (bottom panel) as a function of metallicity for RR~Lyrae stars. We included both the angle and $b/a$ measurements for two sets of conditions including cuts in Galactic latitude (blue circles) or Z coordinate (red squares). The values for individual bins are listed in the Appendix (see Tables~\ref{tab:MetalicityBar} and Table~\ref{tab:MetalicityBar2}).}
\label{fig:BarAngleMET}
\end{figure}

Furthermore, we explore the bar angle's and axis ratio's dependence on the metallicity of RR Lyrae stars. In this examination, we use solely the angle and axis ratio derived from the inertia tensor and its eigenvalues. We use a moving boxcar method to split our metallicity distribution into bins of equal size ($800$ variables each) and step size equal to $300$. For each metallicity bin, we estimated the bar angle and axis ratio and their uncertainties; the dependence found is shown in Figure~\ref{fig:BarAngleMET}.

As we move from the metal-rich to the metal-poor end of our metallicity distribution, we see that the bar prominence decreases. We see that at the most metal-rich bin, we find the bar angle equal to $\iota = 23 \pm 7$\,degrees. Looking at the metal-poor RR~Lyrae stars ([Fe/H]$_{\rm phot} < -1.0$\,dex) the bar angle varied by around $10$\,degrees. To verify that our criteria on $\mathcal{b}$ did not bias our angle estimates, we modified our criteria by replacing the condition on Galactic latitude with a condition on the Cartesian Z coordinate (Z~$= (-0.5, -1.0)$\,kpc, red points in Fig.~\ref{fig:BarAngleMET}). We observe that both conditions on the vertical extent of our dataset yield consistent results, with the prominence of the bar angle increasing when [Fe/H]$_{\mathrm{phot}} > -1.0$\,dex.

On the other hand, for the axis ratio, we observe a different trend. The ratio remains fairly constant around $b/a \approx 0.82$, but at the peak of the metallicity distribution function for bulge RR~Lyrae stars (photometric metallicity, [Fe/H]$_{\rm phot} = -1.4$\,dex), the axis ratio drops to $0.78$. As discussed in Subsection~\ref{subsec:TestingTensor} and shown in Figure~\ref{fig:TestTensor}, this may be partially influenced by the \textit{non-barred} population (e.g., potential halo interlopers). If we follow the results from the bottom panel of Fig.~\ref{fig:TestTensor} and select $R_{\rm cyl}^{\rm lim} = 3.5$\,kpc, we would obtain $b/a$ values between $0.52$ and $0.6$ and thus much closer to the values obtained using red clump stars. We also tested how the dependence of the bar angle on metallicity would appear if we used $R_{\rm cyl}^{\rm lim} = 1.0$\,kpc (based on our test, this radius recovers the true angle in most cases). Using this limiting radius, we obtain, on average, higher values for the bar angle (approximately five degrees higher), but the individual uncertainties on the measured angle also increase to nine degrees on average. 

\section{Galactic bulge in 5D} \label{sec:5DBulge}

This Section delves into the transverse velocities of RR~Lyrae stars and examines how these velocities align with those of the more metal-rich population of red clump stars analyzed by \citet{Sanders2019} toward the Galactic bulge \citep[see also][]{Clarke2019}. The metal-rich red-clump population seems to follow a more barred spatial distribution than here analyzed RR~Lyrae stars \citep[e.g.,][]{Zoccali2017,Lim2021}. Our findings are also compared to \citet{Du2020}'s study, which investigates RR~Lyrae stars using a different distance estimation method and preceding \textit{Gaia} data release (\textit{Gaia} DR2).

\subsection{Galactic bulge in 5D - transverse velocities} \label{subsec:vtTransverse}

In this subsection, we follow the innovative analysis by \citet{Du2020} and examine the rotation of RR~Lyrae stars using the transverse velocities calculated with our distances and updated and improved proper motions from the \textit{Gaia} DR3. As noted in subsection~\ref{subsec:Gaia} all of the proper motions for our RR~Lyrae sample come from the \textit{Gaia} astrometric catalog. To assess the reliability of these proper motions, we employed the following three cuts:
\begin{gather} \label{eq:PMcut}
\sqrt{ \sum \mathbf{V}^{\ast2} / \text{Tr}(\mathbf{\Sigma^{\ast}}) } > 5.0 ,\\
\hspace{0.3cm} \text{and} \hspace{0.5cm} \text{RUWE} < 1.4  ,\\
\hspace{0.3cm} \text{and} \hspace{0.5cm} \texttt{ipd\_frac\_multi\_peak} < 5 ,
\end{gather}
where the first condition is composed of the vector $\mathbf{V}$ containing proper motions in right ascension and declination ($\mu_{\alpha^{\ast}}$, $\mu_{\delta}$) and their covariance matrix $\mathbf{\Sigma}$ (with uncertanities $\sigma_{\mu_{\alpha}^{\ast}}$, $\sigma_{\mu_{\delta}}$, and correlation $\rho_{\mu_{\alpha}^{\ast},~\mu_{\delta}}$). The covariance matrix was scaled by the RUWE coefficient and diagonalized, yielding a transformed covariance matrix $\Sigma^{\ast}$ to remove the correlation between proper motions that would otherwise affect the cut in the first condition in Eq.~\ref{eq:PMcut}. For the ratio between the transformed vector $\mathbf{V}^{\ast}$ and covariance matrix $\Sigma^{\ast}$, we required at least five $\sigma$ confidence on the transformed proper motions. The second and third conditions match those in Eq.~\ref{eq:Condition1}. 

To compute transverse velocities we use the definition for $v_{\mathcal{l}}$ and $v_{\mathcal{b}}$ from \citet{Koppelman2021}:
\begin{gather}
v_{\mathcal{l}} = 4.74057 \cdot \mu_{\mathcal{l}} \cdot d - U_{\odot}\sin\mathcal{l} + \left ( V_{\odot} + v_{\rm LSR} \right )\cos\mathcal{l} , \\
v_{\mathcal{b}} = 4.74057 \cdot \mu_{\mathcal{b}} \cdot d \\ \hspace{0.6cm} + W_{\odot}\cos\mathcal{b} - \sin\mathcal{b} \cdot  \left ( U_{\odot}\cos\mathcal{l} + \left ( V_{\odot} + v_{\rm LSR} \right )\cdot \sin\mathcal{l} \right ) \hspace{0.2cm} ,
\end{gather}
where $U_{\odot}, V_{\odot}, W_{\odot} = \left (11.1,~12.24,~7.25\right )$\,km\,s$^{-1}$ represent the Solar motion \citep{Schonrich2010} and $v_{\rm LSR} = 230$\,km\,s$^{-1}$ is the velocity of the local standard of rest \citep{Eilers2019}. To estimate transverse velocities and preserve the correlation in conversion from proper motion vectors in equatorial ($\mu_{\alpha}^{\ast},~\mu_{\delta}$) to Galactic coordinates ($\mu_{\mathcal{l}}$, $\mu_{\mathcal{b}}$) we constructed for each RR~Lyrae star a multivariate normal distribution where the random values were drawn from the covariance matrix that contained products obtained from \textit{Gaia}. Since we derived distances independently, we chose to assume negligible correlations between astrometric products and our distances. Using the multivariate distribution, we also estimated Cartesian coordinates (with the Sun at $(\text{X, Y, Z}) = (0,0,0)$\,kpc).

\begin{figure*}
\includegraphics[width=2\columnwidth]{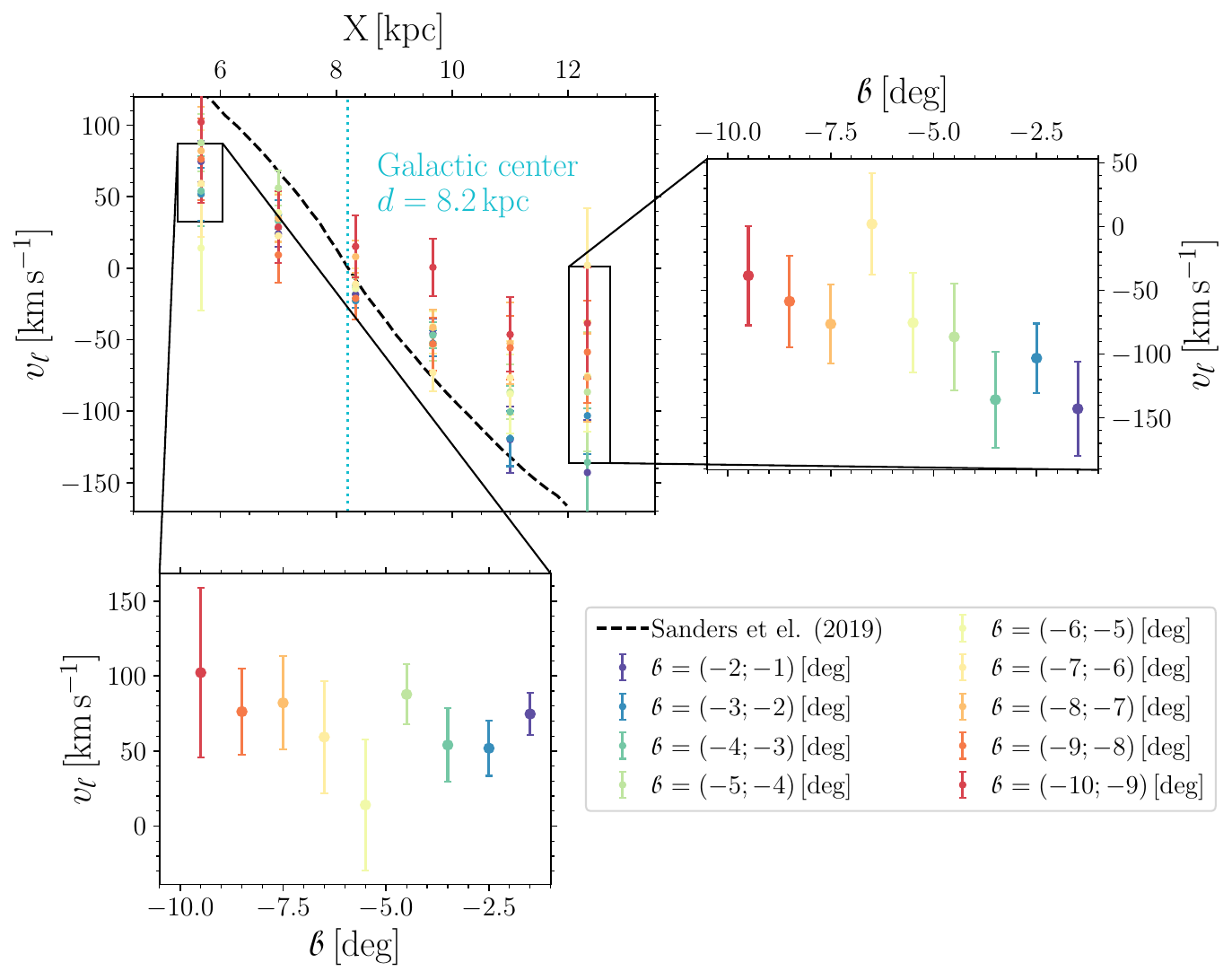}
\caption{The distribution of $v_{\mathcal{l}}$ in different distances and Galactic latitudes below the Galactic plane. The top left-hand panel shows the dependence of $v_{\mathcal{l}}$ on the distance toward the Galactic bulge color-coded according to on different Galactic latitudes. The dashed black line represents the approximate distribution of $v_{\mathcal{l}}$ from \citet{Sanders2019}, and the dotted light-blue line represents the distance to the Galactic center. The top right-hand and bottom left-hand panels show insets for two bins from the top left-hand plot, particularly for X bins at $5.67$\,kpc and $12.33$\,kpc, respectively, with the same color-coding as in the top left-hand panel. All stars were selected within $ \left|\mathcal{l}\right| < 2$\,deg. For the same Figure but for stars above the Galactic plane see Fig.~\ref{fig:RotationVtAbovePlane}.}
\label{fig:RotationVtBelowPlane}
\end{figure*}

Figures~\ref{fig:RotationVtBelowPlane} and \ref{fig:RotationVtAbovePlane} show the results for $v_{\mathcal{l}}$ with respect to the X coordinate below the Galactic plane with individual points representing a different slice in the Galactic latitude. We binned the data in X (seven bins between $5$ and $13$\,kpc, all within $ \left|\mathcal{l}\right| < 2$\,deg) and Galactic latitude (bins from $-1$ to $-10$\,deg with $1$\,deg step size) in the top left-hand panels. We see clear signs that the $v_{\mathcal{l}}$ varies with the Galactic latitude (at the further side of the bulge), yet still lags behind the rotation of the more metal-rich population measured by \citet{Sanders2019}. A similar effect was also observed using $N$-body+smooth particle hydrodynamics simulation \citep{GoughKelly2022} where the older population (tracing a weaker bar) was rotating slower than the younger \citep[as expected from kinematic fractionation][]{Debattista2017}. The dispersion of average $v_{\mathcal{l}}$ is expected to be high on the foreground and background periphery of the Galactic bulge (the most distant and closest bins), where the signs of rotation would be the strongest. As we move toward the Galactic center (here set at $d=8.2$\,kpc), any sign of rotation gets weaker and merges with the interloper contamination (halo and disk stars, assuming it is isotropic and thus centered at $v_{\mathcal{l}}=0$\,km\,s$^{-1}$). 

The insets of both aforementioned Figures highlight the variation in the Galactic latitude in the foreground and background edges of the Galactic bulge. The distribution of points across different Galactic latitudes could be interpreted as a sign of differential rotation. Here, we want to emphasize that the prominence of this effect strongly depends on the selected bins, particularly the foreground bin at X~$=5.67$\,kpc where the size of this bin changes the trend in Galactic latitude. If we choose a different set of bins, this sign of rotation decreases, but the point closest to the Galactic plane, $\mathcal{b}=(-2;-1)$\,deg, nearly always remains with the highest $v_{\mathcal{l}}$, and the lag behind the $v_{\mathcal{l}}$ rotation measured with red clump stars \citep{Sanders2019} is always prevalent.

\subsection{Metallicity effect} \label{subsec:5dContamin}

\begin{figure}
\includegraphics[width=\columnwidth]{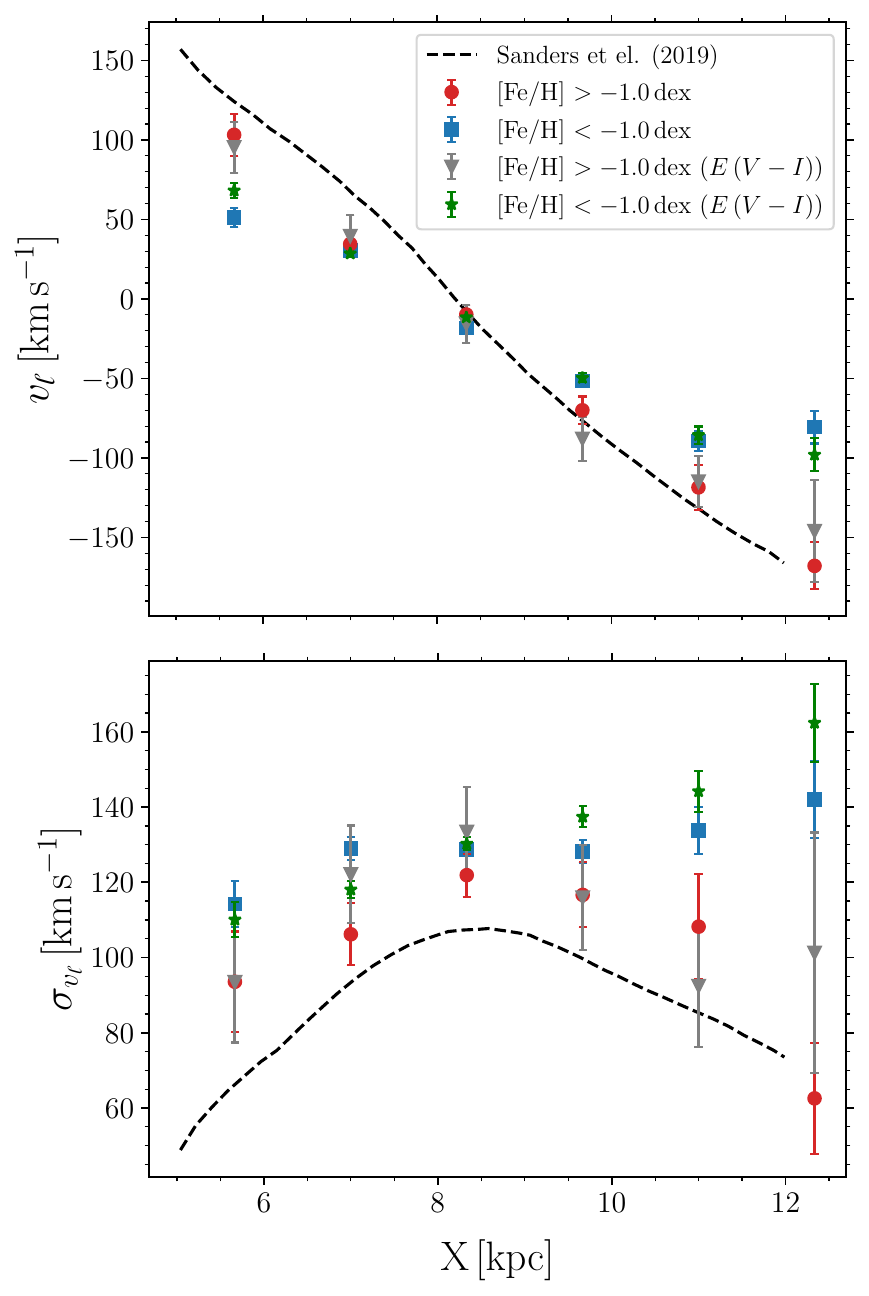}
\caption{The distribution of $v_{\mathcal{l}}$ and $\sigma_{v_{\mathcal{l}}}$ vs. X coordinate. The top panel shows how $v_{\mathcal{l}}$ changes with different distances from the Sun for two metallicity bins, metal-rich (denoted with red points) and metal-poor (represented by blue squares). The black dashed line represents results from \citet{Sanders2019} for red clump stars. The bottom panel presents a dispersion profile of $v_{\mathcal{l}}$ over the same distance range with the same two metallicity samples. The black line again represents the results from \citet{Sanders2019} for the dispersion of red clump giants. The grey and green markers denote the same metallicity bins (metal-rich and metal-poor) but for transverse velocities, dispersions, and X-coordinates estimated using distance based on visual passbands, showing that the transverse velocities do not significantly depend on the extinction map used. Similarly to Figure~\ref{fig:RotationVtBelowPlane}, all stars were selected within $ \left|\mathcal{l}\right| < 2$\,deg.}
\label{fig:TransverseVelMetalDist}
\end{figure}

In the following, we examine the transverse velocities as a function of metallicity, similar to the 3D analysis performed in Section~\ref{subsection:metal3d}. Figure~\ref{fig:TransverseVelMetalDist} provides the same RR~Lyrae properties (and the same binning in X coordinate) as in the previous Figure~\ref{fig:RotationVtBelowPlane} but divided based on photometric metallicities. In this case, we did not bin our data based on Galactic latitude but used all RR~Lyrae stars in this analysis. We divided our data set into two metallicity bins in the same way as in Section~\ref{subsection:metal3d}, with [Fe/H]~$>-1.0$\,dex, and [Fe/H]~$<-1.0$\,dex. We see that the metal-rich part of our data set almost precisely follows the rotation pattern from \citet{Sanders2019} also seen in \citet[][see their Fig.~7]{Du2020}. Particularly on the far side of the Galactic bulge, we see a drastic difference between the rotation of the metal-rich and metal-poor RR~Lyrae population. On the other hand, the metal-poor part follows the general RR~Lyrae trend seen in Fig.~\ref{fig:RotationVtBelowPlane}. In the bottom panel of Fig~\ref{fig:TransverseVelMetalDist}, we see that the dispersion in $v_{\mathcal{l}}$ ($\sigma_{v_{\mathcal{l}}}$) decreases for the metal-rich bin but does not reach the values expected from the younger, rotating population. This is probably due to contamination from RR~Lyrae stars that are associated with the Galactic disk \citep[see, e.g.,][]{Layden1996,Zinn2020,Prudil2020Disk,Iorio2021}. It is important to note that the increase in $\sigma_{v_{\mathcal{l}}}$ on the far side of the bulge is likely unphysical, as we would expect symmetry with respect to the Galactic center. This increase is probably due to greater uncertainties in distance measurements and significant errors in proper motions. Additionally, the cone-shaped view of our dataset includes more stars at higher Z on the far side of the bulge, which may lead to a higher inclusion of halo stars.

Similarly to Section~\ref{subsection:metal3d}, Figure~\ref{fig:TransverseVelMetalDist} presents a comparison of results and distances based on visual passbands ($E(V - I)$, and $m_{I}$ mean intensity magnitudes). We utilized the same metallicity bins as before and examined $v_{\mathcal{l}}$ and $\sigma_{v_{\mathcal{l}}}$. As with the previous case, metal-rich RR~Lyrae stars more closely mimic red clump stars' rotation and dispersion patterns than metal-poor ones. From Figure~\ref{fig:TransverseVelMetalDist}, it is evident that trends in $v_{\mathcal{l}}$ and $\sigma_{v_{\mathcal{l}}}$ remain consistent across different distance estimation methods. Thus, the variance between visual and near-infrared distance assessments, discussed in Section~\ref{sec:bulge3D}, has a negligible impact on the analysis of transverse velocities and rotation trends of RR~Lyrae stars towards the Galactic bulge.

\begin{figure}
\includegraphics[width=\columnwidth]{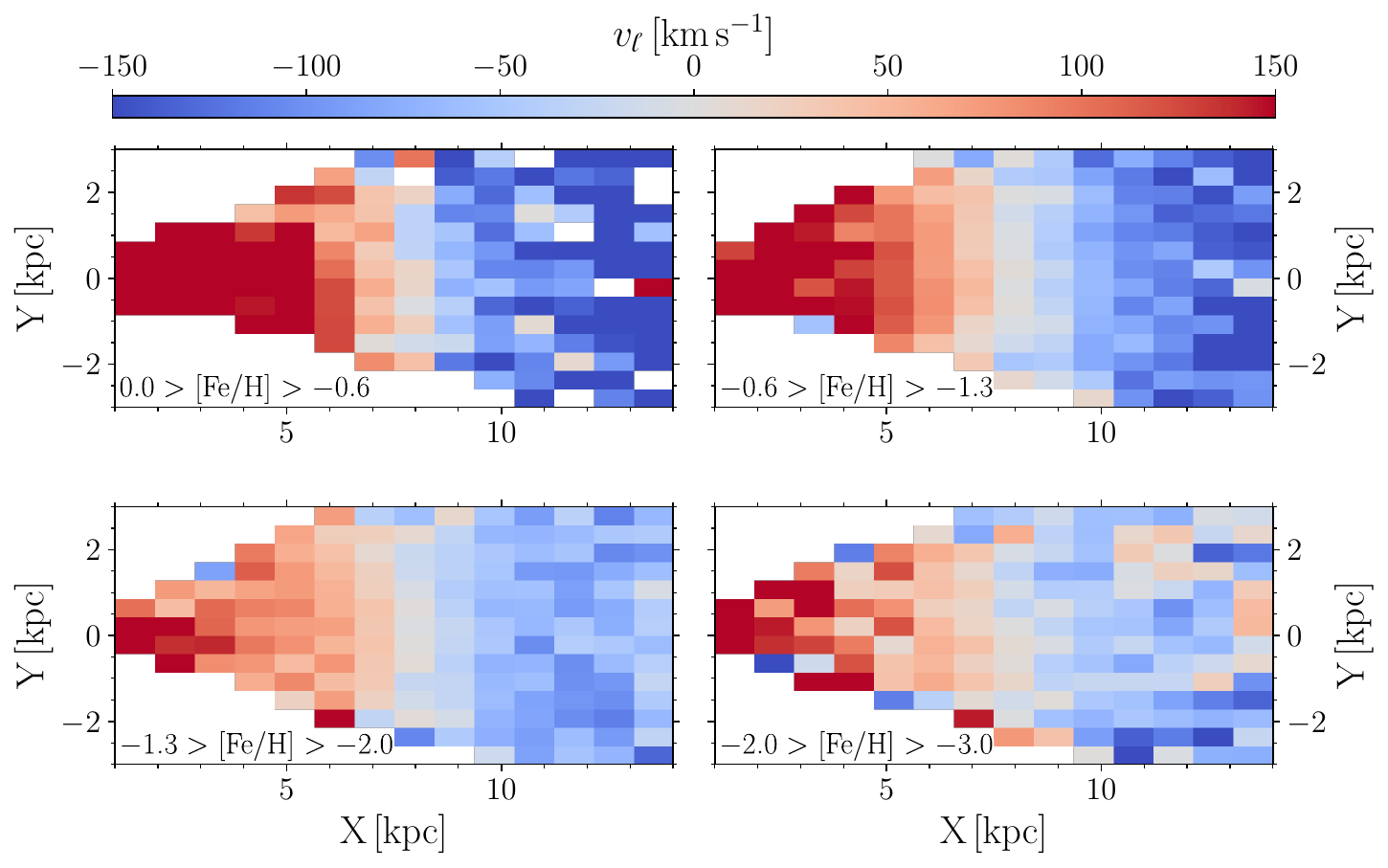}
\caption{The four panels, divided based on photometric metallicity going from metal-rich to metal-poor RR~Lyrae variables, show the binned spatial distribution in Y vs. X with color-coding based on the median value of $v_{\mathcal{l}}$ in a given bin. The Sun is located at X~$ = 0$ and Y~$ = 0$\,kpc.}
\label{fig:TransverseVelMetalDist2Dhist}
\end{figure}

In Figure~\ref{fig:TransverseVelMetalDist2Dhist}, we present the spatial properties of our RR~Lyrae dataset, using color coding to represent the values of $v_{\mathcal{l}}$ across four metallicity bins. This approach allows us to investigate further how $v_{\mathcal{l}}$ varies with metallicity. Our observations indicate a clear trend: for metal-poor RR~Lyrae stars, there is a discernible decrease in the absolute value of $v_{\mathcal{l}}$. This effect is at least partly due to halo interlopers, which can slightly influence the observed rotation signal, as demonstrated by RR~Lyrae kinematics \citep[e.g.,][]{Kunder2020,OlivaresCarvajal2024}. In contrast, metal-rich RR~Lyrae stars exhibit higher absolute values of $v_{\mathcal{l}}$, especially in the expected regions, in the foreground and background to the Galactic bulge.

%------------------------------------------------------------------------------------

\section{Disscusion} \label{sec:Discusion}

Since the discovery of a large numbers of RR~Lyrae stars toward the Galactic bulge, there has been an ongoing debate whether these old stars trace the barred bulge or if they instead belong to a spheroidal inner halo or bulge. The first MACHO survey results showed that the bulge RR~Lyrae stars do not follow the barred distribution that the MACHO red clump stars show \citep{Minniti1998}. Subsequent MACHO RR~Lyrae papers all agreed with this analysis \citep[e.g.,][]{Alcock1998,Kunder2008MACHO}, suggesting the older inner Galaxy RR~Lyrae population may have a different origin than the younger bulge/bar. In contrast, the first OGLE survey results showed the bulge RR~Lyrae stars do trace a barred distribution \citep{Collinge2006}, and subsequent OGLE RR~Lyrae papers continue to find a barred signature in their bulge RR~Lyrae samples \citep[e.g.,][]{Pietrukowicz2015}. The advent of IR surveys targeting the bulge, particularly the VVV survey, may be able to help resolve this discrepancy. Some IR photometry agrees with the absence of the barred nature of the bulge RR~Lyrae stars \citep{Dekany2013}, while other studies argue that the bulge RR~Lyrae trace the bulge/bar \citep[e.g.,][]{Molnar2022}. 

The tens of thousands of bulge RR~Lyrae stars with precisely measured pulsation properties from OGLE combined with the bulge RR~Lyrae stars with near-infrared photometry open the possibility of taking a closer look at the effects of the extinction toward the line-of-sight of the RR~Lyrae star. We also can now use new, more precise calibrations of RR~Lyrae star PMZ relations that are made possible thanks to \textit{Gaia} parallaxes as well as improved and homogeneous RR~Lyrae metallicities \citep{Bhardwaj2023,Prudil2024} to probe the 3D and 5D spatial distribution of the bulge RR~Lyrae stars. 

We find that the major cause for the debate surrounding the spatial distribution of the bulge RR~Lyrae stars arises from variations in dust properties. A barred RR~Lyrae signature naturally arises due to an extinction gradient (mainly in visual passbands) between positive and negative Galactic longitudes, as shown in Section~\ref{sec:RedMapsComp}. In particular, Figures~\ref{fig:AICompThisStudy}, \ref{fig:AICompSurot}, and \ref{fig:AICompGonza} illustrate that whereas the specific amplitude of the reddening gradient varies depending on the reddening map used, the overall trend remains consistent. In regions of positive $\mathcal{l}$, extinction calculated from $E\left ( V - I \right )$ is larger and thus overestimated compared to negative $\mathcal{l}$. This contributes up to a $1$\,kpc difference in distance estimates based on visual and near-infrared passbands, when compared to distances inferred from $E\left ( J - K_{\rm s} \right )$ values derived here and in the literature. The reddening gradient seen in $E\left ( V - I \right )$ disappears when using redder passbands (see the right panels of these figures).

Therefore, in Section~\ref{sec:bulge3D}, we show it is straightforward to recover the bar and its angle from RR~Lyrae distances based on visual passbands, but when using distances based on near-infrared distances, the bar and its associated tilt are almost negligible (Figure~\ref{fig:SpatialComparisonCNG}). We quantify the $R_{VI}$ and $R_{JK_{\rm s}}$ deviations in the extinction law across the three different lines of sight. As expected, the visual spectrum is significantly more impacted (by more than a factor of three) as compared to the infrared spectrum. That reddening law variations exist across the bulge is not new \citep[e.g.,][show variations in extinction ratios toward the bulge can exceed $20$ percent]{Nataf2016}, but we show how this directly affects the appearance of the spatial distribution of the inner Galaxy RR~Lyrae. Figure~\ref{fig:SpatialComparisonWithDiffRedd} demonstrates that the severe variation in the extinction law toward the bulge can be mitigated by employing the reddening law appropriate for the sight-line toward an RR~Lyrae star \citep[for example, Table~3 in][]{Nataf2013}. Applying the variations in the interstellar extinction curve reduces the prominence of the bar, but it does not completely remove it for $E\left ( V - I \right )$.

\begin{figure*}
\includegraphics[width=2\columnwidth]{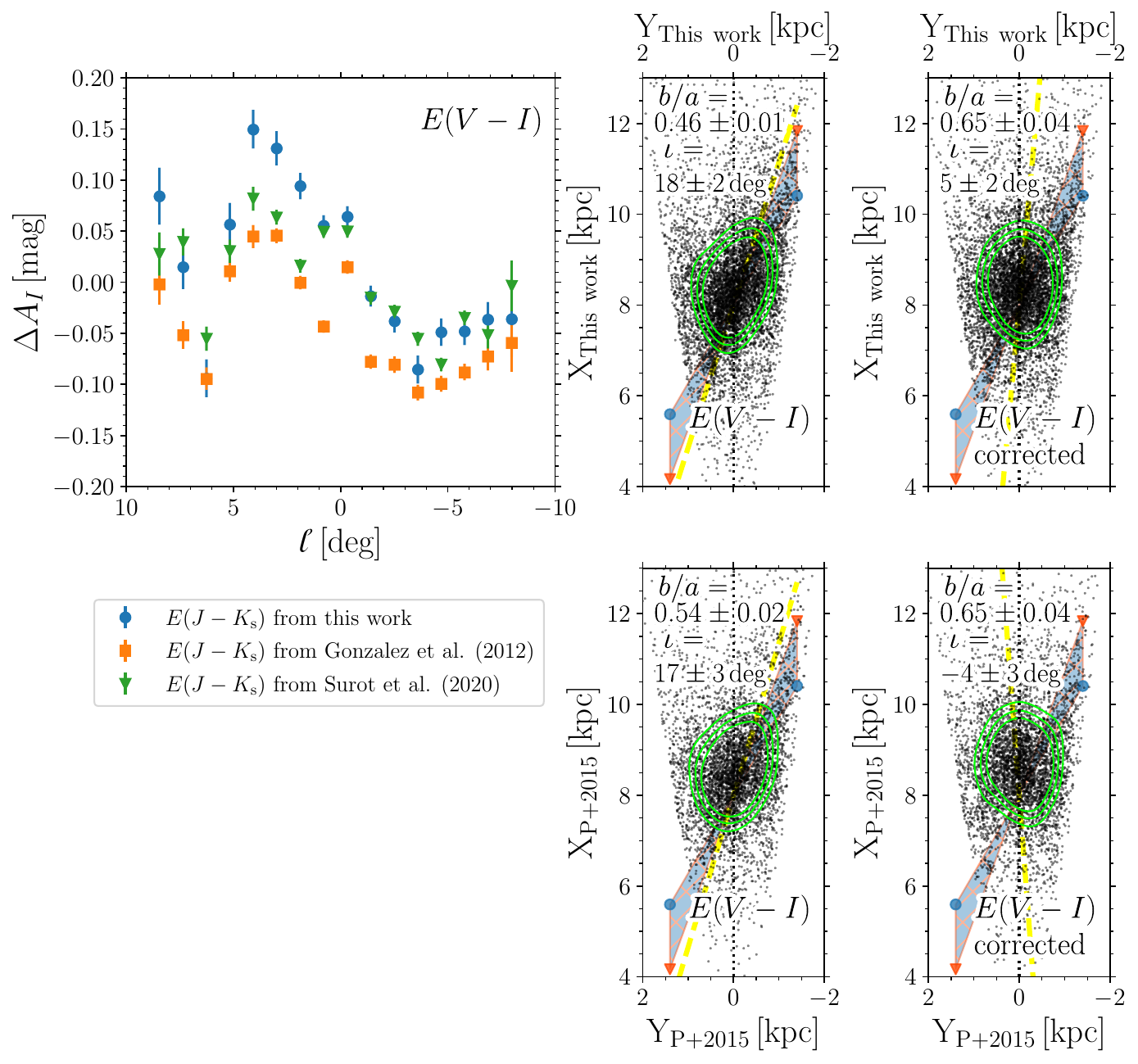}
\caption{A comparison of spatial distribution estimated through visual passbands in this work (top middle and right-hand panels) and using the method described in \citet[][marked as P+2015]{Pietrukowicz2015} (bottom middle and right-hand panels) also based partially on visual bands. The right-hand top panel shows a variation in $A_{I}$ extinction as a function of the Galactic longitude, estimated using three sources for $E\left ( J - K_{\rm s} \right )$ from this work (blue points), form \citet[][orange squares]{Gonzalez2012}, and from \citet[][green triangles]{Surot2020}. The red triangles and blue circles represent two bar angles $20$ and $30$\,degrees, respectively, with shading representing the angles in between. Same as in Fig.~\ref{fig:EllipseAnalysis}, the yellow dashed and black dotted lines represent the measured and zero angles, respectively. Lastly, the $\chi^{2}$ of ellipse fits from the top middle panel to the bottom-right hand plot are $0.38$, $0.31$, $0.25$, and $0.18$.}
\label{fig:GradientExplore}
\end{figure*}

\begin{figure}
\includegraphics[width=\columnwidth]{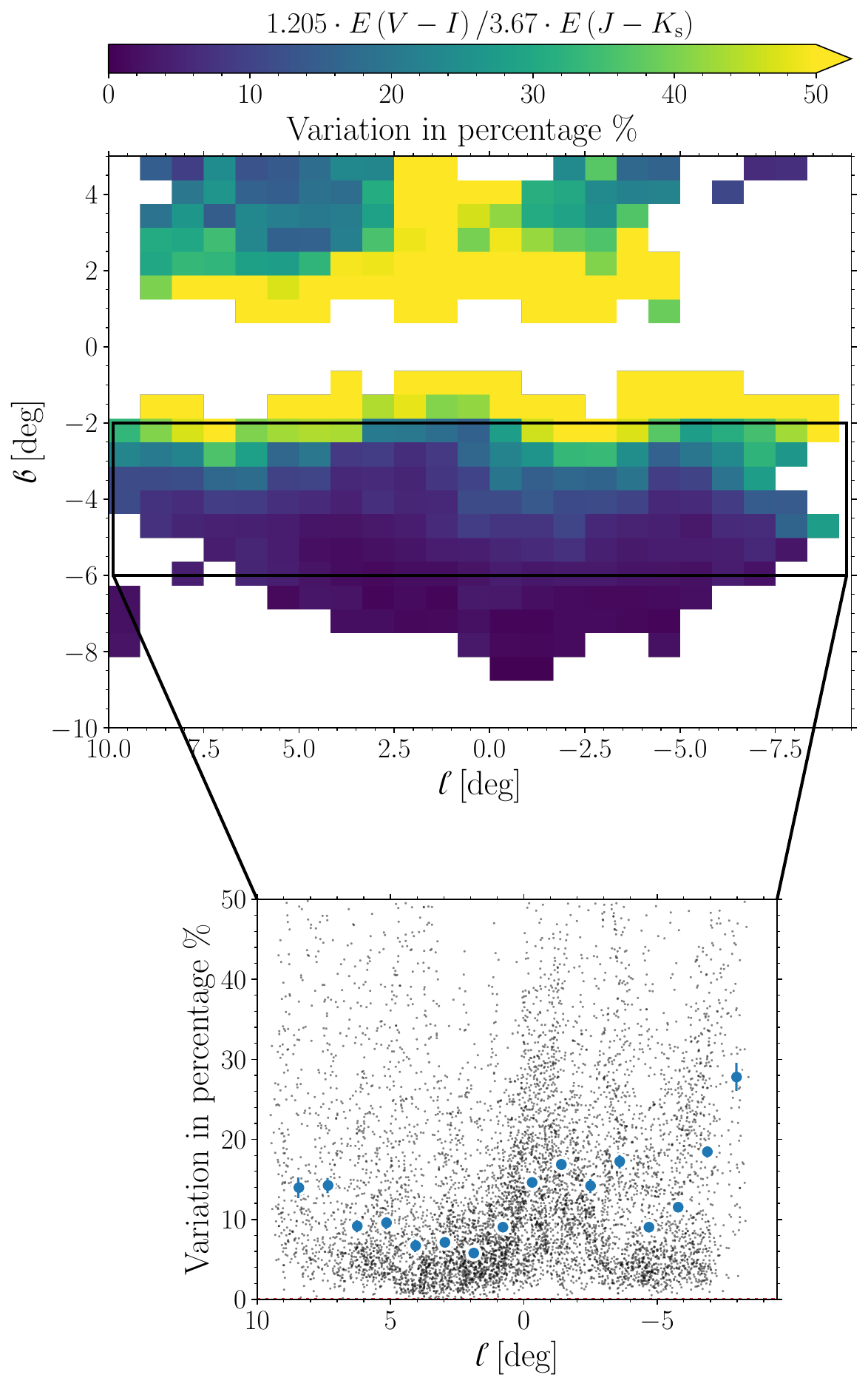}
\caption{Similar to Figure~\ref{fig:AICompThisStudy} but for ratio between extinction in $A_{I}$ estimated from $E\left ( V - I \right )$ and $E\left ( J - K_{\rm s} \right )$, respectively.}
\label{fig:RatioExtinc}
\end{figure}

Attempting to remove the effects of the reddening gradient using corrections based on the reddening maps derived here ($E\left ( J - K_{\rm s} \right )$), we estimate the bar angle $\iota = 5 \pm 2$\,deg (for our estimates of $E\left ( V - I \right )$), resulting in a non-barred spatial distribution of RR~Lyrae stars (Figure~\ref{fig:GradientExplore}). However, we do detect the signature of the barred bulge in the more metal-rich RR~Lyrae stars ([Fe/H]$_{\rm phot} > -1.0$\,dex) with $\iota = 18 \pm 5$\,deg, suggesting that not all inner Galaxy RR~Lyrae stars belong to the same substructures and therefore suggesting the different epoch of formation of the Milky Way's bar/bulge. In Figure~\ref{fig:RatioExtinc} we illustrate the extinction variation as a function of Galactic longitude derived from visual and near-infrared passbands. Significant fluctuations in both Galactic longitude and latitude are apparent. Specifically, in Galactic longitude, the average change in extinction across the selected regions is about $13$ percent. However, there is a noticeable disparity between positive and negative longitude in the median values. For positive Galactic longitudes, the median variation is smaller, around nine percent, while for negative Galactic longitudes, it increases to approximately $16$ percent. Above the Galactic plane, the variation in derived extinction is markedly higher, with median values reaching nearly $50$ percent. This variation is particularly pronounced in the top panel of Figure~\ref{fig:RatioExtinc}, where we observe a distinct difference in extinction variation above the Galactic plane between negative and positive $\mathcal{l}$, with the negative values being more significantly affected. Thus, we report a larger extinction variation than the one noted by \citet{Sanders2022}. However, in our case, we cover a larger area, both in Galactic longitude and latitude, than in the aforementioned study.

The variations in the extinction law along Galactic longitude and latitude directly contribute to the inconsistent findings regarding the prominence of the bar observed using RR~Lyrae variables. Relying on color excess values derived from visual passbands will lead to discrepancies in distance estimates if variations in the reddening law are not taken into account. Although near-infrared colors also exhibit variations, these have a smaller impact and are, therefore, more accommodating to the systematic uncertainties than when using visual passbands.

When examining the transverse kinematics of RR~Lyrae variables towards the Galactic bulge, the choice between $E\left ( V - I \right )$ or $E\left ( J - K_{\rm s} \right )$ for distance estimation does not significantly alter the outcomes. Utilizing both approaches, we reach a consistent conclusion: metal-rich RR~Lyrae stars demonstrate transverse velocities similar to those estimated using red clump stars, as reported in \citet{Sanders2019}, while metal-poor RR~Lyrae stars rotate more slowly than red clump giants. Whether the same holds true for kinematics based on line-of-sight velocities, we will examine in future work.

% ----------------------- ############# -----------------------
\section{Conclusions} \label{sec:Summary}

In this study, we analyzed the spatial and transverse velocity distribution of RR~Lyrae stars toward the Galactic bulge. We utilized data from OGLE-IV, \textit{Gaia}, VVV, and VHS surveys to create reddening maps, derive reddening laws, and estimate distances tailored to each given RR~Lyrae variable identified in the Galactic bulge (covering approximately $20 < \mathcal{l} < -15$\,deg and $-15 < \mathcal{b} < 15$\,deg area). Our reddening and distances range from visual to near-infrared passbands. Using the near-infrared passbands, we estimated the distance to the Galactic center as $d_{JK_{\rm s}}^{\rm cen} = 8217 \pm 1\,({\rm stat}) \pm 528\,({\rm sys})$\,pc, after applying a correction for the cone-effect. This value is in excellent agreement with measurements from GRAVITY \citep{DistantoSMBH2019,GRAVITY2021}.

We confirm and examine the discrepancy in the spatial distribution observed in inner Galaxy RR~Lyrae star studies \citep[e.g.,][]{Minniti1998,Collinge2006,Kunder2008MACHO,Dekany2013,Pietrukowicz2015}. These studies come to different conclusions on whether or not RR~Lyrae pulsators trace the bar or not. After careful examination of the reddening maps and their variation, we believe the main driver behind this inconsistency lies in the variation of the extinction law toward Galactic bulge sight lines. The bar signature in the spatial distribution of RR~Lyrae stars disappears in the following cases:
\begin{itemize}
  \item Using a single reddening law for near-infrared passbands, $E\left( I - K_{\rm s}\right)$ and $E\left ( J - K_{\rm s} \right )$, derived in this study. 
  \item By utilizing $E\left ( J - K_{\rm s} \right )$ from reddening maps by \citet{Gonzalez2012} and \citet{Surot2020} together with a single reddening law.
  \item For $E\left( V - I \right)$ when we apply variations in the reddening law, $R_{VI}$, for example, measured by \citet{Nataf2013}.
\end{itemize}
We also observe a gradient in extinction differences estimated using visual and near-infrared passbands along the Galactic longitudes and latitudes, where the former is most likely responsible for the appearance of the bar in spatial properties in visual passbands-based distances.

The purpose of this study was not to measure the extent and angle of the bar since the density of RR~Lyrae stars is lower than that of red clump giants. Despite that, we provide a first-order estimate for the bar angle using distance estimates based on $E\left ( J - K_{\rm s} \right )$ for metal-rich RR~Lyrae stars ([Fe/H]$_{\rm phot} > -1.0$\,dex), $\iota = 18 \pm 5$\,deg. This is well within values estimated in the literature \citep[][and references therein]{Bland-Hawthorn2016}. The bar's recovered prominence and angle in the metal-rich RR~Lyrae population point tentatively toward their younger age ($\approx$~10\,Gyr) than the bulk of the RR~Lyrae bulge population. We also estimated the bar angle as a function of metallicity for our RR~Lyrae data using their near-infrared distances. We found a clear dependence of the bar angle on metallicity, such that as we move toward the metal-rich end of our metallicity distribution, the bar tilt becomes more prominent. A similar trend was observed using Mira variables \citep[e.g.,][]{Catchpole2016,Grady2020}, where young Miras exhibit a triaxial distribution while older Miras show a more spheroidal spatial distribution. Finally, metal-poor RR~Lyrae stars ([Fe/H]$_{\rm phot} \leq -1.0$\,dex) exhibit a smaller bar angle compared to their metal-rich counterparts. There may be two possible explanations for the smaller bar angle observed on the metal-poor side of the RRLyrae metallicity distribution function. First, a high number of interlopers and RRLyrae variables not on bar-like orbits could result in a lower estimated bar angle. Second, it might indicate that RR~Lyrae stars follow a nearly end-on barred distribution in the Galactic bulge. Resolving between these two possibilities will require sufficient systemic velocity measurements for the bulge RRLyrae population, which will be the focus of our next study.

The 5D kinematical analysis relies mainly on the transverse velocities for Galactic longitude. We report that the average velocities derived for RR~Lyrae stars lag in rotation when compared to average velocities estimated for red clump giants \citep{Sanders2019}. Analogously to the 3D analysis, we recover a clear rotation pattern similar to the red clump stars for metal-rich RR~Lyrae stars ([Fe/H]$_{\rm phot} > -1.0$\,dex). Irrespective of the method to estimate distances toward the RR~Lyrae population in the Galactic bulge, our 5D kinematical results remain the same. This means that despite the variation in extinction law, the conclusions in transverse velocities remain the same both for visual and near-infrared-based distance estimates.

For a more conclusive answer to the cause of the lag observed in 5D, we would need to include additional dimensions from the line-of-sight velocities. Moreover, the additional dimension will allow the analysis of whether metal-poor RR~Lyrae variables ([Fe/H]$_{\rm phot} < -1.0$\,dex) exhibit the bar-like tilt and rotation. The full 6D analysis will be the aim of a subsequent study.

\begin{acknowledgements}
Z.~P. is grateful to Francesca Fragkoudi, for valuable discussions. LBeS acknowledges the support provided by the Heising Simons Foundation through the Barbara Pichardo Future Faculty Fellowship from grant \# 2022-3927. 

This research made use of the following Python packages: \texttt{Astropy} \citep{astropy2013,astropy2018}, \texttt{emcee} \citep{Foreman-Mackey2013}, \texttt{IPython} \citep{ipython}, \texttt{Matplotlib} \citep{matplotlib}, \texttt{NumPy} \citep{numpy} and \texttt{SciPy} \citep{scipy}. 

Z.P. would like to dedicate this work to his daughter, Sofie P.

\end{acknowledgements}

\bibliographystyle{aa}
\bibliography{biby} 

\begin{appendix} 

\section{Verification of our distances using Galactic bulge globular clusters} \label{sec:VerfDistGC}

\begin{figure}
\includegraphics[width=\columnwidth]{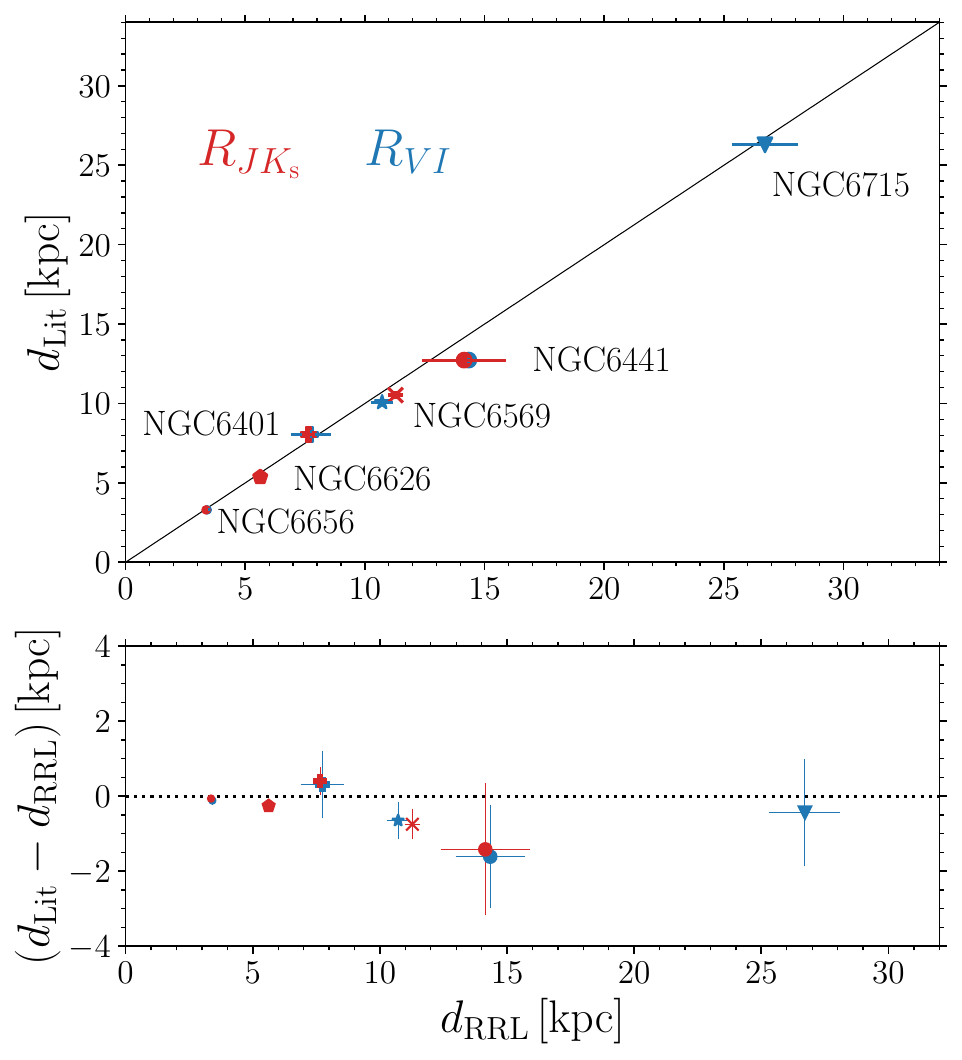}
\caption{Comparison of distances derived in this work (based on two reddening laws $R_{VI}$ and $R_{JK_{\rm s}}$) for RR~Lyrae stars associated with several MW globular clusters, their literature distances from \citet{Vasiliev2021}. The black line in the top panel represents the identity line. Uncertainties on the difference between literature and our distances were estimated by adding our uncertainties and uncertainties on distance from literature in quadrature.}
\label{fig:CompDist}
\end{figure}

A comparison of the calculated RR~Lyrae distances with globular cluster distances toward the Galactic bulge is depicted in Figure~\ref{fig:CompDist}. We used approximate associations of RR~Lyrae stars to globular clusters done by the OGLE team \citep{Soszynski2014BulgeRRlyr,Soszynski2019Disk}, their proper motions \citep[using data from the $Gaia$ space telescope][see Subsection~\ref{subsec:Gaia} for more details]{GaiaDR32023} together with distances and proper motions from \citet{Vasiliev2021}. We selected those RR~Lyrae stars whose proper motions matched the proper motions of globular clusters. We included only those clusters where we could associate at least seven RR~Lyrae pulsators (except NGC6453 for passband combination $I, G_{\rm BP}-I$), and as a distance of a cluster, we selected the weighted average of linked RR~Lyrae variables. The listed uncertainties in Table~\ref{tab:DistToGC} and displayed in Figure~\ref{fig:CompDist} represent weighted standard deviations of the given distance distribution.  

As a final value for distance to an individual RR~Lyrae variable, we preferentially selected measurements derived based on the near-infrared passbands $K_{\rm s}$-band (with $E\left ( J - K_{\rm s} \right )$ reddening), or if the $J$-band was not available we used $E\left ( I - K_{\rm s} \right )$ reddening. In all other cases, we resort to distances based on $I$-band and $E\left ( V - I \right )$ or $E\left ( G_{\rm BP} - I \right )$ reddening, respectively. 

\setlength{\tabcolsep}{2.pt}
\begin{table}
\caption{List of globular clusters toward the Galactic bulge containing a sufficient number of RR~Lyrae stars. The first column represents filters used for mean intensity magnitudes and reddening. The second lists globular cluster identifiers. The following two columns contain our derived distance and literature distances and their listed uncertainties. The last column contains the number of RR~Lyrae pulsators used in the distance estimate.}
\label{tab:DistToGC}
\begin{tabular}{l|c|c|c|r}
\hline \hline 
Band Comb. & GC & $d_{\rm RRL}$\,[kpc] & $d_{\rm Lit}$\,[kpc] & Num. \\ \hline
$m_{I}, E\left ( G_{\rm BP} - I \right )$ & \multirow{4}{*}{NGC6401} & \phantom{0}$8.53 \pm 1.87$ & \multirow{4}{*}{\phantom{0}$8.06 \pm 0.24$} & 14  \\
$m_{I}, E\left ( V - I \right )$ &  & \phantom{0}$7.74 \pm 0.85$ &  & 22  \\
$m_{K_{\rm s}}, E\left ( I - K_{\rm s} \right )$ &  & \phantom{0}$7.22 \pm 1.01$ &  & 20  \\
$m_{K_{\rm s}}, E\left ( J - K_{\rm s} \right )$ &  & \phantom{0}$7.65 \pm 0.28$ &  & 18  \\ \hline 
$m_{I}, E\left ( G_{\rm BP} - I \right )$ & \multirow{4}{*}{NGC6441} & $15.03 \pm 1.65$ & \multirow{4}{*}{$12.73 \pm 0.16$} & 15  \\
$m_{I}, E\left ( V - I \right )$ &  & $14.34 \pm 1.36$ & & 20  \\
$m_{K_{\rm s}}, E\left ( I - K_{\rm s} \right )$ &  & $14.18 \pm 1.65$ & & 15  \\
$m_{K_{\rm s}}, E\left ( J - K_{\rm s} \right )$ &  & $14.15 \pm 1.75$ & & 15  \\ \hline 
$m_{I}, E\left ( G_{\rm BP} - I \right )$ & \multirow{2}{*}{NGC6453} & $12.26 \pm 3.82$ & \multirow{2}{*}{$10.07 \pm 0.22$} & 5  \\
$m_{I}, E\left ( V - I \right )$ &  & $10.72 \pm 0.45$ & & 7  \\ \hline 
$m_{K_{\rm s}}, E\left ( I - K_{\rm s} \right )$ & \multirow{2}{*}{NGC6569} & \phantom{0}$9.90 \pm 2.23$ & \multirow{2}{*}{$10.53 \pm 0.26$} & 12  \\
$m_{K_{\rm s}}, E\left ( J - K_{\rm s} \right )$ &  & $11.28 \pm 0.31$ & & 10  \\ \hline 
$m_{K_{\rm s}}, E\left ( I - K_{\rm s} \right )$ & \multirow{2}{*}{NGC6626} & \phantom{0}$5.65 \pm 0.07$ & \multirow{2}{*}{\phantom{0}$5.37 \pm 0.10$} & 7  \\
$m_{K_{\rm s}}, E\left ( J - K_{\rm s} \right )$ &  & \phantom{0}$5.63 \pm 0.11$ &  & 7  \\ \hline 
$m_{I}, E\left ( G_{\rm BP} - I \right )$ & \multirow{4}{*}{NGC6656} & \phantom{0}$3.20 \pm 0.16$ & \multirow{4}{*}{\phantom{0}$3.30 \pm 0.04$} & 18  \\
$m_{I}, E\left ( V - I \right )$ &  & \phantom{0}$3.41 \pm 0.11$ & & 19  \\
$m_{K_{\rm s}}, E\left ( I - K_{\rm s} \right )$ &  & \phantom{0}$3.39 \pm 0.06$ & & 19  \\
$m_{K_{\rm s}}, E\left ( J - K_{\rm s} \right )$ &  & \phantom{0}$3.36 \pm 0.07$ & & 18  \\ \hline 
$m_{I}, E\left ( G_{\rm BP} - I \right )$ & \multirow{2}{*}{NGC6715} & $27.29 \pm 3.19$ & \multirow{2}{*}{$26.28 \pm 0.33$} & 27  \\
$m_{I}, E\left ( V - I \right )$ &  & $26.71 \pm 1.40$ & & 22  \\ \hline
\end{tabular}
\end{table}

\section{Oosterhoff dichotomy and the MW bar}

The MW globular clusters containing RR~Lyrae stars exhibit a bifurcation in periods and metallicities known as the Oosterhoff dichotomy \citep{Oosterhoff1939,Oosterhoff1944}. The RR~Lyrae population in the Galactic bulge has been observed to host at least two Oosterhoff groups \citep[see, e.g.,][]{Prudil2019OOspat}, which were previously identified and studied among MW globular clusters \citep[see][for references]{Catelan2009} and the MW halo \citep{Fabrizio2019,Fabrizio2021,Luongo2024}. Recently, it has been shown that estimates of photometric metallicities for RR~Lyrae variables associated with the Oosterhoff type~II (Oo\,II) group are less reliable than for those of Oosterhoff type~I (Oo\,I) \citep[see][]{Hajdu2018,Jurcsik2021,Jurcsik2023}.

\begin{figure}
\includegraphics[width=\columnwidth]{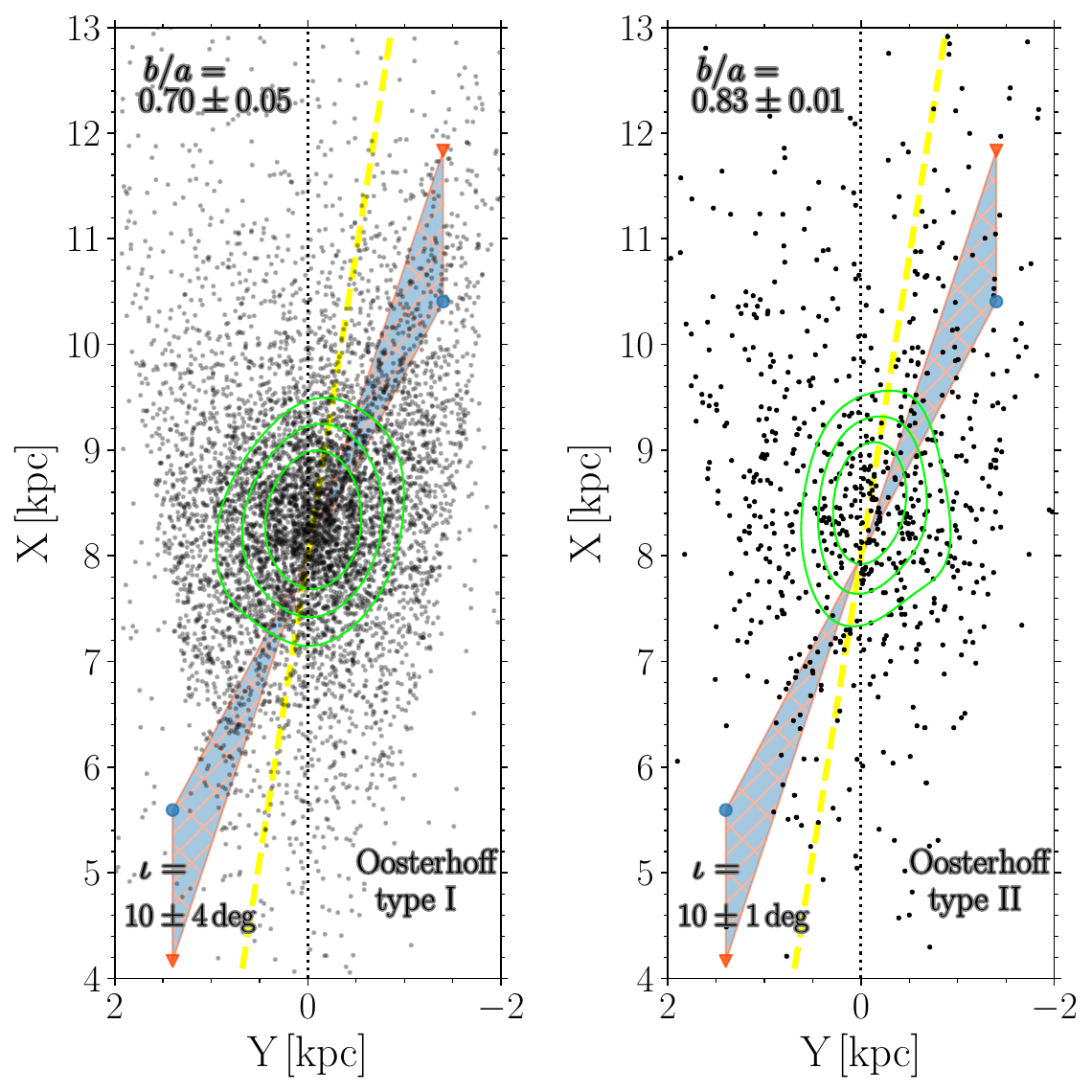}
\caption{The spatial distribution for RR~Lyrae stars toward the Galactic bulge for Oo\,I (left-hand panel) and Oo\,II (right-hand panel) groups, respectively. As Fig.~\ref{fig:EllipseAnalysis}, the measured and zero angles are marked with yellow dashed and black dotted lines, respectively. The quality of the ellipse fit for the left-hand panel is $\chi^{2} = 0.25$.}
\label{fig:BarOO}
\end{figure}

To verify our analysis of the MW bar, we examined the spatial distribution in the X and Y planes for both Oosterhoff types to ensure that our conclusions (e.g., in Section~\ref{sec:bulge3D}) regarding the bar prominence remain consistent for both groups. We utilize the technique outlined in \citep{Prudil2019OOspat} to differentiate between Oosterhoff types for fundamental RR~Lyrae pulsators and illustrated their spatial distribution in Figure~\ref{fig:BarOO} (similar to Figures~\ref{fig:SpatialComparisonCNG}, \ref{fig:SpatialComparisonWithDiffRedd}, \ref{fig:SpatialFEHcomparison}), where we presented spatial distributions for Oo\,I and Oo\,II groups. We observe that both Oosterhoff groups yielded the same result regarding the bar prominence in RR~Lyrae stars. Neither Oo\,I nor Oo\,II exhibit an expected tilt with the bar. Therefore, we conclude that RR Lyrae variables associated with the Oo\,II group offer the same insights despite their less precise photometric metallicities.

We note that the Oo\,II type RRab stars represent a minority (approximately $17$ percent) in our dataset. The reported offset in photometric metallicities \citep[$\approx 0.1$\,dex][]{Jurcsik2021} affects the derived distances for the majority of our sample only negligibly due to the use of near-infrared passbands (resulting in approximately a $50$\,pc difference at the distance of the Galactic bulge). This same insignificant impact also applies to the estimates of reddening and extinction for the near-infrared photometry. Therefore, the results presented in this work appear to be robust despite the caveats in photometric metallicities for both Oosterhoff groups.

\section{Additional tables} \label{sec:AppTables}

\setlength{\tabcolsep}{1pt}
\begin{landscape}
\begin{table}[]
\label{tab:CompleteList}
\caption{Table for estimated photometric properties in our study. The first four columns contain OGLE identifier, equatorial coordinates, and \textit{Gaia} \texttt{source\_id}. Columns five and six list photometric metallicities. The subsequent ten columns contain absolute magnitudes and their uncertainties. Additional eight columns list color excesses estimated in this study. The last two columns provide distances and their uncertainties for individual stars. The entire table is provided in the supplementary material for this paper.}
\scriptsize
\begin{tabular}{llllllllllllllllllllllllllllllllll}
\hline \hline
ID & ra & dec & \texttt{source\_id} & [Fe/H]$_{\rm phot}$ & $\sigma_{\text{[Fe/H]}_{\rm phot}}$ & $M_{K_{\rm s}}$ & $\sigma_{M_{K_{\rm s}}}$ & $M_{J}$ & $\sigma_{M_{J}}$ & $M_{I}$ & $\sigma_{M_{I}}$ & $M_{V}$ & $\sigma_{M_{V}}$ & $M_{G_{\rm BP}}$ & $\sigma_{M_{G_{\rm BP}}}$ & $E\left ( J - K_{\rm s} \right )$ & $\sigma_{E\left ( J - K_{\rm s} \right )}$ & $E\left ( I - K_{\rm s} \right )$ & $\sigma_{E\left ( I - K_{\rm s} \right )}$ & $E\left ( V - I \right )$ & $\sigma_{E\left ( V - I \right )}$ & $E\left ( G_{\rm BP} - I \right )$ & $\sigma_{E\left ( G_{\rm BP} - I \right )}$ & $d$ & $\sigma_{d}$ \\  
 -- & [deg] & [deg] & -- & [dex] & [dex] & [mag] & [mag]  & [mag]  & [mag]  & [mag]  & [mag]  & [mag]  & [mag]  & [mag]  & [mag]  & [mag]  & [mag]  & [mag]  & [mag]  & [mag]  & [mag]  & [mag]  & [mag]  & [kpc] & [kpc] \\ \hline
00002 & 256.283708 & -32.943917 & 5980012953523019648 & -1.31 & 0.18 & -0.06 & 0.10 & 0.12 & 0.17 & 0.45 & 0.12 & 0.83 & 0.15 & 0.85 & 0.13 & 0.20 & 0.20 & 0.53 & 0.15 & -- & -- & 0.68 & 0.18 & 8.97 & 0.56 \\
00003 & 256.291167 & -32.664667 & 5980034497087159424 & -1.34 & 0.07 & -0.31 & 0.09 & -0.07 & 0.17 & 0.31 & 0.11 & 0.76 & 0.14 & 0.78 & 0.13 & 0.28 & 0.19 & 0.66 & 0.15 & -- & -- & 0.77 & 0.17 & 11.68 & 0.71 \\
00004 & 256.313417 & -32.837000 & 5980014735918979840 & -1.90 & 0.19 & -0.63 & 0.10 & -0.35 & 0.17 & 0.06 & 0.12 & 0.57 & 0.15 & 0.59 & 0.13 & 0.35 & 0.20 & 0.64 & 0.15 & -- & -- & 0.79 & 0.18 & 10.42 & 0.66 \\
00005 & 256.341583 & -32.661806 & 5980035184282597120 & -2.46 & 0.59 & -0.56 & 0.12 & -0.33 & 0.20 & 0.03 & 0.16 & 0.48 & 0.19 & 0.50 & 0.18 & 0.39 & 0.23 & 0.72 & 0.20 & -- & -- & 0.82 & 0.25 & 7.50 & 0.58 \\
$\dots$ & $\dots$ & $\dots$ & $\dots$ & $\dots$ & $\dots$ & $\dots$ & $\dots$ & $\dots$ & $\dots$ & $\dots$ & $\dots$ & $\dots$ & $\dots$ & $\dots$ & $\dots$ & $\dots$ & $\dots$ & $\dots$ & $\dots$ & $\dots$ & $\dots$ & $\dots$ & $\dots$ & $\dots$ & $\dots$ \\ \hline
\end{tabular}
\end{table}
\end{landscape}

\begin{table}[]
\caption{List of metallicity bins and measured bar angle and axis ratio for RR~Lyrae spatial distribution estimated using near-infrared passbands (using a condition on $\mathcal{b}$).}
\label{tab:MetalicityBar}
\begin{tabular}{l|c|c}
\hline \hline
[Fe/H] bin & $\iota$ & $b/a$ \\
\phantom{-}[dex] & [deg] & \\ \hline 
$-2.01$ & $12.13 \pm 4.57$ & $0.83 \pm 0.02$ \\
$-1.90$ & $7.16 \pm 4.00$ & $0.84 \pm 0.02$ \\
$-1.81$ & $9.36 \pm 4.44$ & $0.82 \pm 0.02$ \\
$-1.73$ & $7.91 \pm 4.62$ & $0.84 \pm 0.02$ \\
$-1.67$ & $10.49 \pm 4.32$ & $0.83 \pm 0.02$ \\
$-1.62$ & $6.75 \pm 3.69$ & $0.84 \pm 0.02$ \\
$-1.58$ & $7.88 \pm 4.03$ & $0.83 \pm 0.02$ \\
$-1.54$ & $10.13 \pm 4.53$ & $0.84 \pm 0.02$ \\
$-1.51$ & $10.04 \pm 3.93$ & $0.84 \pm 0.02$ \\
$-1.48$ & $13.56 \pm 4.78$ & $0.85 \pm 0.02$ \\
$-1.46$ & $11.42 \pm 4.56$ & $0.85 \pm 0.02$ \\
$-1.44$ & $14.27 \pm 3.51$ & $0.80 \pm 0.02$ \\
$-1.42$ & $11.98 \pm 3.48$ & $0.80 \pm 0.02$ \\
$-1.41$ & $8.32 \pm 2.90$ & $0.77 \pm 0.02$ \\
$-1.39$ & $8.96 \pm 2.46$ & $0.77 \pm 0.02$ \\
$-1.38$ & $5.71 \pm 2.97$ & $0.78 \pm 0.02$ \\
$-1.37$ & $7.46 \pm 3.03$ & $0.79 \pm 0.02$ \\
$-1.36$ & $5.63 \pm 3.06$ & $0.79 \pm 0.02$ \\
$-1.35$ & $5.69 \pm 3.03$ & $0.78 \pm 0.02$ \\
$-1.33$ & $6.87 \pm 2.85$ & $0.78 \pm 0.02$ \\
$-1.32$ & $9.88 \pm 3.67$ & $0.80 \pm 0.02$ \\
$-1.30$ & $14.18 \pm 2.99$ & $0.79 \pm 0.02$ \\
$-1.29$ & $9.68 \pm 3.76$ & $0.82 \pm 0.02$ \\
$-1.27$ & $4.94 \pm 2.55$ & $0.80 \pm 0.02$ \\
$-1.25$ & $3.26 \pm 2.71$ & $0.80 \pm 0.02$ \\
$-1.21$ & $5.68 \pm 2.68$ & $0.79 \pm 0.02$ \\
$-1.16$ & $10.29 \pm 4.28$ & $0.84 \pm 0.02$ \\
$-1.09$ & $7.64 \pm 3.60$ & $0.83 \pm 0.02$ \\
$-1.00$ & $13.24 \pm 4.96$ & $0.84 \pm 0.02$ \\
$-0.85$ & $14.59 \pm 4.28$ & $0.84 \pm 0.02$ \\
$-0.66$ & $18.64 \pm 4.62$ & $0.84 \pm 0.02$ \\
$-0.49$ & $18.36 \pm 4.73$ & $0.83 \pm 0.02$ \\
$-0.25$ & $22.78 \pm 7.18$ & $0.84 \pm 0.03$ \\ \hline  
\end{tabular}
\end{table}

\begin{table}[]
\caption{Same as Table~\ref{tab:MetalicityBar} but for bar angles estimated using a condition on Z coordinates.}
\label{tab:MetalicityBar2}
\begin{tabular}{l|c|c}
\hline \hline
[Fe/H] bin & $\iota$ & $b/a$ \\
\phantom{-}[dex] & [deg] & \\ \hline 
$-1.89$ & $10.13 \pm 4.30$ & $0.81 \pm 0.02$ \\
$-1.75$ & $14.00 \pm 4.90$ & $0.83 \pm 0.02$ \\
$-1.67$ & $11.64 \pm 4.78$ & $0.85 \pm 0.02$ \\
$-1.59$ & $11.42 \pm 5.25$ & $0.87 \pm 0.02$ \\
$-1.53$ & $12.07 \pm 4.67$ & $0.86 \pm 0.02$ \\
$-1.49$ & $11.70 \pm 4.40$ & $0.85 \pm 0.02$ \\
$-1.45$ & $11.31 \pm 4.23$ & $0.83 \pm 0.02$ \\
$-1.43$ & $13.55 \pm 3.71$ & $0.81 \pm 0.02$ \\
$-1.40$ & $12.16 \pm 2.79$ & $0.77 \pm 0.02$ \\
$-1.38$ & $9.15 \pm 3.35$ & $0.79 \pm 0.02$ \\
$-1.37$ & $4.32 \pm 2.91$ & $0.83 \pm 0.02$ \\
$-1.35$ & $6.35 \pm 3.54$ & $0.82 \pm 0.02$ \\
$-1.33$ & $8.19 \pm 3.82$ & $0.83 \pm 0.02$ \\
$-1.30$ & $15.14 \pm 3.35$ & $0.81 \pm 0.02$ \\
$-1.28$ & $13.27 \pm 3.17$ & $0.81 \pm 0.02$ \\
$-1.24$ & $11.69 \pm 3.30$ & $0.79 \pm 0.02$ \\
$-1.17$ & $13.11 \pm 4.40$ & $0.83 \pm 0.02$ \\
$-1.05$ & $18.93 \pm 3.90$ & $0.83 \pm 0.02$ \\
$-0.84$ & $20.24 \pm 3.98$ & $0.83 \pm 0.02$ \\
$-0.65$ & $20.01 \pm 3.28$ & $0.80 \pm 0.02$ \\
$-0.39$ & $21.07 \pm 5.82$ & $0.81 \pm 0.03$ \\ \hline 
\end{tabular}
\end{table}

\section{Additional figures} \label{sec:AppFigures}
\begin{figure}
\includegraphics[width=\columnwidth]{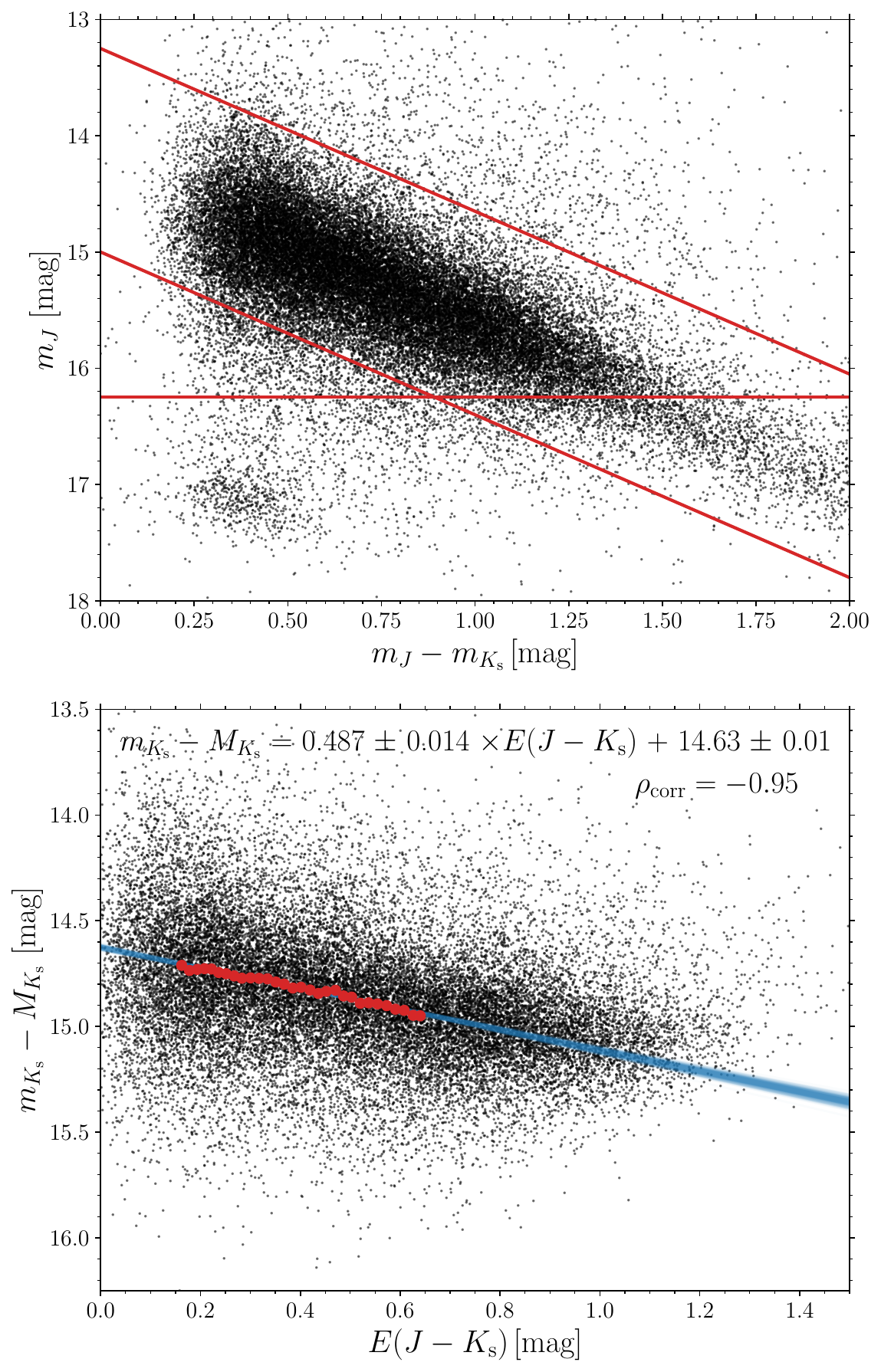}
\caption{The color-magnitude diagram (top panel) and linear dependence between color excess $E\left ( J - K_{\rm s} \right )$ (lower panel) versus the difference of mean intensity and absolute magnitudes. In the top panel, we plot all sample stars. The red lines denote conditions listed in Table~\ref{tab:CutsOnMagAndColor} for the selection of bulge RR~Lyrae sub-sample. In the bottom panel, the black dots represent selected stars from our sample, blue lines represent the reddening vector, and red circles depict individual bins.}
\label{fig:ReddLawJK}
\end{figure}

\begin{figure}
\includegraphics[width=\columnwidth]{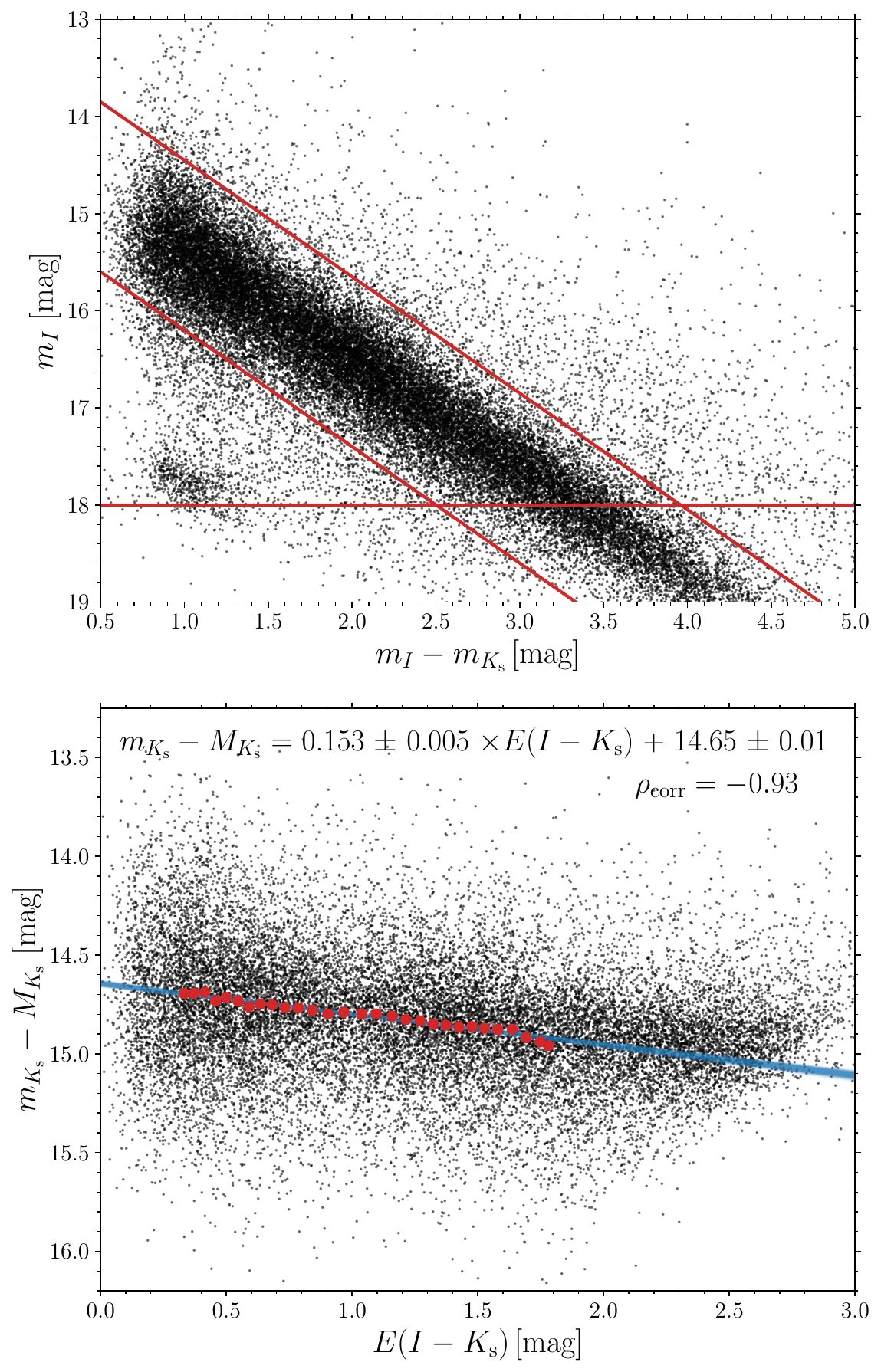}
\caption{Same as Figure~\ref{fig:ReddLawJK} but for $I$ and $K_{\rm s}$ passbands.}
\label{fig:ReddLawIK}
\end{figure}

\begin{figure}
\includegraphics[width=\columnwidth]{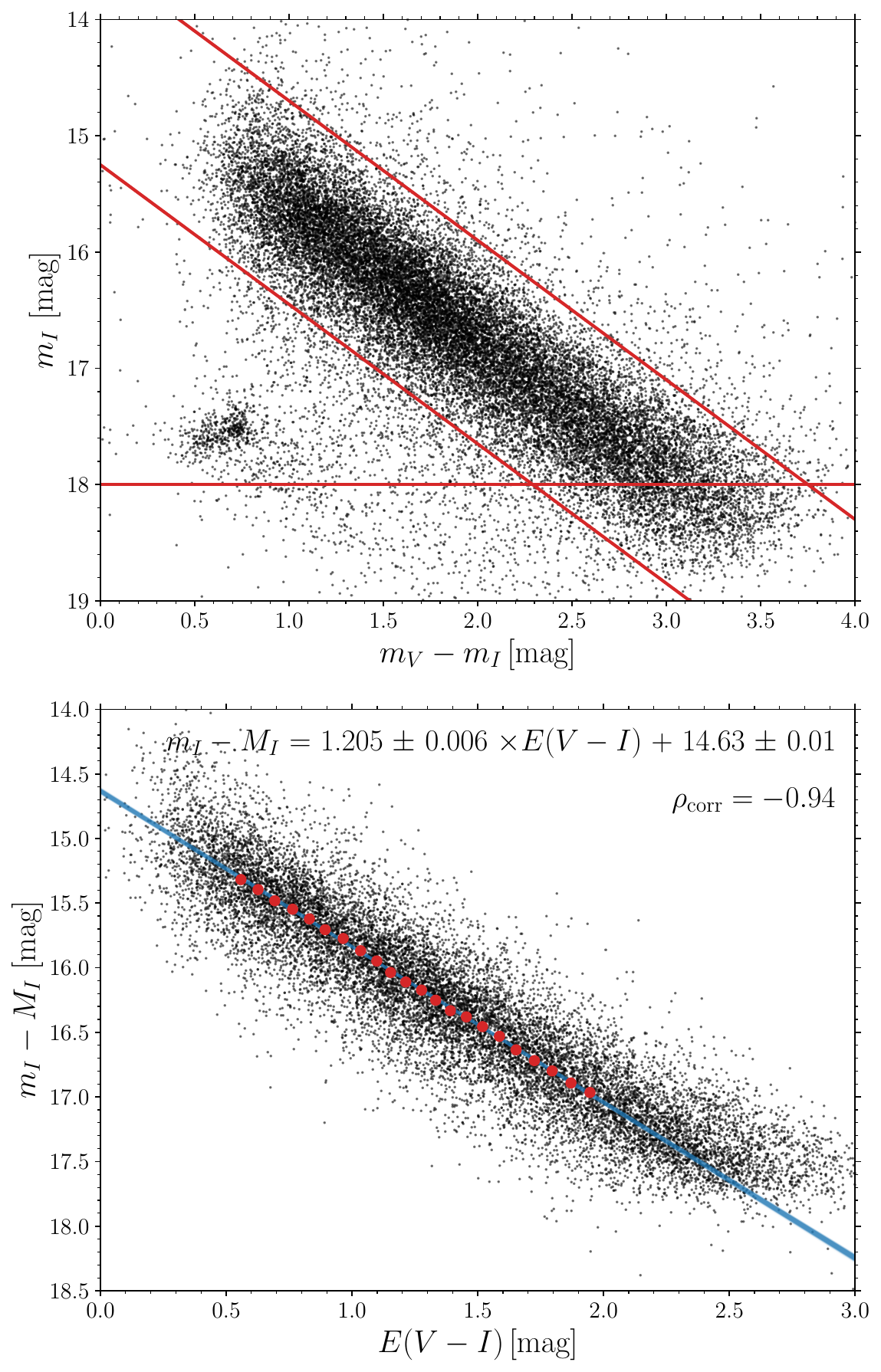}
\caption{Same as Figure~\ref{fig:ReddLawIK} but for $V$ and $I$ passbands.}
\label{fig:ReddLawVI}
\end{figure}

\begin{figure}
\includegraphics[width=\columnwidth]{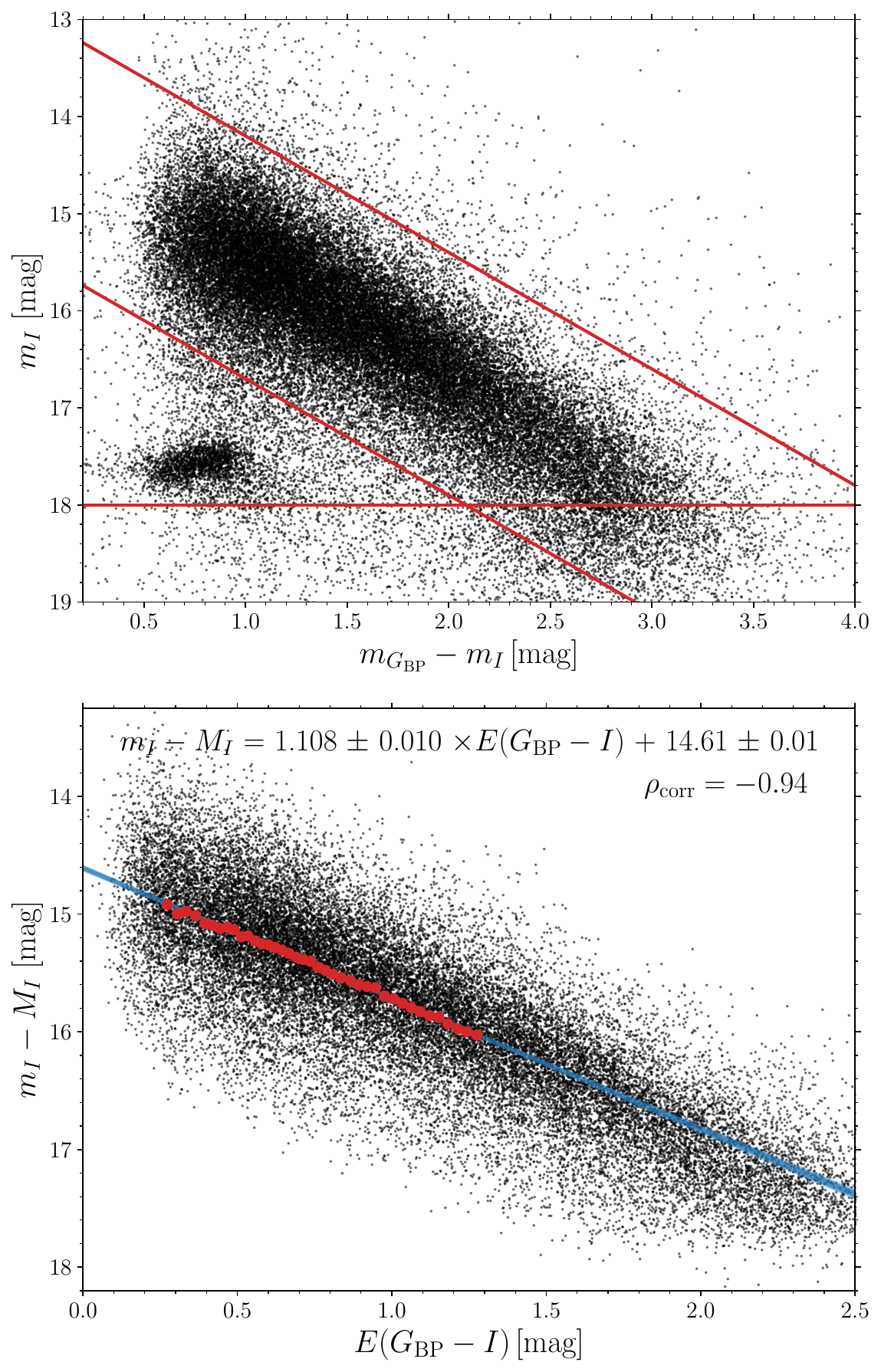}
\caption{Same as Figure~\ref{fig:ReddLawIK} but for $G_{\rm BP}$ and $I$ passbands.}
\label{fig:ReddLawGbpI}
\end{figure}

The additional displayed Figures provide the necessary context for estimation of the reddening law (Fig.~\ref{fig:ReddLawJK},~\ref{fig:ReddLawIK},~\ref{fig:ReddLawVI}, and \ref{fig:ReddLawGbpI}), particularly to selected regions in magnitude space for RR~Lyrae sample. We also include further analysis of the variation in the reddening law and test of the estimated color-excesses (Figures~\ref{fig:AICompGonza},~\ref{fig:ReddLawModified}, and~\ref{fig:ComparisonRedd}). Lastly, we include an analysis of transverse velocities for RR~Lyrae variables above the Galactic plane.

\begin{figure*}
\includegraphics[width=2\columnwidth]{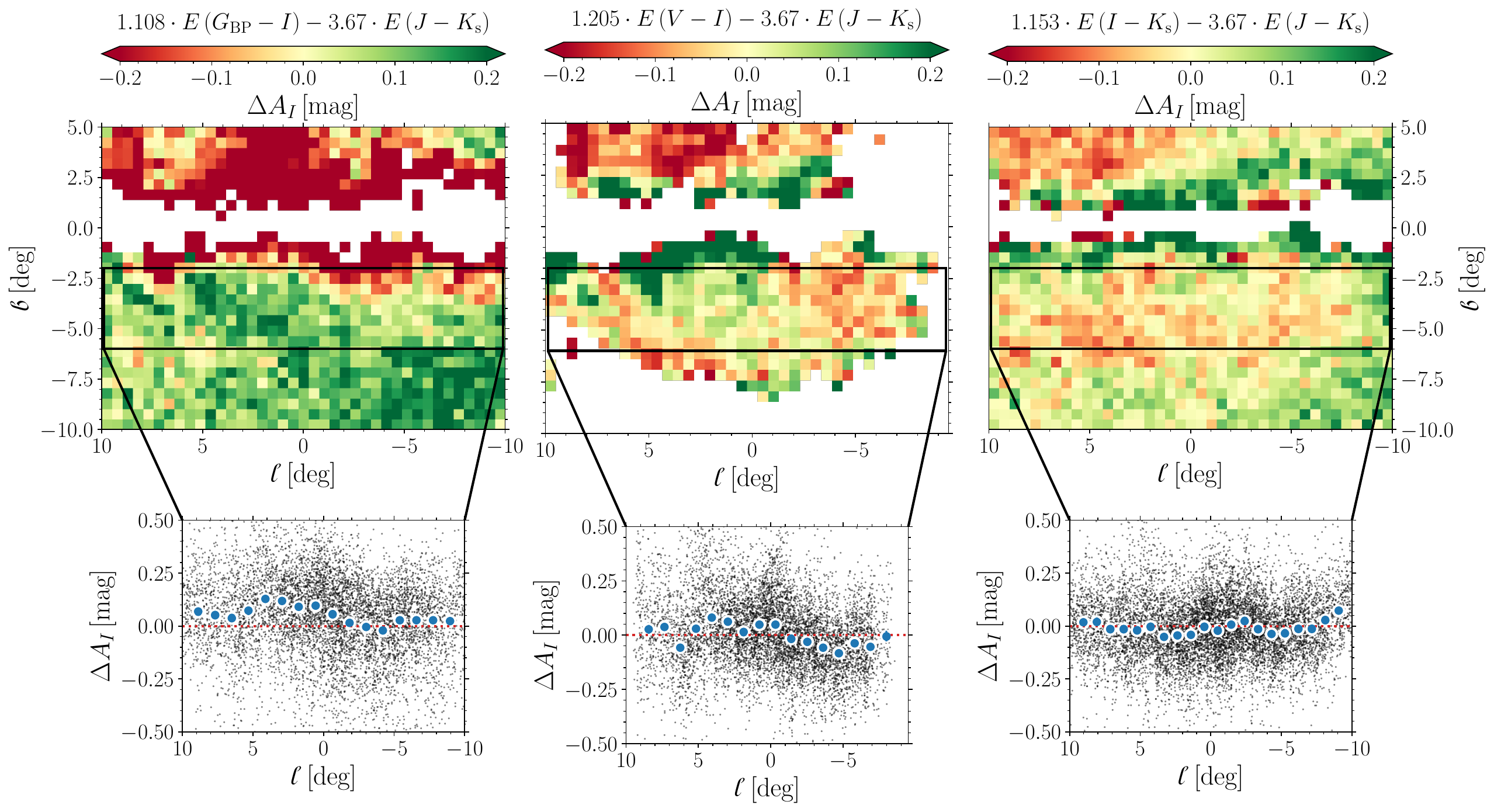}
\caption{Same as Figure~\ref{fig:AICompThisStudy} but for $E\left ( J - K_{\rm s} \right )$ from \citet{Surot2020}.}
\label{fig:AICompSurot}
\end{figure*}

\begin{figure*}
\includegraphics[width=2\columnwidth]{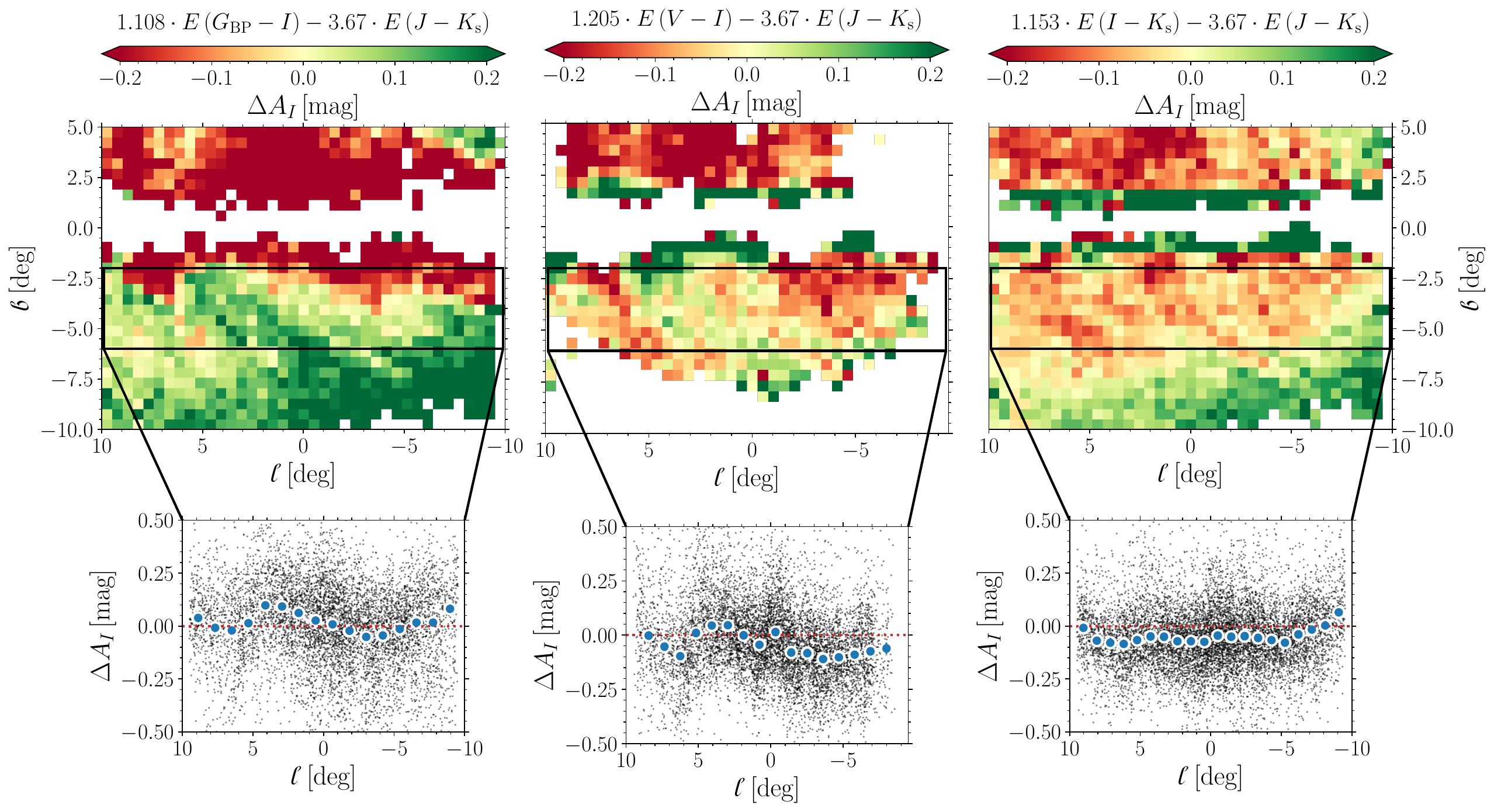}
\caption{Same as Figure~\ref{fig:AICompThisStudy} but for $E\left ( J - K_{\rm s} \right )$ from \citet{Gonzalez2012}.}
\label{fig:AICompGonza}
\end{figure*}

\begin{figure*}
\includegraphics[width=2\columnwidth]{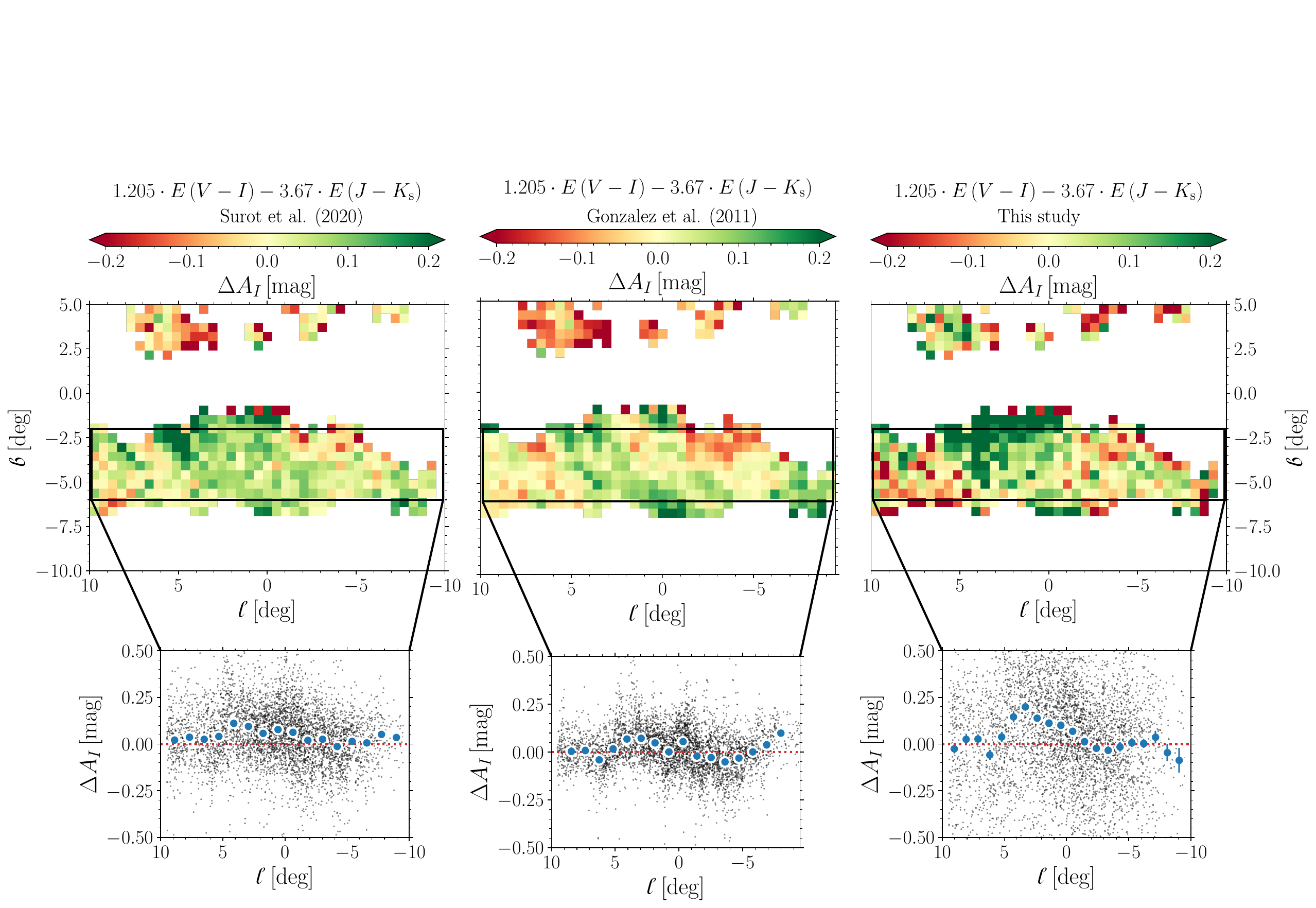}
\caption{Same as Figure~\ref{fig:AICompThisStudy} but for $E\left ( V - I \right )$ from \citet{Nataf2013}, and $E\left ( J - K_{\rm s} \right )$ estimated in this work, and from reddening maps from \citet{Surot2020} and \citet{Gonzalez2012}.}
\label{fig:AICompNataf}
\end{figure*}

\begin{figure*}
\includegraphics[width=2\columnwidth]{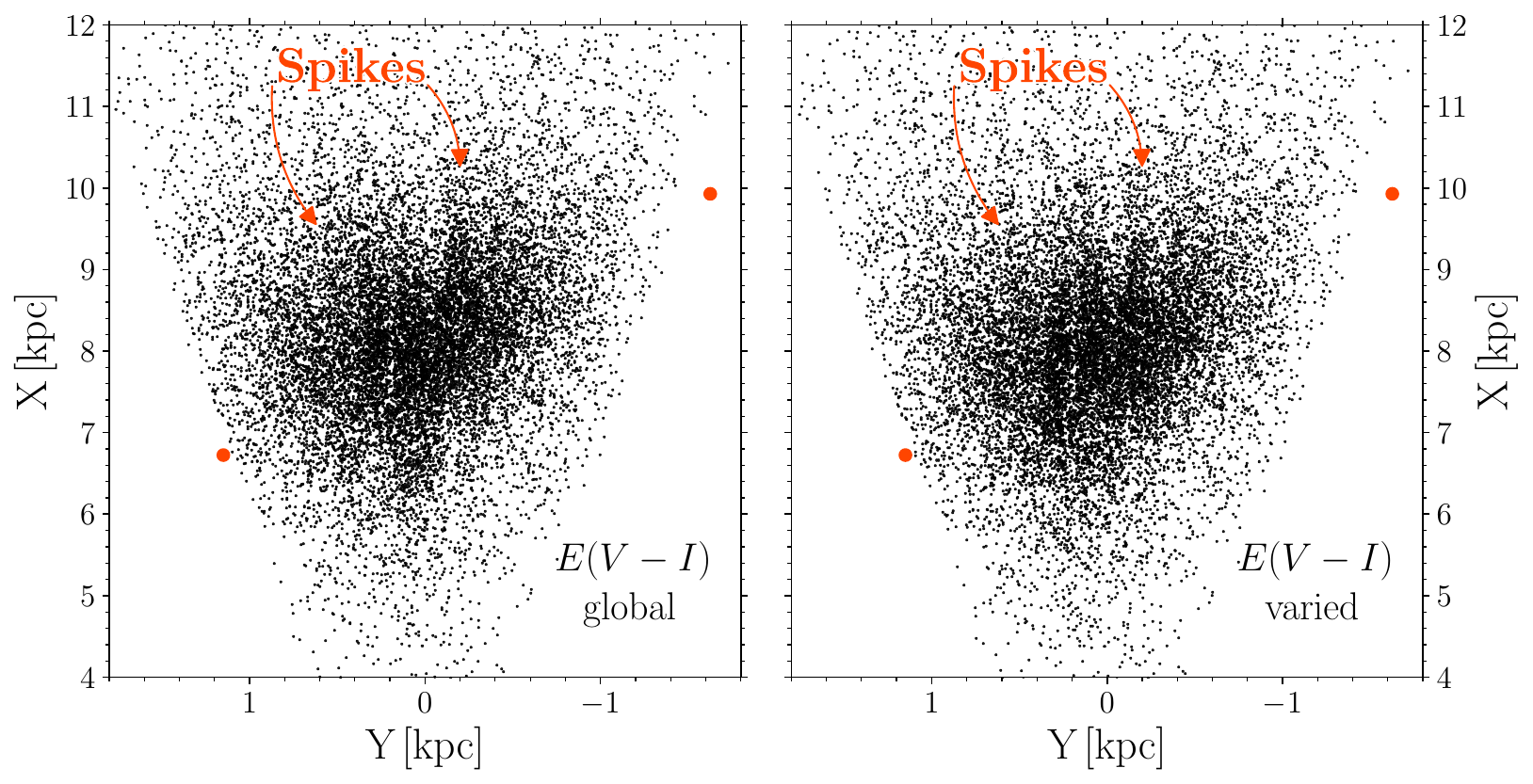}
\caption{The spatial distribution of RR~Lyrae pulsators in the Cartesian space for two versions of reddening law. The left and right-hand panels show spatial distribution with reddening determined through a single and varied (see Table~\ref{tab:ModifyRedLaw}) reddening law, respectively. We note that we used the same data set as for Figure~\ref{fig:SpatialComparisonCNG}. The two red dots represent the first and last position of the bar \citep{Gonzalez2011}.}
\label{fig:ReddLawModified}
\end{figure*}

\begin{figure*}
\includegraphics[width=2\columnwidth]{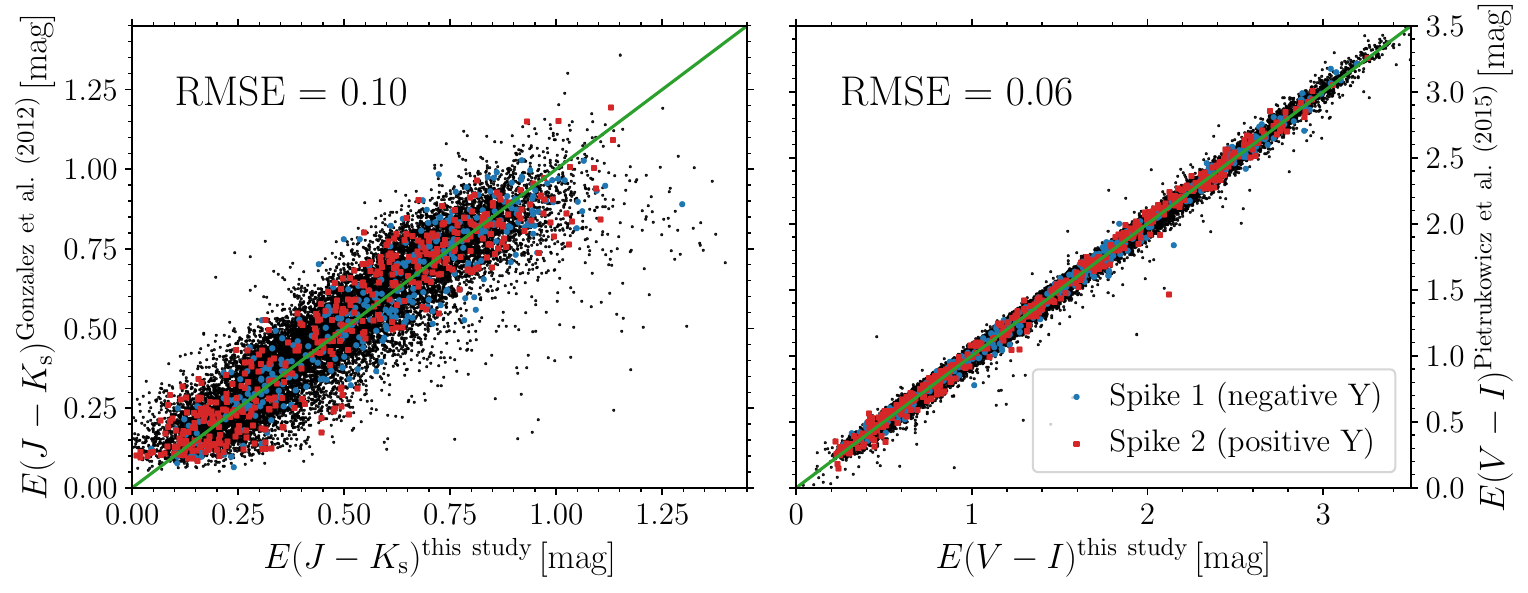}
\caption{Comparison of reddening $E(J-K_{\rm s})$ (left-hand panel) and $E(V-I)$ (right-hand panel) determined in this study and $E(J-K_{\rm s})$ vs $E(V-I)$ obtained based on procedure from \citet{Pietrukowicz2015}. The merged data set used in this Section is displayed with black dots. RR~Lyrae variables approximately located in the X and Y plane at the position of spikes (see bottom panels of Fig.~\ref{fig:SpatialComparisonCNG}) are marked with red and blue markers. The green line represents the identity line.}
\label{fig:ComparisonRedd}
\end{figure*}

\begin{figure*}
\includegraphics[width=2\columnwidth]{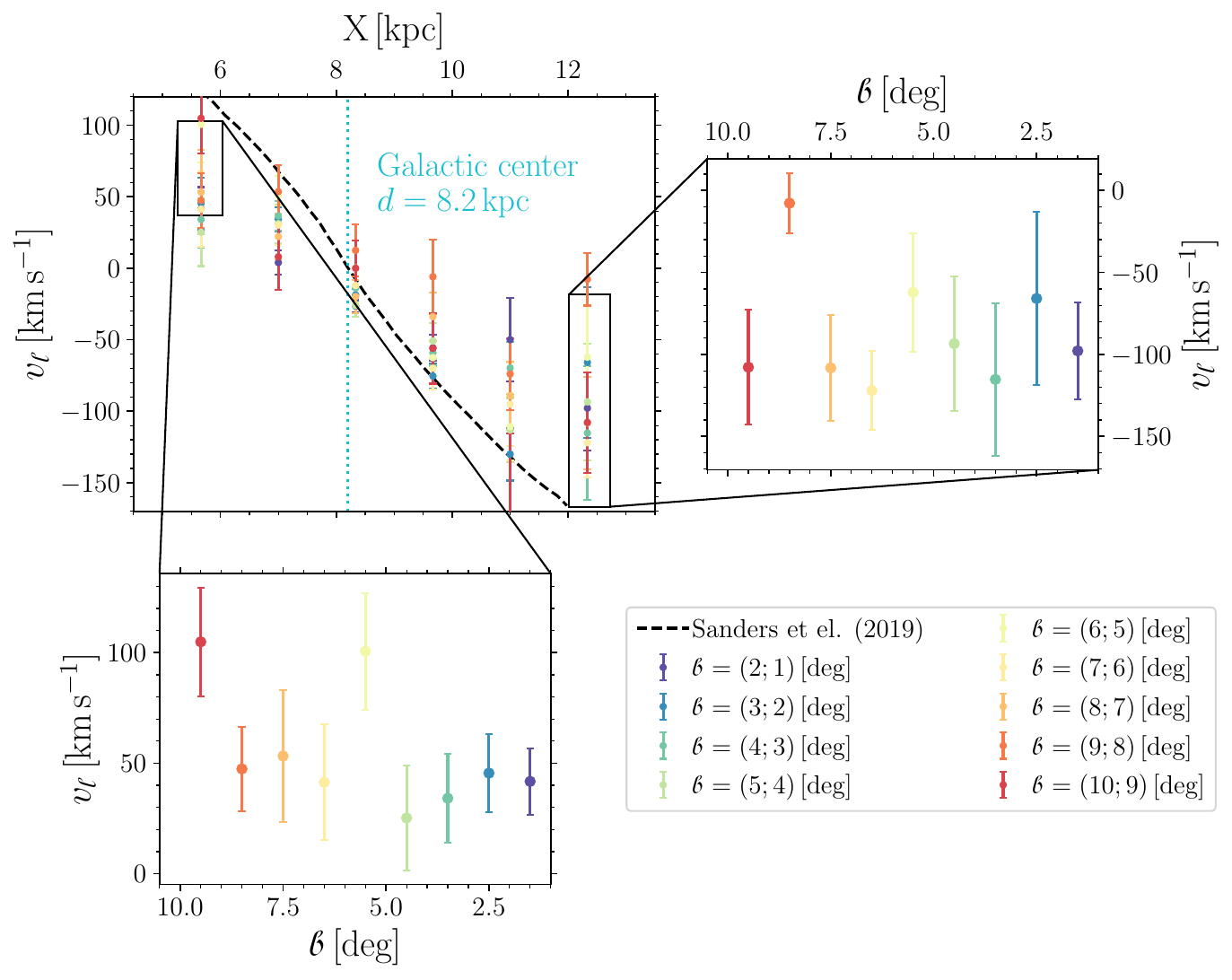}
\caption{Same as Figure~\ref{fig:RotationVtBelowPlane} but for region above the Galactic plane ($b$~$> 0$\,deg).}
\label{fig:RotationVtAbovePlane}
\end{figure*}

\end{appendix}
\end{document}